\documentclass[pra,aps, twocolumn, nobalancelastpage, superscriptaddress,nofootinbib, linenumbers]{revtex4-2}
\usepackage[utf8]{inputenc}
\usepackage{array, colortbl, xcolor} 
\usepackage[utf8]{inputenc}
\usepackage{bbold}
\usepackage{geometry}
\usepackage{float}
\usepackage{graphicx}
\usepackage{braket}
\usepackage[normalem]{ulem}
\usepackage[T1]{fontenc}
\usepackage{gensymb}
\usepackage{amsmath,amsfonts,amssymb}
\usepackage{amsmath}
\usepackage{amsthm}
\usepackage{xfrac}
\usepackage{comment}
\usepackage{standalone}
\usepackage{dashbox}
\usepackage{nicefrac}
\usepackage[colorlinks=true]{hyperref}
\hypersetup{citecolor=blue, urlcolor=blue, linkcolor=black}

\newcommand{\fst}{$\rm 1^{st}$ }
\newcommand{\scnd}{$\rm2^{nd}$ }

\newcommand{\be}{\begin{equation}}
\newcommand{\ee}{\end{equation}}
\newcommand{\ba}{\begin{aligned}}
\newcommand{\ea}{\end{aligned}}

\usepackage[utf8]{inputenc}
\usepackage[T1]{fontenc} 
\usepackage{microtype} 
\usepackage{graphicx}
\usepackage{dcolumn}
\usepackage{enumerate,amsthm,amssymb,color}
\usepackage{amsfonts}
\usepackage{amsmath}
\usepackage{mathtools}
\usepackage{dsfont}
\usepackage[separate-uncertainty=true]{siunitx} 
\usepackage{enumitem}  

\usepackage{subcaption}
\usepackage{ragged2e}
\DeclareCaptionJustification{justified}{\justifying}
\usepackage{bbm}
\usepackage{bm}
\usepackage{synttree}
\usepackage{float}
\usepackage{bbold}
\usepackage{braket}
\usepackage{xcolor}
\usepackage{gensymb}

\usepackage{booktabs}
\usepackage{upgreek, eurosym}

\newfloat{Box}{tbp}{lop}

\begin{document}



\nolinenumbers
\title{Asymmetric two-photon response of an incoherently driven quantum emitter}

\author{Lennart Jehle}\thanks{Address all correspondence to lennart.jehle@univie.ac.at}
\affiliation{University of Vienna, Faculty of Physics, Vienna Center for Quantum Science and Technology (VCQ), 1090 Vienna, Austria}
\affiliation{Christian Doppler Laboratory for Photonic Quantum Computer, University of Vienna, Faculty of Physics, 1090 Vienna, Austria}

\author{Lena M. Hansen}
\affiliation{University of Vienna, Faculty of Physics, Vienna Center for Quantum Science and Technology (VCQ), 1090 Vienna, Austria}
\affiliation{Christian Doppler Laboratory for Photonic Quantum Computer, University of Vienna, Faculty of Physics, 1090 Vienna, Austria}

\author{Patrik I. Sund}
\affiliation{University of Vienna, Faculty of Physics, Vienna Center for Quantum Science and Technology (VCQ), 1090 Vienna, Austria}
\affiliation{Christian Doppler Laboratory for Photonic Quantum Computer, University of Vienna, Faculty of Physics, 1090 Vienna, Austria}

\author{Thomas W. Sand{\o}}
\affiliation{University of Vienna, Faculty of Physics, Vienna Center for Quantum Science and Technology (VCQ), 1090 Vienna, Austria}
\affiliation{Christian Doppler Laboratory for Photonic Quantum Computer, University of Vienna, Faculty of Physics, 1090 Vienna, Austria}

\author{Raphael Joos}
\affiliation{Institut f\"ur Halbleiteroptik und Funktionelle Grenzfl\"achen, \\Center for Integrated Quantum Science and Technology (IQ\textsuperscript{ST}) and SCoPE, \\University of Stuttgart, Allmandring 3, 70569 Stuttgart, Germany}

\author{Michael Jetter}
\affiliation{Institut f\"ur Halbleiteroptik und Funktionelle Grenzfl\"achen, \\Center for Integrated Quantum Science and Technology (IQ\textsuperscript{ST}) and SCoPE, \\University of Stuttgart, Allmandring 3, 70569 Stuttgart, Germany}

\author{Simone L. Portalupi}
\affiliation{Institut f\"ur Halbleiteroptik und Funktionelle Grenzfl\"achen, \\Center for Integrated Quantum Science and Technology (IQ\textsuperscript{ST}) and SCoPE, \\University of Stuttgart, Allmandring 3, 70569 Stuttgart, Germany}

\author{Mathieu Bozzio}
\affiliation{University of Vienna, Faculty of Physics, Vienna Center for Quantum Science and Technology (VCQ), 1090 Vienna, Austria}

\author{Peter Michler}
\affiliation{Institut f\"ur Halbleiteroptik und Funktionelle Grenzfl\"achen, \\Center for Integrated Quantum Science and Technology (IQ\textsuperscript{ST}) and SCoPE, \\University of Stuttgart, Allmandring 3, 70569 Stuttgart, Germany}

\author{Philip Walther}
\affiliation{University of Vienna, Faculty of Physics, Vienna Center for Quantum Science and Technology (VCQ), 1090 Vienna, Austria}
\affiliation{Christian Doppler Laboratory for Photonic Quantum Computer, University of Vienna, Faculty of Physics, 1090 Vienna, Austria}
\affiliation{Institute for Quantum Optics and Quantum Information (IQOQI) Vienna, Austrian Academy of Sciences, Vienna, Austria}


\begin{abstract}

Quantum emitters promise to emit exactly one photon with high probability when pumped by a laser pulse.
However, even in ideal systems, re-excitation during a laser pulse causes the consecutive emission of two photons, thus limiting the single-photon purity.
Although the probability and properties of re-excitation are largely determined by the optical excitation method, until now only resonant driving has been studied.
Here, we demonstrate qualitative differences in the process arising from phonon-assisted excitation---a scheme standing out for its robustness and straightforward spectral suppression of scattered laser light while preserving highly indistinguishable emission.
In contrast to previous studies under resonant driving, we measure not only the $g^{(2)}(0)$ as a function of pulse length but also resolve the distinct temporal and spectral shape of each of the photons, report an asymmetric two-photon spectrum and uncover correlations between the emission time and wavelength, which are unique to phonon-assisted pumping.
On the fundamental side, we show how the spectrum stemming from re-excitation provides direct access to the Rabi frequency of an incoherently driven quantum dot.
On the application side, we use the asymmetric spectral response to selectively suppress multiphoton noise from re-excitation, ensuring a high-single photon purity regardless of the laser pulse length and thus enhancing implementations across quantum cryptography and quantum computing.

\end{abstract}

\maketitle

\noindent 
Studying the rich physics of light--matter interaction not only provides valuable insights into fundamental quantum processes but also inspires the design of applications that benefit from these processes. Among the variety of quantum emitters that allow such studies, quantum dots (QDs) have proven to be a powerful platform for foundational experiments in quantum optics \cite{JTC:NatPhys2022, TMN:NatPhys2023, TAD:Science2023, HGJ:arxiv2024} and applications in both quantum communication \cite{ZDL:PRL2025, SVN:arxiv2024, LRB:arxiv2024, VKD:arxiv2024, YLL:NNano2023} and computing \cite{CHG:PRL2024, CSK:NPhot2023, MCN:NComms2024, UMZ:NNano2021},
thanks to simultaneously high single-photon efficiencies \cite{DGX:NPhot2025}, low multiphoton contribution \cite{SJZ:APL2024, HF:npj2018} and highly indistinguishable emission \cite{TJN:NNano2021, HBK:arxiv2025}.
A large part of the success stems from the high level of control gained by embedding these artificial atoms into a semiconductor host structure, which allows the integration of photonic cavities to improve collection efficiencies and provide Purcell enhancement, as well as electrical diodes \cite{NPG:NComms2014, BPS:APL2010} to stabilize the charge environment and tune the emission wavelength.
However, the benefits of stable semiconductor integration also come with a coupling to the surrounding environment---most notably to charge noise, spin reservoirs, and the phonon bath.
When unmanaged, such interactions can impair the properties of the emitter, as evidenced by damped Rabi oscillations \cite{HBS:arxiv2024}, diminished photon indistinguishability \cite{TDF:PRL2018, LKD:PRAppl2018}, or reduced spin coherence times \cite{SGM:NComms2016, YTO:NNano2018}.
Yet, when understood and used to our advantage, these very interactions also offer opportunities, including spin refocussing \cite{ZSB:NNano2023}, control over the charge state \cite{BPS:APL2010}, or phonon-mediated schemes to incoherently excite the QD \cite{RWH:PRB2019}.

In recent years, longitudinal acoustic (LA) phonon-assisted pumping \cite{GBA:PRL2013} has attracted significant interest \cite{VJN:APL23, LHS:arxiv2025, BVN:npj2022, MPB:arxiv2024} due to its intrinsic robustness against laser instabilities \cite{VJN:APL23} and the straightforward suppression of laser background via spectral filtering.
The latter renders LA excitation ideal for generating polarization-entangled linear cluster states from QDs \cite{LHS:arxiv2025, CFB:NPhot2023}, while applications in quantum communication additionally benefit from the phase randomization required in many cryptographic protocols and provided by the coupling to the phonon bath \cite{BVN:npj2022}.


\begin{figure*}
	\begin{center}
		 \includegraphics[width=178mm]{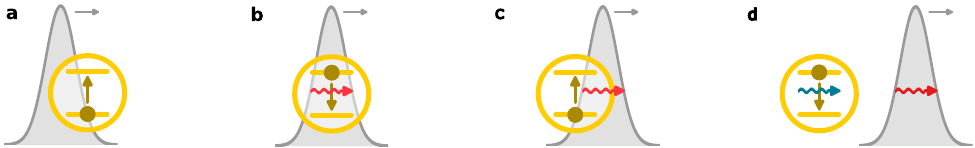}
		\caption{\textbf{Schematic time series of a re-excitation event.} Modelling the QD as a two-level system, the re-excitation process from a single laser pulse is described using a representative quantum trajectory, broken down into four distinct events.
        \textbf{(a)} Early during the pulse interaction with the laser pulse, the QD is excited and \textbf{(b)} decays shortly after, emitting a \fst single photon (red arrow). 
        \textbf{(c)} Towards the end of the interaction with the laser pulse, the QD is excited a second time. 
        \textbf{(d)} The laser pulse and the \fst photon have passed, when the QD spontaneously decays again and emits the \scnd photon (blue arrow).
        Thus, the \fst and \scnd photon are emitted consecutively in a strict time-ordered manner.}
		\label{fig1}
	\end{center}
\end{figure*}


As a result of continuous improvements in QD fabrication, fundamental challenges---rather than technical imperfections---are now beginning to limit the achievable performance \cite{SSB:PRL2022, SSH:PRL2020}.
Among these, the impact of re-excitation, where a single laser pulse may induce the emission of multiple photons, is particularly severe as it compromises the single-photon purity of the source.
In resonantly driven two-level systems, the re-excitation probability scales proportionally to the laser pulse length and the radiative decay rate \cite{HF:npj2018}.
However, except for initial theoretical work that focussed on adiabatic undressing \cite{CUC:PRL2019}, the phenomenon is largely unexplored for phonon-assisted pumping. 
At the same time, a comprehensive understanding of the re-excitation dynamics is becoming increasingly important as recent cavity designs yield Purcell factors exceeding 25 \cite{RVH:ACSPhot2025}, leading to emitter decay times $<\SI{30}{\ps}$.
While the accelerated decay improves the photon indistinguishability and supports faster clock rates, it also increases the multiphoton emission caused by re-excitation.


Here, we present experimental results exploring both the temporal and spectral signatures of the re-excitation process under LA phonon-assisted pumping.
We individually resolve the time and frequency modes of the first and second emitted photon and reveal significant differences compared to the resonant counterpart.
Importantly, we observe a spectral shift for the first photon that is unique to phonon-assisted driving and can be interpreted as dynamic Stark tuning caused by the excitation laser.
The shift allows us to assess the Rabi frequency of an incoherently driven QD. 
Moreover, we demonstrate how these Stark-induced time-frequency correlations can be exploited to limit the multiphoton contribution irrespective of the laser pulse length only using spectral filtering.





\begin{figure*}
	\begin{center}
		 \includegraphics[width=178mm]{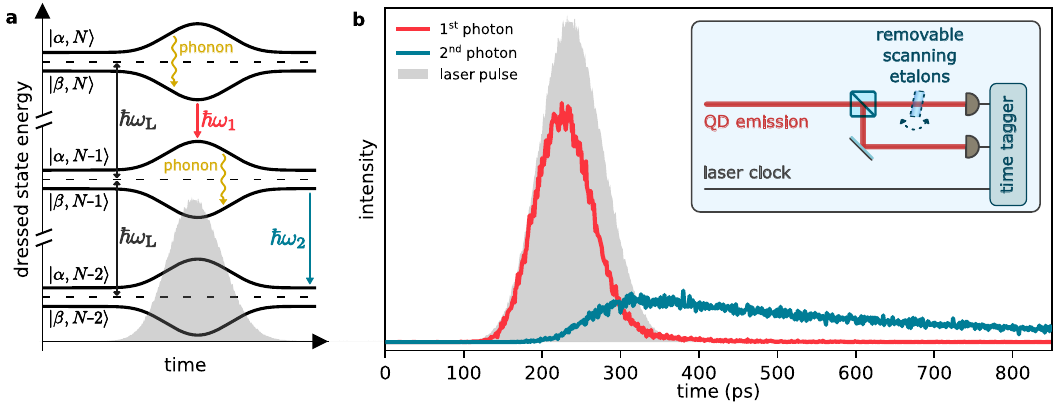}
		\caption{\textbf{Re-excitation under phonon-assisted driving.}
        \textbf{(a)} For one exemplary quantum trajectory, the re-excitation process under LA driving is schematically depicted in the energy diagram of the dynamical dressed states reduced to three manifolds of the laser field.
        The laser pulse is shown in gray.
        The emission of a first phonon sets the system into the lower dressed state $\ket{\beta, N}$ and, shortly after, but when the system is still dressed, a \fst photon is emitted at $\omega_1$ by decaying into $\ket{\alpha, N-1}$. 
        Thereafter, another phonon emission brings the system into the lower dressed state of this manifold ($\ket{\beta, N-1}$) which is adiabatically transformed into the exciton state once the laser pulse exits. 
        The emission of a \scnd photon at $\omega_2$ completes the process.
        The phonon coupling efficiency is governed by the phonon spectral density evaluated at the instantaneous Rabi frequency \cite{BLV:PRB2016}.
        \textbf{(b)} Time-histogram of the \fst and \scnd photon showing the different temporal shapes.
        The laser pulse in gray is displayed for time reference. Inset: schematic detection setup in HBT configuration, additionally equipped with the electronic laser clock for advanced correlations.
        The scanning etalons indicated with dashed lines are only used for the spectrally resolved measurements.}
		\label{fig2}
	\end{center}
\end{figure*}

\bigskip
\begin{center}
    \textbf{RESULTS}\\
\end{center}

\noindent\textbf{Re-excitation under phonon-assisted driving}\\
The re-excitation process is conceptually explained in Fig.~\ref{fig1} for one exemplary quantum trajectory.
Independent of the pumping scheme, when driving the QD with a short laser pulse, there is a chance that it is excited early during the interaction (a) and quickly decays by emitting a \fst photon (b) such that the QD can be excited again by the same pulse (c) and emit a \scnd photon at a later time (d).
The causality of the process requires that whenever two photons were created from one laser pulse, the \fst photon must have been emitted while the pulse was still interacting with the QD (by referring to the \fst or \scnd photon, we imply that indeed two photons were detected).
Following this reasoning, temporal filtering has been proposed to restore a high single-photon purity \cite{KHR:npjQI2020, GZJ:arxiv2025} but is limited to post-selection, because the fast time scales are infeasible for current electro-optic modulators.

To understand re-excitation in the phonon-assisted scheme, we briefly recall the dressed-state picture, where the QD is treated as a two-level system with transition frequency $\omega_0$, and the laser frequency $\omega_L$ is detuned by $\delta\omega_L=\omega_L - \omega_0$ (here, we limit ourselves to $\delta\omega_L>0$).
The interaction of the QD with the external laser field is captured by a light--matter Hamiltonian in the rotating frame, which gives rise to a new set of eigenstates, the so-called dressed states $\ket{\alpha, N}$ and $\ket{\beta, N}$, where $N$ denotes the number of laser photons \cite{CR:JPhysB1977}.
At the beginning of the pulse, the ground state coincides with $\ket{\alpha, N}$ but as the laser field increases, it admixes an excitonic (ground state) component to $\ket{\alpha, N}$ ($\ket{\beta, N}$), and thereby activates phonon interactions between the dressed states \cite{QBL:PRL2015}.
When the level splitting of the dressed states---determined by the effective Rabi frequency $\rm\Omega_{eff}(t)=\sqrt{\Omega^2(t) + \delta\omega_L^2}$---matches the spectral phonon density, $\ket{\beta, N}$ can be efficiently populated through phonon emission \cite{GBA:PRL2013}.
As the laser pulse exits the system, $\ket{\beta, N}$ is adiabatically transformed back to the exciton state, which completes population inversion.
Recombination at a later time then produces a single photon at $\omega_0$ \cite{QBL:PRL2015}.
However, if a photon is emitted while the laser pulse is still present, we must consider the transition $\ket{\beta, N}\rightarrow\ket{\alpha, N-1}$ in the dressed-state picture.

The complete LA re-excitation process is schematically outlined in Fig.~\ref{fig2}a and we note two major differences compared to previously studied resonant systems. 
First, the excitation requires not only the absorption of a laser photon but also the emission of a phonon, which (depending on the phonon decay rate) might delay the excitation and thereby lower the re-excitation probability.
Second, the eigenenergies of $\ket{\alpha}$ and $\ket{\beta}$ change antisymmetrically with the instantaneous laser field strength such that the transition frequency inherits the time dependence of the driving pulse 
\begin{equation}
    \omega_\text{QD}(t) = \omega_L - \Omega_\text{eff}(t)\;.
\label{eq:emission_freq}
\end{equation}
This time-varying frequency shift is reminiscent of the dynamic Stark effect \cite{UML:PRL2004}, where a laser pulse alters the energy structure of a few-level system. 
However, whereas in previous works the Stark shift was purposefully induced using a secondary pump laser \cite{UML:PRL2004, MFL:PRL2008, MFL:PRL2009}, in the LA scheme the excitation laser itself causes the red-shift of the QD emission. 

\bigskip
\noindent\textbf{Temporal and spectral signatures of re-excitation}\\
\begin{figure}
	\begin{center}
    	\includegraphics[width=83mm]{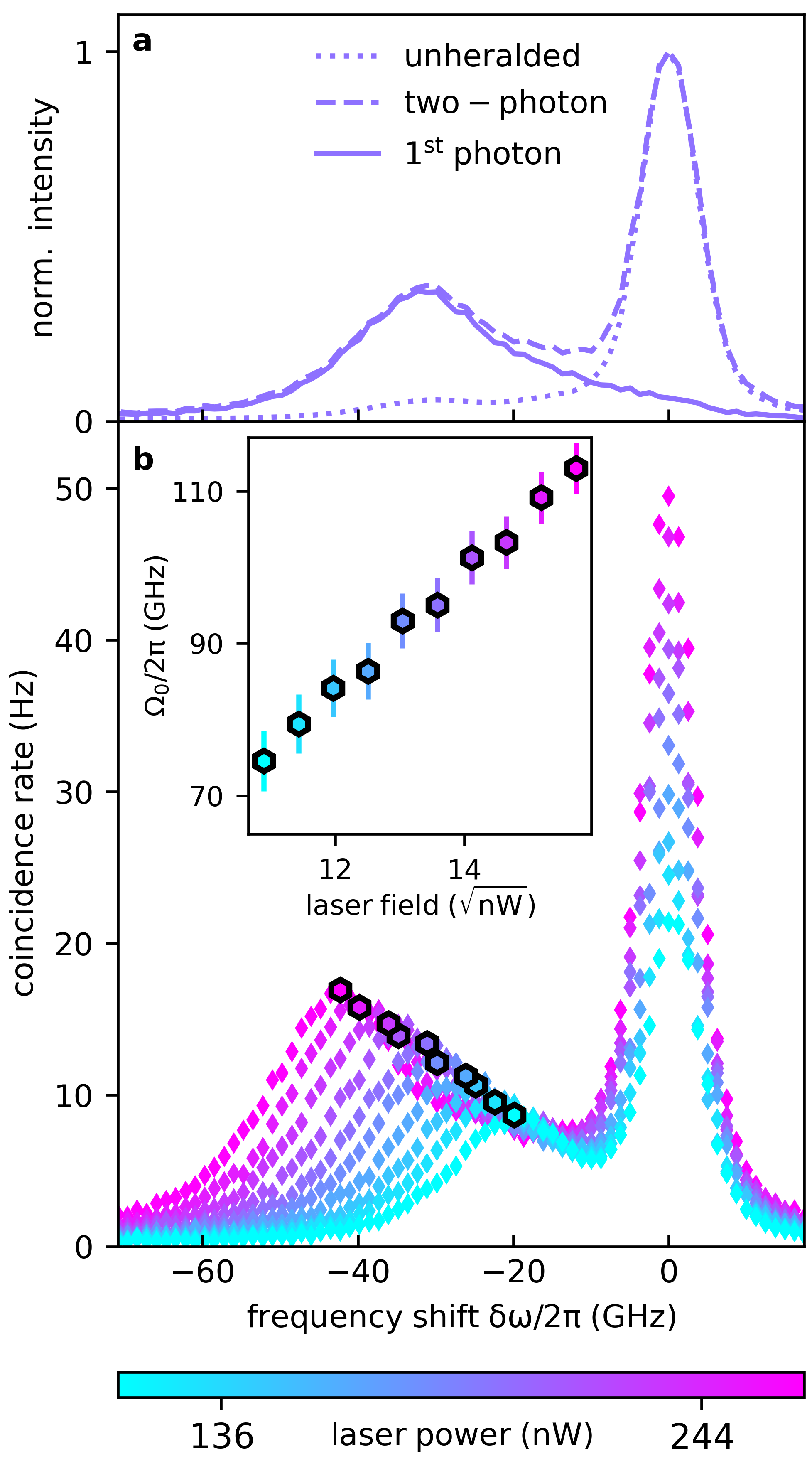}
        \caption{
        \textbf{Re-excitation spectrum for varying pump power.}
        A scanning frequency filter ($\Delta \omega_f/2\pi=\SI{5.3\pm0.3}{\GHz}$ is used to measure the QD spectrum based on different gating logics. 
        Unheralded: no gating, all events detected on the spectrally filtered channel.
        Two-photon: all events on the spectrally filtered channel coinciding with a heralding photon within $\Delta\tau_c=\SI{3}{\ns}$.
        \fst photon: the heralding event is used only if it is registered $>\SI{350}{\ps}$ after the laser clock.
        The laser pulse length is $\Delta t_L=\SI{80\pm1}{\ps}$.
        The displayed data is broadened by the transfer function of the frequency filter.
        \textbf{(a)} The laser power is set to $\SI{185}{\nW}$ and individual data points have been omitted for visual clarity.
        \textbf{(b)} The laser power is varied (see color bar) and the two-photon spectrum (as above) is measured.
        The spectral position of the side peak is marked for every power and used to calculate the Rabi frequency according to Eq.~\ref{eq:rabi0}, which is shown in the inset as function of the laser field.}
        \label{fig3}
	\end{center}
\end{figure}
In the experiment, we use an epitaxially grown In(Ga)As QD emitting in the telecom C-band, that is embedded in a circular Bragg grating cavity with wide-band enhancement and excited using a blue-detuned laser with adjustable pulse length. 
The QD transition is attributed to a charged exciton with a lifetime of $\tau_{QD}=\SI{465\pm1}{\ps}$, indicating a Purcell factor of $\SI{4.3\pm0.8}{}$ \cite{JBK:NanoLett24}.
More details on the sample and laser preparation can be found in the Methods. 


Starting with the temporal analysis of the re-excitation signal, we excite the QD with $\SI{80\pm1}{\ps}$ long Gaussian laser pulses, suppress the scattered laser and isolate the QD line from broadband background noise.
The emission is detected in a Hanbury-Brown and Twiss (HBT) configuration and correlated with the laser clock (see inset Fig.~\ref{fig2}b).
Instead of performing a standard second-order autocorrelation $g^{(2)}(\tau)$, we check for coincidences between the two optical channels and record the timestamp of the early and late click for each coincidence. 
Sorting this way, we separate events caused by \fst photons from those caused by \scnd photons, allowing us to resolve the dynamics of the two-photon emission individually.
The resulting histograms, displayed in Fig.~\ref{fig2}b, represent the temporal shape of the \fst and \scnd photon and thus provide the first direct experimental confirmation of the temporal ordering associated with the re-excitation process.
Since the \fst photon must be emitted when the laser is still present, its temporal profile strongly resembles that of the laser. 
The \scnd photon, on the other hand, can be emitted during the full lifetime which leads to the Gaussian-broadened exponential decay expected from spontaneous emission.
Even though the probability distributions overlap in time, two photons originating from one laser pulse are always emitted into orthogonal time modes (see Supplementary Information) and only averaging over many repetitions induces the statistical overlap.

Turning to the spectral analysis, we employ a tunable filter that enables us to frequency-resolve either the overall emission or, in a heralded manner, the two-photon signal (see inset Fig.~\ref{fig2}b and Methods).
Recording the count rate of the filtered path while scanning the filter frequency reconstructs the spectrum averaged over all photons, whereas using the second detector as herald post-selects on two-photon events, effectively measuring the two-photon spectrum (see Fig.~\ref{fig3}a).
The unheralded data features a prominent peak at $\omega_0$ representing undisturbed single-photon emission.
However, because of the length of the laser pulses, the re-excitation effect is already noticeable without any heralding and shows itself in the broad shoulder on the low-frequency side, which we attribute to the \fst photon.
In the heralded configuration, where the spectra of the \fst and \scnd photon contribute equally, the shoulder then becomes a second but broader peak.
To further support the assertion of the low-frequency peak to the \fst photon, we refine the gating logic and ignore all heralding events occurring before $t=\SI{350}{\ps}$  (compare Fig.~\ref{fig2}b), that is, all photons emitted within the laser pulse duration. 
Counting coincidences in this way ensures that the photon in the filtered path must have been the \fst photon.
The resulting spectrum overlaps perfectly with the broad side peak observed in the two-photon spectrum (see Fig.~\ref{fig3}a). 
The increased bandwidth of the \fst photon is a consequence of its narrower temporal shape (recall Fig.~\ref{fig2}b) and is also predicted for resonant driving \cite{FHW:NPhys2017}.
The shift from the natural frequency $\omega_0$, on the other hand, is a signature of the transition between the dressed states $\ket{\beta, N}$ and $\ket{\alpha, N-1}$ and is unique to re-excitation under phonon-assisted pumping.
Note that the cavity's influence on these measurements is minimal, since its mode is much wider than the observed frequency shift. 

\bigskip
\noindent\textbf{Measuring the Rabi frequency for incoherent excitation}\\
Following Eq.~\ref{eq:emission_freq}, we can express the maximum frequency shift as
$\delta\omega_\text{max} =  \delta\omega_L - \sqrt{\Omega_0^2+\delta\omega_L^2}$.
As expected from the dynamic Stark effect, the shift increases for stronger laser--QD interactions, that is, larger external fields or less detuning.
Rearranging the equation as
\begin{equation}
    \Omega_0 = \sqrt{\delta\omega_\text{max}(\delta\omega_\text{max}-2\delta\omega_L)}
\label{eq:rabi0}
\end{equation}
yields a remarkable result: we can infer the Rabi frequency---fundamental to coherent light--matter interaction in two-level systems---from an incoherent process.
Instead of relying on Rabi rotations or coherent two-photon scattering (the origin of the Mollow-triplet \cite{Mollow:PRL1969}), here we exploit the signatures of the phonon-assisted re-excitation spectrum.

\begin{figure*}
	\begin{center}
		 \includegraphics[width=178mm]{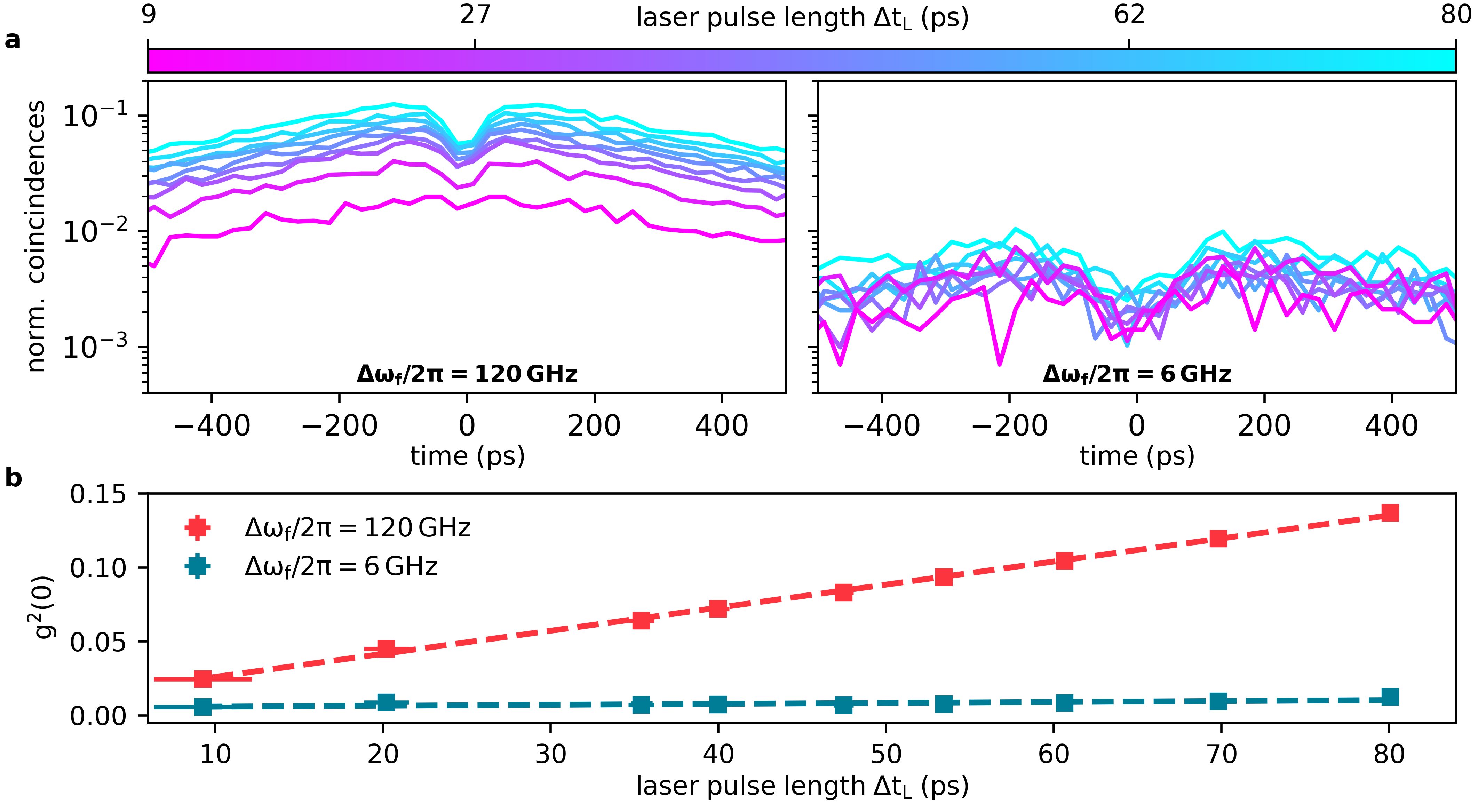}
		\caption{\textbf{Reducing the multiphoton contribution by frequency filtering.}
        For a constant laser power of $\SI{185}{\nW}$, the pulse length is varied and the second-order autocorrelation $g^{(2)}(\tau)$ is measured for each step.
        \textbf{(a)} The zoomed-in central peak of the coincidence counts normalized to Poissonian level is shown as function of the delay $\tau$ for a $\SI{120\pm1}{\GHz}$ VBG filter (left panel) and a $\SI{6.0\pm0.3}{\GHz}$ etalon (right panel).
        \textbf{(b)} The $g^{(2)}(0)$ is evaluated by dividing the area of the correlated peak by that of the averaged uncorrelated peaks.
        The error bars are derived from Poissonian statistics and are too small to be visible. 
        The uncertainty in laser pulse length arises from the deconvolution with the system response function of $\SI{26\pm1}{\ps}$ (see Methods).
        The linear fit $g^{(2)}(0,\Delta t_L)=A\cdot\Delta t_L/\tau_{QD}$ yields $\rm A_{120GHz}=0.72\pm0.01$ and $\rm A_{6GHz}=0.03\pm0.01$.}
		\label{fig4}
	\end{center}
\end{figure*}

To experimentally verify these findings, we measure the two-photon spectrum for increasing laser power.
Our results, presented in Fig.~\ref{fig3}b, clearly demonstrate that the red-shift of the \fst photon intensifies with power, while the spectral position of the \scnd photon is unchanged.
In the parameter regime studied here, and in agreement with the dressed-state model, the shift increases approximately linearly with the external laser field.
Extracting $\Omega_0$ from the induced red-shift according to Eq.~\ref{eq:rabi0} and plotting it against the applied laser field shows an excellent linear trend (see inset Fig.~\ref{fig3}b), confirming our understanding of the process.
Note that coherent two-photon scattering, as recently shown also for pulsed excitation \cite{BKB:PRL2024, LGL:NPhot2024}, would produce a similar, but importantly, symmetric spectral response.
And although both processes could coexist, we show in the Supplementary Information that for the used laser parameters, efficient phonon coupling is ensured and the incoherent process dominates.

Lastly, we stress that the spectral overlap of the two photons reduces with increasing laser power.
This is particularly helpful when frequency filtering the QD emission to reduce the multiphoton contribution, a method we introduce in the next section.

\bigskip
\noindent\textbf{Restoring high single-photon purities}\\
Recalling that for applications of QDs as single-photon sources the presence of an additional \fst photon diminishes the single-photon purity, we now leverage the strong spectro-temporal correlations found in Fig.~\ref{fig3} to our advantage. 
Using a frequency filter slightly wider than the spontaneous emission line at $\omega_0$, we expect to greatly suppress the \fst photon.
To ensure that we do not simply reduce the multiphoton noise caused by other spectrally distinct processes (such as neighboring QDs), we vary the laser pulse length across an order of magnitude while maintaining a constant laser power of $\SI{185}{\nW}$.
In this way, we tune the re-excitation probability but leave other processes unaffected.

Fig.~\ref{fig4}a compares the zoomed-in, correlated peaks around $\tau=0$ of a standard $g^{(2)}(\tau)$ for various pulse lengths.
The QD signal shown in the left panel is only isolated from broadband background, but for the measurements displayed in the right panel a $\SI{6\pm0.3}{\GHz}$ etalon centered on $\omega_0$ is added.
Unfiltered, the multiphoton probability clearly increases for longer pulses and the $g^{(2)}(\tau)$ also features the characteristic re-excitation dip at $\tau=0$.
The depth is limited by the system response function of the detection stage (see Methods).
On the other hand, the filtered emission exhibits a drastically reduced multiphoton response and, importantly, is independent of the laser pulse length. 
Only for the longest pulse, a slight increase is notable. 
Since we maintain a constant laser power throughout the scan, the effective Rabi frequency decreases for longer pulses as $\Omega_\text{eff}\propto1/\sqrt{\Delta t_L}$, which results in a smaller red-shift of the \fst photon and therefore in a larger spectral overlap with the spontaneous emission line.
Interestingly, we also observe a (significantly wider) dip in the $g^{(2)}(\tau)$ for the filtered signal, which could be caused by charge-carrier refilling \cite{OKB:APL2017} or some residual transmission of the \fst photon.
This attribution is supported by time-resolved measurements (see Supplementary Information), which show that \fst photons transmitted through the etalon are emitted at the very beginning of the laser pulse (when the dynamic frequency shift is still small), resulting in a larger delay between the \fst and \scnd photon.

The quantitative comparison of the unfiltered and filtered multiphoton probability, displayed in Fig.~\ref{fig4}b, demonstrates the stark difference in $g^{(2)}(0)$ when the laser pulse length is tuned.
The multiphoton component for the filtered emission is virtually unaffected for all pulse durations studied here, reaching more than $\nicefrac{1}{6}$ of the QD decay time.
On the other hand, the $g^{(2)}(0)$ of the unfiltered signal increases linearly with $\Delta t_L$, indicating that re-excitation dominates the multiphoton probability, which is in good agreement with work on resonantly driven two-level systems \cite{HF:npj2018}.

Finally, we note that the multiphoton response related to re-excitation scales with $\Delta t_L/\tau_{QD}$, whereas the efficiency of the proposed filtering method depends on the frequency shift of the \fst photon, which increases for shorter pulses (see above).
Consequently, the filtering of multiphoton noise becomes more effective, benefiting sources with fast decay times in particular.

\bigskip
\begin{center}
    \textbf{DISCUSSION}\\
\end{center}


We have presented an experimental study that provides nuanced understanding of the re-excitation dynamics and, at the same time, constitutes the first work on the phenomenon under phonon-assisted driving.
In addition to verifying the strict temporal ordering of the emission predicted by theory \cite{FHW:NPhys2017}, we find strong spectro-temporal correlations for the \fst emitted photon, which are induced by the dynamic Stark effect and lead to an asymmetric two-photon spectrum unique to phonon-assisted pumping.
The implications of our results are both fundamental and practical.
We have shown that the Rabi frequency of an incoherently driven QD can be extracted from the re-excitation spectrum, and on the application side, we have demonstrated how the spectral distinguishability of the two photons can be leveraged to efficiently remove multiphoton noise.

As advancements of high-Purcell cavities continue to reduce QD lifetimes, managing the re-excitation signal with conventional methods becomes increasingly challenging, and the frequency filtering introduced here represents a compelling new approach.
In applications, the laser pulses are significantly shorter than those used in this fundamental study, inducing a larger spectral shift, which allows for higher transmission without compromising the single-photon purity.
Moreover, our findings reveal an additional intrinsic advantage of phonon-assisted driving when used with narrow-band cavities.
During the laser--QD interaction, the dynamic frequency shift detunes the QD transition from the cavity mode, effectively canceling the Purcell enhancement.
This mechanism delays photon emission until the laser pulse subsides and greatly suppresses re-excitation.


We have shown that a deeper understanding of phonon-assisted excitation offers unique advantages for applications; however, other key aspects of the QD--phonon interactions are yet to be explored.
In particular, the impact of the phonon spectral density, which could be tailored \cite{RPZ:NNano2023} using phononic cavities, presents a promising research direction that can create new opportunities in quantum communication and photonic quantum computing.

During the submission process, we became aware of related work focusing on re-excitation under resonant excitation \cite{SCS:arxiv2025}.

\bigskip
\begin{center}
    \textbf{METHODS}\\
\end{center}

\noindent\textbf{Quantum dot sample}\\
The single-photon source is based on an epitaxially grown In(Ga)As quantum dot coupled to a circular Bragg grating cavity with a FWHM of $\SI{\approx1250}{\GHz}$.
Starting from a GaAs substrate, a non-linear metamorphic buffer layer \cite{SNK:Nanophot2022} is deposited.
Reducing the lattice mismatch to the subsequently grown InAs QD layer enables emission in the telecom C-band.
In a next step, circular Bragg grating cavities are etched into the sample in a non-deterministic process.
More details on the fabrication and the exact geometry of the final structure can be found in \cite{NJK:AdvQTech2023} and the corresponding supplementary material, following the same design principles.
The QD transition is attributed to a charged exciton that features highly polarized emission ($>95\%$) \cite{JBK:NanoLett24} at $\lambda\,=\,\SI{1552.25\pm0.03}{\nm}$.
Under optimized LA phonon excitation conditions, we achieve a polarized and fiber-coupled single-photon generation efficiency of $\SI{5.7\pm0.2}{\percent}$ at $g^{(2)}(0)=0.013\pm0.001$.
The spontaneous emission line is subject to spectral diffusion and not Fourier-transform limited.
The inhomogeneous broadening results in a predominantly Gaussian line shape.
The measurement data used to infer the lifetime and cavity mode width can be found in the Supplementary Information.

\bigskip
\noindent\textbf{Temporal measurements and autocorrelation}\\
The system response function (SRF) characterizes how the measurement apparatus blurs the detection time of an infinitely short pulse.
In our setup, the jitter introduced by the superconducting nanowire single-photon detectors (SNSPDs, Single Quantum EOS) dominates compared to the time-tagging device (Swabian Time Tagger X).
From a correlation between the electrical laser clock and an SNSPD channel exposed to an attenuated $\SI{1.8}{\ps}$ long laser pulse, we measure the SRF as full width at half maximum (FWHM), yielding $\SI{26\pm1}{\ps}$ for the first and $\SI{29\pm1}{\ps}$ for the second channel.
The timing jitter for an autocorrelation between both optical channels is $\SI{39\pm2}{\ps}$.

From the second-order autocorrelation functions $g^{(2)}(\tau)$ shown in Fig.~\ref{fig4}a we evaluate the multiphoton probability $g^{(2)}(0)$ presented in Fig.~\ref{fig4}b by dividing the area of the peak around $\tau=0$ by the averaged area of the neighboring peaks at $\tau=\pm\SI{13.17}{\ns}$.
Due to electronic or optical signal reflections observed at $\approx\pm\SI{5}{\ns}$ that should not be included in the multiphoton estimation, we use an integration window of $\SI{6.58}{\ns}$, which corresponds to half the repetition interval of the experiment. 
We note that this integration time still captures $>\SI{99.9}{\percent}$ of the emitted photons.

\bigskip
\noindent\textbf{Laser pulse generation}\\
Starting from $\SI{9\pm3}{\ps}$ long approximately Gaussian laser pulses generated at a repetition rate of $\SI{75.95}{\MHz}$ by a fiber-based laser (Pritel UOC), we use a tuneable width frequency filter (EXFO XTM-50) to reduce the spectral width, thereby stretching the pulse.
The smallest possible bandwidth of the filter is measured using a tuneable cw-laser (Santec TSL550) and is $\SI{5.7\pm0.3}{\GHz}$.
For each bandwidth setting, the resulting pulse length is measured by sending a strongly attenuated pulse to the SNSPD and deconvolving the detected signal using the SRF.

\bigskip
\noindent\textbf{Spectral measurements}\\
The QD emission is passed through a volume Bragg grating (VBG) that reflects the scattered laser light with an FWHM of $\SI{120\pm1}{\GHz}$ and OD6. 
A second identical VBG, centered at $\approx\SI{-40}{\GHz}$ with respect to the spontaneous emission frequency $\omega_0$, selects the QD transition under study and isolates it from any remaining broadband background. 

Requiring low-dark count noise and a gating logic, the analysis setup is based on a scanning frequency filter and photon detection using SNSPDs and a time tagger. 
The scanning filter consists of a cascaded pair of etalons with a free spectral range (FSR) of $\SI{125}{\GHz}$ ($\SI{292}{\GHz}$) and FWHM of $\SI{5.5\pm0.3}{\GHz}$ ($\SI{16.4\pm0.3}{\GHz}$).
Using piezomotor-controlled angle tuning, the resonance frequencies are overlapped and then jointly scanned. 
The spectral resolution of the setup of $\SI{5.3\pm0.3}{\GHz}$ is dominated by the narrow-band etalon. 
The two-photon spectrum is measured by correlating (coincidence window $\Delta\tau_c=\SI{3}{\ns}$) the filtered and unfiltered channel while tuning the central frequency stepwise. 
In this way, plotting the coincidence count rate as a function of the filter position reconstructs the two-photon spectrum.
The \fst photon spectrum is acquired in the same way, but the heralding channel is additionally gated in postselection, such that only clicks registered within $\SI{350}{\ps}<t<\SI{3000}{\ps}$ after the laser clock signal are accepted and used for the subsequent correlation with the filtered optical channel.

\bigskip
\begin{center}
    \noindent{\bf AUTHOR CONTRIBUTIONS}\\
\end{center}
L.J. conceived the project and designed the experimental setup. L.J. performed the measurements with participation of T.W.S. The data was processed and analyzed by L.J. with input from L.M.H and P.I.S.
M.B. and P.W. supervised the project. The quantum dot sample was provided and characterized by R.J., M.J., S.L.P. and P.M. Finally, the manuscript was written by L.J, with input from all other authors. \\

\begin{center}
    \noindent\textbf{ACKNOWLEDGMENTS}\\
\end{center}
\noindent
The Authors would like to thank Paul Hagen for discussions regarding the phonon spectral density and Michal Vyvlecka for the early design of the QD excitation setup.

L.J., L.M.H, P.I.S, T.W.S, M.B. and P.W. acknowledge funding from the European Union’s Horizon 2020 and Horizon Europe Research and Innovation programme under Grant Agreement and No. 899368 (EPIQUS), from the Marie Skłodowska-Curie Grant Agreement No. 956071 (AppQInfo).
R. J., M.J., S. L. P. and P.M. greatly acknowledge the German Federal Ministry of Education and Research (BMBF) via the Project QR.X (No.16KISQ013) and QR.N (16KIS2207).
Funding was also provided via the project EQSOTIC. (QuantERA II Programme that received funding from the EU’s H2020 research and innovation programme under the GA No 101017733, and respective BMBF project number 16KIS2060K).
The financial support by the Austrian Federal Ministry for Digital and Economic Affairs, the National Foundation for Research, Technology and Development and the Christian Doppler Research Association is gratefully acknowledged.\\

\begin{center}
    \noindent{\bf COMPETING INTERESTS}\\
\end{center}
The authors declare no conflict of interest.\\




\nolinenumbers
\bibliography{References}

\begin{thebibliography}{8}%
\makeatletter
\providecommand \@ifxundefined [1]{%
 \@ifx{#1\undefined}
}%
\providecommand \@ifnum [1]{%
 \ifnum #1\expandafter \@firstoftwo
 \else \expandafter \@secondoftwo
 \fi
}%
\providecommand \@ifx [1]{%
 \ifx #1\expandafter \@firstoftwo
 \else \expandafter \@secondoftwo
 \fi
}%
\providecommand \natexlab [1]{#1}%
\providecommand \enquote  [1]{``#1''}%
\providecommand \bibnamefont  [1]{#1}%
\providecommand \bibfnamefont [1]{#1}%
\providecommand \citenamefont [1]{#1}%
\providecommand \href@noop [0]{\@secondoftwo}%
\providecommand \href [0]{\begingroup \@sanitize@url \@href}%
\providecommand \@href[1]{\@@startlink{#1}\@@href}%
\providecommand \@@href[1]{\endgroup#1\@@endlink}%
\providecommand \@sanitize@url [0]{\catcode `\\12\catcode `\$12\catcode `\&12\catcode `\#12\catcode `\^12\catcode `\_12\catcode `\%12\relax}%
\providecommand \@@startlink[1]{}%
\providecommand \@@endlink[0]{}%
\providecommand \url  [0]{\begingroup\@sanitize@url \@url }%
\providecommand \@url [1]{\endgroup\@href {#1}{\urlprefix }}%
\providecommand \urlprefix  [0]{URL }%
\providecommand \Eprint [0]{\href }%
\providecommand \doibase [0]{https://doi.org/}%
\providecommand \selectlanguage [0]{\@gobble}%
\providecommand \bibinfo  [0]{\@secondoftwo}%
\providecommand \bibfield  [0]{\@secondoftwo}%
\providecommand \translation [1]{[#1]}%
\providecommand \BibitemOpen [0]{}%
\providecommand \bibitemStop [0]{}%
\providecommand \bibitemNoStop [0]{.\EOS\space}%
\providecommand \EOS [0]{\spacefactor3000\relax}%
\providecommand \BibitemShut  [1]{\csname bibitem#1\endcsname}%
\let\auto@bib@innerbib\@empty
\bibitem [{\citenamefont {Boos}\ \emph {et~al.}(2024)\citenamefont {Boos}, \citenamefont {Kim}, \citenamefont {Bracht}, \citenamefont {Sbresny}, \citenamefont {Kaspari}, \citenamefont {Cygorek}, \citenamefont {Riedl}, \citenamefont {Bopp}, \citenamefont {Rauhaus}, \citenamefont {Calcagno}, \citenamefont {Finley}, \citenamefont {Reiter},\ and\ \citenamefont {M\"uller}}]{BKB:PRL2024}%
  \BibitemOpen
  \bibfield  {author} {\bibinfo {author} {\bibfnamefont {K.}~\bibnamefont {Boos}}, \bibinfo {author} {\bibfnamefont {S.~K.}\ \bibnamefont {Kim}}, \bibinfo {author} {\bibfnamefont {T.}~\bibnamefont {Bracht}}, \bibinfo {author} {\bibfnamefont {F.}~\bibnamefont {Sbresny}}, \bibinfo {author} {\bibfnamefont {J.~M.}\ \bibnamefont {Kaspari}}, \bibinfo {author} {\bibfnamefont {M.}~\bibnamefont {Cygorek}}, \bibinfo {author} {\bibfnamefont {H.}~\bibnamefont {Riedl}}, \bibinfo {author} {\bibfnamefont {F.~W.}\ \bibnamefont {Bopp}}, \bibinfo {author} {\bibfnamefont {W.}~\bibnamefont {Rauhaus}}, \bibinfo {author} {\bibfnamefont {C.}~\bibnamefont {Calcagno}}, \bibinfo {author} {\bibfnamefont {J.~J.}\ \bibnamefont {Finley}}, \bibinfo {author} {\bibfnamefont {D.~E.}\ \bibnamefont {Reiter}},\ and\ \bibinfo {author} {\bibfnamefont {K.}~\bibnamefont {M\"uller}},\ }\bibfield  {title} {\bibinfo {title} {Signatures of dynamically dressed states},\ }\href {https://doi.org/10.1103/PhysRevLett.132.053602} {\bibfield  {journal}
  {\bibinfo  {journal} {Phys. Rev. Lett.}\ }\textbf {\bibinfo {volume} {132}},\ \bibinfo {pages} {053602} (\bibinfo {year} {2024})}\BibitemShut {NoStop}%
\bibitem [{\citenamefont {Liu}\ \emph {et~al.}(2024)\citenamefont {Liu}, \citenamefont {Gustin}, \citenamefont {Liu}, \citenamefont {Li}, \citenamefont {Yu}, \citenamefont {Ni}, \citenamefont {Niu}, \citenamefont {Hughes}, \citenamefont {Wang},\ and\ \citenamefont {Liu}}]{LGL:NPhot2024}%
  \BibitemOpen
  \bibfield  {author} {\bibinfo {author} {\bibfnamefont {S.}~\bibnamefont {Liu}}, \bibinfo {author} {\bibfnamefont {C.}~\bibnamefont {Gustin}}, \bibinfo {author} {\bibfnamefont {H.}~\bibnamefont {Liu}}, \bibinfo {author} {\bibfnamefont {X.}~\bibnamefont {Li}}, \bibinfo {author} {\bibfnamefont {Y.}~\bibnamefont {Yu}}, \bibinfo {author} {\bibfnamefont {H.}~\bibnamefont {Ni}}, \bibinfo {author} {\bibfnamefont {Z.}~\bibnamefont {Niu}}, \bibinfo {author} {\bibfnamefont {S.}~\bibnamefont {Hughes}}, \bibinfo {author} {\bibfnamefont {X.}~\bibnamefont {Wang}},\ and\ \bibinfo {author} {\bibfnamefont {J.}~\bibnamefont {Liu}},\ }\bibfield  {title} {\bibinfo {title} {Dynamic resonance fluorescence in solid-state cavity quantum electrodynamics},\ }\href {https://doi.org/10.1038/s41566-023-01359-x} {\bibfield  {journal} {\bibinfo  {journal} {Nature Photonics}\ }\textbf {\bibinfo {volume} {18}},\ \bibinfo {pages} {318} (\bibinfo {year} {2024})}\BibitemShut {NoStop}%
\bibitem [{\citenamefont {Mollow}(1969)}]{Mollow:PRL1969}%
  \BibitemOpen
  \bibfield  {author} {\bibinfo {author} {\bibfnamefont {B.~R.}\ \bibnamefont {Mollow}},\ }\bibfield  {title} {\bibinfo {title} {Power spectrum of light scattered by two-level systems},\ }\href {https://doi.org/10.1103/PhysRev.188.1969} {\bibfield  {journal} {\bibinfo  {journal} {Physical Review Letters}\ }\textbf {\bibinfo {volume} {188}} (\bibinfo {year} {1969})}\BibitemShut {NoStop}%
\bibitem [{CT:(1998)}]{CT:AtomPhotonBook}%
  \BibitemOpen
  \bibinfo {title} {The dressed atom approach},\ in\ \href {https://doi.org/https://doi.org/10.1002/9783527617197.ch6} {\emph {\bibinfo {booktitle} {Atom—Photon Interactions}}}\ (\bibinfo  {publisher} {John Wiley \& Sons, Ltd},\ \bibinfo {year} {1998})\ Chap.~\bibinfo {chapter} {6}, pp.\ \bibinfo {pages} {407--514}\BibitemShut {NoStop}%
\bibitem [{\citenamefont {Vagov}\ \emph {et~al.}(2011)\citenamefont {Vagov}, \citenamefont {Croitoru}, \citenamefont {Gl{\"a}ssl}, \citenamefont {Axt},\ and\ \citenamefont {Kuhn}}]{VCG:PRB2011}%
  \BibitemOpen
  \bibfield  {author} {\bibinfo {author} {\bibfnamefont {A.}~\bibnamefont {Vagov}}, \bibinfo {author} {\bibfnamefont {M.~D.}\ \bibnamefont {Croitoru}}, \bibinfo {author} {\bibfnamefont {M.}~\bibnamefont {Gl{\"a}ssl}}, \bibinfo {author} {\bibfnamefont {V.~M.}\ \bibnamefont {Axt}},\ and\ \bibinfo {author} {\bibfnamefont {T.}~\bibnamefont {Kuhn}},\ }\bibfield  {title} {\bibinfo {title} {Real-time path integrals for quantum dots: Quantum dissipative dynamics with superohmic environment coupling},\ }\href {https://doi.org/10.1103/PhysRevB.83.094303} {\bibfield  {journal} {\bibinfo  {journal} {Phys. Rev. B}\ }\textbf {\bibinfo {volume} {83}} (\bibinfo {year} {2011})}\BibitemShut {NoStop}%
\bibitem [{\citenamefont {Gustin}\ and\ \citenamefont {Hughes}(2019)}]{GH:AdvQTech2019}%
  \BibitemOpen
  \bibfield  {author} {\bibinfo {author} {\bibfnamefont {C.}~\bibnamefont {Gustin}}\ and\ \bibinfo {author} {\bibfnamefont {S.}~\bibnamefont {Hughes}},\ }\bibfield  {title} {\bibinfo {title} {Efficient pulse--excitation techniques for single photon sources from quantum dots in optical cavities},\ }\href {https://doi.org/10.1002/qute.201900073} {\bibfield  {journal} {\bibinfo  {journal} {Advanced Quantum Technologies}\ }\textbf {\bibinfo {volume} {3}},\ \bibinfo {pages} {1900073} (\bibinfo {year} {2019})}\BibitemShut {NoStop}%
\bibitem [{\citenamefont {Barth}\ \emph {et~al.}(2016)\citenamefont {Barth}, \citenamefont {L\"uker}, \citenamefont {Vagov}, \citenamefont {Reiter}, \citenamefont {Kuhn},\ and\ \citenamefont {Axt}}]{BLV:PRB2016}%
  \BibitemOpen
  \bibfield  {author} {\bibinfo {author} {\bibfnamefont {A.~M.}\ \bibnamefont {Barth}}, \bibinfo {author} {\bibfnamefont {S.}~\bibnamefont {L\"uker}}, \bibinfo {author} {\bibfnamefont {A.}~\bibnamefont {Vagov}}, \bibinfo {author} {\bibfnamefont {D.~E.}\ \bibnamefont {Reiter}}, \bibinfo {author} {\bibfnamefont {T.}~\bibnamefont {Kuhn}},\ and\ \bibinfo {author} {\bibfnamefont {V.~M.}\ \bibnamefont {Axt}},\ }\bibfield  {title} {\bibinfo {title} {Fast and selective phonon-assisted state preparation of a quantum dot by adiabatic undressing},\ }\href {https://doi.org/10.1103/PhysRevB.94.045306} {\bibfield  {journal} {\bibinfo  {journal} {Phys. Rev. B}\ }\textbf {\bibinfo {volume} {94}},\ \bibinfo {pages} {045306} (\bibinfo {year} {2016})}\BibitemShut {NoStop}%
\bibitem [{\citenamefont {Hanschke}\ \emph {et~al.}()\citenamefont {Hanschke}, \citenamefont {Bracht}, \citenamefont {Schöll}, \citenamefont {Bauch}, \citenamefont {Berger}, \citenamefont {Kallert}, \citenamefont {Peter}, \citenamefont {Jr.}, \citenamefont {da~Silva}, \citenamefont {Manna}, \citenamefont {Rastelli}, \citenamefont {Schumacher}, \citenamefont {Reiter},\ and\ \citenamefont {Jöns}}]{HBS:arxiv2024}%
  \BibitemOpen
  \bibfield  {author} {\bibinfo {author} {\bibfnamefont {L.}~\bibnamefont {Hanschke}}, \bibinfo {author} {\bibfnamefont {T.~K.}\ \bibnamefont {Bracht}}, \bibinfo {author} {\bibfnamefont {E.}~\bibnamefont {Schöll}}, \bibinfo {author} {\bibfnamefont {D.}~\bibnamefont {Bauch}}, \bibinfo {author} {\bibfnamefont {E.}~\bibnamefont {Berger}}, \bibinfo {author} {\bibfnamefont {P.}~\bibnamefont {Kallert}}, \bibinfo {author} {\bibfnamefont {M.}~\bibnamefont {Peter}}, \bibinfo {author} {\bibfnamefont {A.~J.~G.}\ \bibnamefont {Jr.}}, \bibinfo {author} {\bibfnamefont {S.~F.~C.}\ \bibnamefont {da~Silva}}, \bibinfo {author} {\bibfnamefont {S.}~\bibnamefont {Manna}}, \bibinfo {author} {\bibfnamefont {A.}~\bibnamefont {Rastelli}}, \bibinfo {author} {\bibfnamefont {S.}~\bibnamefont {Schumacher}}, \bibinfo {author} {\bibfnamefont {D.~E.}\ \bibnamefont {Reiter}},\ and\ \bibinfo {author} {\bibfnamefont {K.~D.}\ \bibnamefont {Jöns}},\ }\href@noop {} {\bibinfo {title} {Experimental measurement of the reappearance of rabi
  rotations in semiconductor quantum dots}},\ \Eprint {https://arxiv.org/abs/2409.19167} {arXiv:2409.19167 [cond-mat.mes-hall]} \BibitemShut {NoStop}%
\end{thebibliography}%


\begin{thebibliography}{55}%
\makeatletter
\providecommand \@ifxundefined [1]{%
 \@ifx{#1\undefined}
}%
\providecommand \@ifnum [1]{%
 \ifnum #1\expandafter \@firstoftwo
 \else \expandafter \@secondoftwo
 \fi
}%
\providecommand \@ifx [1]{%
 \ifx #1\expandafter \@firstoftwo
 \else \expandafter \@secondoftwo
 \fi
}%
\providecommand \natexlab [1]{#1}%
\providecommand \enquote  [1]{``#1''}%
\providecommand \bibnamefont  [1]{#1}%
\providecommand \bibfnamefont [1]{#1}%
\providecommand \citenamefont [1]{#1}%
\providecommand \href@noop [0]{\@secondoftwo}%
\providecommand \href [0]{\begingroup \@sanitize@url \@href}%
\providecommand \@href[1]{\@@startlink{#1}\@@href}%
\providecommand \@@href[1]{\endgroup#1\@@endlink}%
\providecommand \@sanitize@url [0]{\catcode `\\12\catcode `\$12\catcode `\&12\catcode `\#12\catcode `\^12\catcode `\_12\catcode `\%12\relax}%
\providecommand \@@startlink[1]{}%
\providecommand \@@endlink[0]{}%
\providecommand \url  [0]{\begingroup\@sanitize@url \@url }%
\providecommand \@url [1]{\endgroup\@href {#1}{\urlprefix }}%
\providecommand \urlprefix  [0]{URL }%
\providecommand \Eprint [0]{\href }%
\providecommand \doibase [0]{https://doi.org/}%
\providecommand \selectlanguage [0]{\@gobble}%
\providecommand \bibinfo  [0]{\@secondoftwo}%
\providecommand \bibfield  [0]{\@secondoftwo}%
\providecommand \translation [1]{[#1]}%
\providecommand \BibitemOpen [0]{}%
\providecommand \bibitemStop [0]{}%
\providecommand \bibitemNoStop [0]{.\EOS\space}%
\providecommand \EOS [0]{\spacefactor3000\relax}%
\providecommand \BibitemShut  [1]{\csname bibitem#1\endcsname}%
\let\auto@bib@innerbib\@empty
\bibitem [{\citenamefont {{Le Jeannic}}\ \emph {et~al.}(2022)\citenamefont {{Le Jeannic}}, \citenamefont {Tiranov}, \citenamefont {Carolan}, \citenamefont {Ramos}, \citenamefont {Wang}, \citenamefont {Appel}, \citenamefont {Scholz}, \citenamefont {Wieck}, \citenamefont {Ludwig}, \citenamefont {Rotenberg}, \citenamefont {Midolo}, \citenamefont {Garc{\'i}a-Ripoll}, \citenamefont {S{\o}rensen},\ and\ \citenamefont {Lodahl}}]{JTC:NatPhys2022}%
  \BibitemOpen
  \bibfield  {author} {\bibinfo {author} {\bibfnamefont {H.}~\bibnamefont {{Le Jeannic}}}, \bibinfo {author} {\bibfnamefont {A.}~\bibnamefont {Tiranov}}, \bibinfo {author} {\bibfnamefont {J.}~\bibnamefont {Carolan}}, \bibinfo {author} {\bibfnamefont {T.}~\bibnamefont {Ramos}}, \bibinfo {author} {\bibfnamefont {Y.}~\bibnamefont {Wang}}, \bibinfo {author} {\bibfnamefont {M.~H.}\ \bibnamefont {Appel}}, \bibinfo {author} {\bibfnamefont {S.}~\bibnamefont {Scholz}}, \bibinfo {author} {\bibfnamefont {A.~D.}\ \bibnamefont {Wieck}}, \bibinfo {author} {\bibfnamefont {A.}~\bibnamefont {Ludwig}}, \bibinfo {author} {\bibfnamefont {N.}~\bibnamefont {Rotenberg}}, \bibinfo {author} {\bibfnamefont {L.}~\bibnamefont {Midolo}}, \bibinfo {author} {\bibfnamefont {J.~J.}\ \bibnamefont {Garc{\'i}a-Ripoll}}, \bibinfo {author} {\bibfnamefont {A.~S.}\ \bibnamefont {S{\o}rensen}},\ and\ \bibinfo {author} {\bibfnamefont {P.}~\bibnamefont {Lodahl}},\ }\bibfield  {title} {\bibinfo {title} {Dynamical photon--photon interaction mediated by
  a quantum emitter},\ }\href {https://doi.org/10.1038/s41567-022-01720-x} {\bibfield  {journal} {\bibinfo  {journal} {Nature Physics}\ }\textbf {\bibinfo {volume} {18}},\ \bibinfo {pages} {1191} (\bibinfo {year} {2022})}\BibitemShut {NoStop}%
\bibitem [{\citenamefont {Tomm}\ \emph {et~al.}(2023)\citenamefont {Tomm}, \citenamefont {Mahmoodian}, \citenamefont {Antoniadis}, \citenamefont {Schott}, \citenamefont {Valentin}, \citenamefont {Wieck}, \citenamefont {Ludwig}, \citenamefont {Javadi},\ and\ \citenamefont {Warburton}}]{TMN:NatPhys2023}%
  \BibitemOpen
  \bibfield  {author} {\bibinfo {author} {\bibfnamefont {N.}~\bibnamefont {Tomm}}, \bibinfo {author} {\bibfnamefont {S.}~\bibnamefont {Mahmoodian}}, \bibinfo {author} {\bibfnamefont {N.~O.}\ \bibnamefont {Antoniadis}}, \bibinfo {author} {\bibfnamefont {R.}~\bibnamefont {Schott}}, \bibinfo {author} {\bibfnamefont {S.~R.}\ \bibnamefont {Valentin}}, \bibinfo {author} {\bibfnamefont {A.~D.}\ \bibnamefont {Wieck}}, \bibinfo {author} {\bibfnamefont {A.}~\bibnamefont {Ludwig}}, \bibinfo {author} {\bibfnamefont {A.}~\bibnamefont {Javadi}},\ and\ \bibinfo {author} {\bibfnamefont {R.~J.}\ \bibnamefont {Warburton}},\ }\bibfield  {title} {\bibinfo {title} {Photon bound state dynamics from a single artificial atom},\ }\href {https://doi.org/10.1038/s41567-023-01997-6} {\bibfield  {journal} {\bibinfo  {journal} {Nature Physics}\ }\textbf {\bibinfo {volume} {19}},\ \bibinfo {pages} {857} (\bibinfo {year} {2023})}\BibitemShut {NoStop}%
\bibitem [{\citenamefont {Tiranov}\ \emph {et~al.}(2023)\citenamefont {Tiranov}, \citenamefont {Angelopoulou}, \citenamefont {{van Diepen}}, \citenamefont {Schrinski}, \citenamefont {Sandberg}, \citenamefont {Wang}, \citenamefont {Midolo}, \citenamefont {Scholz}, \citenamefont {Wieck}, \citenamefont {Ludwig}, \citenamefont {S{\o}rensen},\ and\ \citenamefont {Lodahl}}]{TAD:Science2023}%
  \BibitemOpen
  \bibfield  {author} {\bibinfo {author} {\bibfnamefont {A.}~\bibnamefont {Tiranov}}, \bibinfo {author} {\bibfnamefont {V.}~\bibnamefont {Angelopoulou}}, \bibinfo {author} {\bibfnamefont {C.~J.}\ \bibnamefont {{van Diepen}}}, \bibinfo {author} {\bibfnamefont {B.}~\bibnamefont {Schrinski}}, \bibinfo {author} {\bibfnamefont {O.~A.~D.}\ \bibnamefont {Sandberg}}, \bibinfo {author} {\bibfnamefont {Y.}~\bibnamefont {Wang}}, \bibinfo {author} {\bibfnamefont {L.}~\bibnamefont {Midolo}}, \bibinfo {author} {\bibfnamefont {S.}~\bibnamefont {Scholz}}, \bibinfo {author} {\bibfnamefont {A.~D.}\ \bibnamefont {Wieck}}, \bibinfo {author} {\bibfnamefont {A.}~\bibnamefont {Ludwig}}, \bibinfo {author} {\bibfnamefont {A.~S.}\ \bibnamefont {S{\o}rensen}},\ and\ \bibinfo {author} {\bibfnamefont {P.}~\bibnamefont {Lodahl}},\ }\bibfield  {title} {\bibinfo {title} {Collective super- and subradiant dynamics between distant optical quantum emitters},\ }\href {https://doi.org/10.1126/science.ade9324} {\bibfield  {journal} {\bibinfo
  {journal} {Science}\ }\textbf {\bibinfo {volume} {379}},\ \bibinfo {pages} {389} (\bibinfo {year} {2023})}\BibitemShut {NoStop}%
\bibitem [{\citenamefont {Hansen}\ \emph {et~al.}()\citenamefont {Hansen}, \citenamefont {Giorgino}, \citenamefont {Jehle}, \citenamefont {Carosini}, \citenamefont {Carreño}, \citenamefont {Arrazola}, \citenamefont {Walther},\ and\ \citenamefont {Loredo}}]{HGJ:arxiv2024}%
  \BibitemOpen
  \bibfield  {author} {\bibinfo {author} {\bibfnamefont {L.~M.}\ \bibnamefont {Hansen}}, \bibinfo {author} {\bibfnamefont {F.}~\bibnamefont {Giorgino}}, \bibinfo {author} {\bibfnamefont {L.}~\bibnamefont {Jehle}}, \bibinfo {author} {\bibfnamefont {L.}~\bibnamefont {Carosini}}, \bibinfo {author} {\bibfnamefont {J.~C.~L.}\ \bibnamefont {Carreño}}, \bibinfo {author} {\bibfnamefont {I.}~\bibnamefont {Arrazola}}, \bibinfo {author} {\bibfnamefont {P.}~\bibnamefont {Walther}},\ and\ \bibinfo {author} {\bibfnamefont {J.~C.}\ \bibnamefont {Loredo}},\ }\href@noop {} {\bibinfo {title} {Non-classical excitation of a solid-state quantum emitter}},\ \Eprint {https://arxiv.org/abs/2407.20936} {arXiv:2407.20936 [quant-ph]} \BibitemShut {NoStop}%
\bibitem [{\citenamefont {Zhang}\ \emph {et~al.}(2025)\citenamefont {Zhang}, \citenamefont {Ding}, \citenamefont {Li}, \citenamefont {Zhang}, \citenamefont {Guo}, \citenamefont {Wang}, \citenamefont {Ning}, \citenamefont {Xu}, \citenamefont {Liu}, \citenamefont {Zhao}, \citenamefont {Zou}, \citenamefont {Wang}, \citenamefont {Cao}, \citenamefont {He}, \citenamefont {Peng}, \citenamefont {Huo}, \citenamefont {Liao}, \citenamefont {Lu}, \citenamefont {Xu},\ and\ \citenamefont {Pan}}]{ZDL:PRL2025}%
  \BibitemOpen
  \bibfield  {author} {\bibinfo {author} {\bibfnamefont {Y.}~\bibnamefont {Zhang}}, \bibinfo {author} {\bibfnamefont {X.}~\bibnamefont {Ding}}, \bibinfo {author} {\bibfnamefont {Y.}~\bibnamefont {Li}}, \bibinfo {author} {\bibfnamefont {L.}~\bibnamefont {Zhang}}, \bibinfo {author} {\bibfnamefont {Y.-P.}\ \bibnamefont {Guo}}, \bibinfo {author} {\bibfnamefont {G.-Q.}\ \bibnamefont {Wang}}, \bibinfo {author} {\bibfnamefont {Z.}~\bibnamefont {Ning}}, \bibinfo {author} {\bibfnamefont {M.-C.}\ \bibnamefont {Xu}}, \bibinfo {author} {\bibfnamefont {R.-Z.}\ \bibnamefont {Liu}}, \bibinfo {author} {\bibfnamefont {J.-Y.}\ \bibnamefont {Zhao}}, \bibinfo {author} {\bibfnamefont {G.-Y.}\ \bibnamefont {Zou}}, \bibinfo {author} {\bibfnamefont {H.}~\bibnamefont {Wang}}, \bibinfo {author} {\bibfnamefont {Y.}~\bibnamefont {Cao}}, \bibinfo {author} {\bibfnamefont {Y.-M.}\ \bibnamefont {He}}, \bibinfo {author} {\bibfnamefont {C.-Z.}\ \bibnamefont {Peng}}, \bibinfo {author} {\bibfnamefont {Y.-H.}\ \bibnamefont {Huo}}, \bibinfo
  {author} {\bibfnamefont {S.-K.}\ \bibnamefont {Liao}}, \bibinfo {author} {\bibfnamefont {C.-Y.}\ \bibnamefont {Lu}}, \bibinfo {author} {\bibfnamefont {F.}~\bibnamefont {Xu}},\ and\ \bibinfo {author} {\bibfnamefont {J.-W.}\ \bibnamefont {Pan}},\ }\bibfield  {title} {\bibinfo {title} {Experimental single-photon quantum key distribution surpassing the fundamental weak coherent-state rate limit},\ }\href {https://doi.org/10.1103/PhysRevLett.134.210801} {\bibfield  {journal} {\bibinfo  {journal} {Phys. Rev. Lett.}\ }\textbf {\bibinfo {volume} {134}},\ \bibinfo {pages} {210801} (\bibinfo {year} {2025})}\BibitemShut {NoStop}%
\bibitem [{\citenamefont {Strobel}\ \emph {et~al.}()\citenamefont {Strobel}, \citenamefont {Vyvlecka}, \citenamefont {Neureuther}, \citenamefont {Bauer}, \citenamefont {Schäfer}, \citenamefont {Kazmaier}, \citenamefont {Sharma}, \citenamefont {Joos}, \citenamefont {Weber}, \citenamefont {Nawrath}, \citenamefont {Nie}, \citenamefont {Bhayani}, \citenamefont {Hopfmann}, \citenamefont {Becher}, \citenamefont {Michler},\ and\ \citenamefont {Portalupi}}]{SVN:arxiv2024}%
  \BibitemOpen
  \bibfield  {author} {\bibinfo {author} {\bibfnamefont {T.}~\bibnamefont {Strobel}}, \bibinfo {author} {\bibfnamefont {M.}~\bibnamefont {Vyvlecka}}, \bibinfo {author} {\bibfnamefont {I.}~\bibnamefont {Neureuther}}, \bibinfo {author} {\bibfnamefont {T.}~\bibnamefont {Bauer}}, \bibinfo {author} {\bibfnamefont {M.}~\bibnamefont {Schäfer}}, \bibinfo {author} {\bibfnamefont {S.}~\bibnamefont {Kazmaier}}, \bibinfo {author} {\bibfnamefont {N.~L.}\ \bibnamefont {Sharma}}, \bibinfo {author} {\bibfnamefont {R.}~\bibnamefont {Joos}}, \bibinfo {author} {\bibfnamefont {J.~H.}\ \bibnamefont {Weber}}, \bibinfo {author} {\bibfnamefont {C.}~\bibnamefont {Nawrath}}, \bibinfo {author} {\bibfnamefont {W.}~\bibnamefont {Nie}}, \bibinfo {author} {\bibfnamefont {G.}~\bibnamefont {Bhayani}}, \bibinfo {author} {\bibfnamefont {C.}~\bibnamefont {Hopfmann}}, \bibinfo {author} {\bibfnamefont {C.}~\bibnamefont {Becher}}, \bibinfo {author} {\bibfnamefont {P.}~\bibnamefont {Michler}},\ and\ \bibinfo {author} {\bibfnamefont {S.~L.}\
  \bibnamefont {Portalupi}},\ }\href@noop {} {\bibinfo {title} {Quantum teleportation with telecom photons from remote quantum emitters}},\ \Eprint {https://arxiv.org/abs/2411.12904} {arXiv:2411.12904 [quant-ph]} \BibitemShut {NoStop}%
\bibitem [{\citenamefont {Laneve}\ \emph {et~al.}()\citenamefont {Laneve}, \citenamefont {Ronco}, \citenamefont {Beccaceci}, \citenamefont {Barigelli}, \citenamefont {Salusti}, \citenamefont {Claro-Rodriguez}, \citenamefont {Pascalis}, \citenamefont {Suprano}, \citenamefont {Chiaudano}, \citenamefont {Schöll}, \citenamefont {Hanschke}, \citenamefont {Krieger}, \citenamefont {Buchinger}, \citenamefont {da~Silva}, \citenamefont {Neuwirth}, \citenamefont {Stroj}, \citenamefont {Höfling}, \citenamefont {Huber-Loyola}, \citenamefont {Castaneda}, \citenamefont {Carvacho}, \citenamefont {Spagnolo}, \citenamefont {Rota}, \citenamefont {Basset}, \citenamefont {Rastelli}, \citenamefont {Sciarrino}, \citenamefont {Jöns},\ and\ \citenamefont {Trotta}}]{LRB:arxiv2024}%
  \BibitemOpen
  \bibfield  {author} {\bibinfo {author} {\bibfnamefont {A.}~\bibnamefont {Laneve}}, \bibinfo {author} {\bibfnamefont {G.}~\bibnamefont {Ronco}}, \bibinfo {author} {\bibfnamefont {M.}~\bibnamefont {Beccaceci}}, \bibinfo {author} {\bibfnamefont {P.}~\bibnamefont {Barigelli}}, \bibinfo {author} {\bibfnamefont {F.}~\bibnamefont {Salusti}}, \bibinfo {author} {\bibfnamefont {N.}~\bibnamefont {Claro-Rodriguez}}, \bibinfo {author} {\bibfnamefont {G.~D.}\ \bibnamefont {Pascalis}}, \bibinfo {author} {\bibfnamefont {A.}~\bibnamefont {Suprano}}, \bibinfo {author} {\bibfnamefont {L.}~\bibnamefont {Chiaudano}}, \bibinfo {author} {\bibfnamefont {E.}~\bibnamefont {Schöll}}, \bibinfo {author} {\bibfnamefont {L.}~\bibnamefont {Hanschke}}, \bibinfo {author} {\bibfnamefont {T.~M.}\ \bibnamefont {Krieger}}, \bibinfo {author} {\bibfnamefont {Q.}~\bibnamefont {Buchinger}}, \bibinfo {author} {\bibfnamefont {S.~F.~C.}\ \bibnamefont {da~Silva}}, \bibinfo {author} {\bibfnamefont {J.}~\bibnamefont {Neuwirth}}, \bibinfo {author}
  {\bibfnamefont {S.}~\bibnamefont {Stroj}}, \bibinfo {author} {\bibfnamefont {S.}~\bibnamefont {Höfling}}, \bibinfo {author} {\bibfnamefont {T.}~\bibnamefont {Huber-Loyola}}, \bibinfo {author} {\bibfnamefont {M.~A.~U.}\ \bibnamefont {Castaneda}}, \bibinfo {author} {\bibfnamefont {G.}~\bibnamefont {Carvacho}}, \bibinfo {author} {\bibfnamefont {N.}~\bibnamefont {Spagnolo}}, \bibinfo {author} {\bibfnamefont {M.~B.}\ \bibnamefont {Rota}}, \bibinfo {author} {\bibfnamefont {F.~B.}\ \bibnamefont {Basset}}, \bibinfo {author} {\bibfnamefont {A.}~\bibnamefont {Rastelli}}, \bibinfo {author} {\bibfnamefont {F.}~\bibnamefont {Sciarrino}}, \bibinfo {author} {\bibfnamefont {K.}~\bibnamefont {Jöns}},\ and\ \bibinfo {author} {\bibfnamefont {R.}~\bibnamefont {Trotta}},\ }\href@noop {} {\bibinfo {title} {Quantum teleportation with dissimilar quantum dots over a hybrid quantum network}},\ \Eprint {https://arxiv.org/abs/2411.12387} {arXiv:2411.12387 [quant-ph]} \BibitemShut {NoStop}%
\bibitem [{\citenamefont {Vajner}\ \emph {et~al.}()\citenamefont {Vajner}, \citenamefont {Kaymazlar}, \citenamefont {Drauschke}, \citenamefont {Rickert}, \citenamefont {von Helversen}, \citenamefont {Liu}, \citenamefont {Li}, \citenamefont {Ni}, \citenamefont {Niu}, \citenamefont {Pappa},\ and\ \citenamefont {Heindel}}]{VKD:arxiv2024}%
  \BibitemOpen
  \bibfield  {author} {\bibinfo {author} {\bibfnamefont {D.~A.}\ \bibnamefont {Vajner}}, \bibinfo {author} {\bibfnamefont {K.}~\bibnamefont {Kaymazlar}}, \bibinfo {author} {\bibfnamefont {F.}~\bibnamefont {Drauschke}}, \bibinfo {author} {\bibfnamefont {L.}~\bibnamefont {Rickert}}, \bibinfo {author} {\bibfnamefont {M.}~\bibnamefont {von Helversen}}, \bibinfo {author} {\bibfnamefont {H.}~\bibnamefont {Liu}}, \bibinfo {author} {\bibfnamefont {S.}~\bibnamefont {Li}}, \bibinfo {author} {\bibfnamefont {H.}~\bibnamefont {Ni}}, \bibinfo {author} {\bibfnamefont {Z.}~\bibnamefont {Niu}}, \bibinfo {author} {\bibfnamefont {A.}~\bibnamefont {Pappa}},\ and\ \bibinfo {author} {\bibfnamefont {T.}~\bibnamefont {Heindel}},\ }\href@noop {} {\bibinfo {title} {Single-photon advantage in quantum cryptography beyond qkd}},\ \Eprint {https://arxiv.org/abs/2412.14993} {arXiv:2412.14993 [quant-ph]} \BibitemShut {NoStop}%
\bibitem [{\citenamefont {Yu}\ \emph {et~al.}(2023)\citenamefont {Yu}, \citenamefont {Liu}, \citenamefont {Lee}, \citenamefont {Michler}, \citenamefont {Reitzenstein}, \citenamefont {Srinivasan}, \citenamefont {Waks},\ and\ \citenamefont {Liu}}]{YLL:NNano2023}%
  \BibitemOpen
  \bibfield  {author} {\bibinfo {author} {\bibfnamefont {Y.}~\bibnamefont {Yu}}, \bibinfo {author} {\bibfnamefont {S.}~\bibnamefont {Liu}}, \bibinfo {author} {\bibfnamefont {C.-M.}\ \bibnamefont {Lee}}, \bibinfo {author} {\bibfnamefont {P.}~\bibnamefont {Michler}}, \bibinfo {author} {\bibfnamefont {S.}~\bibnamefont {Reitzenstein}}, \bibinfo {author} {\bibfnamefont {K.}~\bibnamefont {Srinivasan}}, \bibinfo {author} {\bibfnamefont {E.}~\bibnamefont {Waks}},\ and\ \bibinfo {author} {\bibfnamefont {J.}~\bibnamefont {Liu}},\ }\bibfield  {title} {\bibinfo {title} {Telecom-band quantum dot technologies for long-distance quantum networks},\ }\href {https://doi.org/10.1038/s41565-023-01528-7} {\bibfield  {journal} {\bibinfo  {journal} {Nature Nanotechnology}\ }\textbf {\bibinfo {volume} {18}},\ \bibinfo {pages} {1389} (\bibinfo {year} {2023})}\BibitemShut {NoStop}%
\bibitem [{\citenamefont {Cao}\ \emph {et~al.}(2024)\citenamefont {Cao}, \citenamefont {Hansen}, \citenamefont {Giorgino}, \citenamefont {Carosini}, \citenamefont {Zah\'alka}, \citenamefont {Zilk}, \citenamefont {Loredo},\ and\ \citenamefont {Walther}}]{CHG:PRL2024}%
  \BibitemOpen
  \bibfield  {author} {\bibinfo {author} {\bibfnamefont {H.}~\bibnamefont {Cao}}, \bibinfo {author} {\bibfnamefont {L.~M.}\ \bibnamefont {Hansen}}, \bibinfo {author} {\bibfnamefont {F.}~\bibnamefont {Giorgino}}, \bibinfo {author} {\bibfnamefont {L.}~\bibnamefont {Carosini}}, \bibinfo {author} {\bibfnamefont {P.}~\bibnamefont {Zah\'alka}}, \bibinfo {author} {\bibfnamefont {F.}~\bibnamefont {Zilk}}, \bibinfo {author} {\bibfnamefont {J.~C.}\ \bibnamefont {Loredo}},\ and\ \bibinfo {author} {\bibfnamefont {P.}~\bibnamefont {Walther}},\ }\bibfield  {title} {\bibinfo {title} {Photonic source of heralded greenberger-horne-zeilinger states},\ }\href {https://doi.org/10.1103/PhysRevLett.132.130604} {\bibfield  {journal} {\bibinfo  {journal} {Phys. Rev. Lett.}\ }\textbf {\bibinfo {volume} {132}},\ \bibinfo {pages} {130604} (\bibinfo {year} {2024})}\BibitemShut {NoStop}%
\bibitem [{\citenamefont {Cogan}\ \emph {et~al.}(2023)\citenamefont {Cogan}, \citenamefont {Su}, \citenamefont {Kenneth},\ and\ \citenamefont {Gershoni}}]{CSK:NPhot2023}%
  \BibitemOpen
  \bibfield  {author} {\bibinfo {author} {\bibfnamefont {D.}~\bibnamefont {Cogan}}, \bibinfo {author} {\bibfnamefont {Z.-E.}\ \bibnamefont {Su}}, \bibinfo {author} {\bibfnamefont {O.}~\bibnamefont {Kenneth}},\ and\ \bibinfo {author} {\bibfnamefont {D.}~\bibnamefont {Gershoni}},\ }\bibfield  {title} {\bibinfo {title} {Deterministic generation of indistinguishable photons in a cluster state},\ }\href {https://doi.org/10.1038/s41566-022-01152-2} {\bibfield  {journal} {\bibinfo  {journal} {Nature Photonics}\ }\textbf {\bibinfo {volume} {17}},\ \bibinfo {pages} {324} (\bibinfo {year} {2023})}\BibitemShut {NoStop}%
\bibitem [{\citenamefont {Meng}\ \emph {et~al.}(2024)\citenamefont {Meng}, \citenamefont {Chan}, \citenamefont {Nielsen}, \citenamefont {Appel}, \citenamefont {Liu}, \citenamefont {Wang}, \citenamefont {Bart}, \citenamefont {Wieck}, \citenamefont {Ludwig}, \citenamefont {Midolo}, \citenamefont {Tiranov}, \citenamefont {S{\o}rensen},\ and\ \citenamefont {Lodahl}}]{MCN:NComms2024}%
  \BibitemOpen
  \bibfield  {author} {\bibinfo {author} {\bibfnamefont {Y.}~\bibnamefont {Meng}}, \bibinfo {author} {\bibfnamefont {M.~L.}\ \bibnamefont {Chan}}, \bibinfo {author} {\bibfnamefont {R.~B.}\ \bibnamefont {Nielsen}}, \bibinfo {author} {\bibfnamefont {M.~H.}\ \bibnamefont {Appel}}, \bibinfo {author} {\bibfnamefont {Z.}~\bibnamefont {Liu}}, \bibinfo {author} {\bibfnamefont {Y.}~\bibnamefont {Wang}}, \bibinfo {author} {\bibfnamefont {N.}~\bibnamefont {Bart}}, \bibinfo {author} {\bibfnamefont {A.~D.}\ \bibnamefont {Wieck}}, \bibinfo {author} {\bibfnamefont {A.}~\bibnamefont {Ludwig}}, \bibinfo {author} {\bibfnamefont {L.}~\bibnamefont {Midolo}}, \bibinfo {author} {\bibfnamefont {A.}~\bibnamefont {Tiranov}}, \bibinfo {author} {\bibfnamefont {A.~S.}\ \bibnamefont {S{\o}rensen}},\ and\ \bibinfo {author} {\bibfnamefont {P.}~\bibnamefont {Lodahl}},\ }\bibfield  {title} {\bibinfo {title} {Deterministic photon source of genuine three-qubit entanglement},\ }\href {https://doi.org/10.1038/s41467-024-52086-y} {\bibfield
  {journal} {\bibinfo  {journal} {Nature Communications}\ }\textbf {\bibinfo {volume} {15}},\ \bibinfo {pages} {7774} (\bibinfo {year} {2024})}\BibitemShut {NoStop}%
\bibitem [{\citenamefont {Uppu}\ \emph {et~al.}(2021)\citenamefont {Uppu}, \citenamefont {Midolo}, \citenamefont {Zhou}, \citenamefont {Carolan},\ and\ \citenamefont {Lodahl}}]{UMZ:NNano2021}%
  \BibitemOpen
  \bibfield  {author} {\bibinfo {author} {\bibfnamefont {R.}~\bibnamefont {Uppu}}, \bibinfo {author} {\bibfnamefont {L.}~\bibnamefont {Midolo}}, \bibinfo {author} {\bibfnamefont {X.}~\bibnamefont {Zhou}}, \bibinfo {author} {\bibfnamefont {J.}~\bibnamefont {Carolan}},\ and\ \bibinfo {author} {\bibfnamefont {P.}~\bibnamefont {Lodahl}},\ }\bibfield  {title} {\bibinfo {title} {Quantum-dot-based deterministic photon-emitter interfaces for scalable photonic quantum technology},\ }\href {https://doi.org/10.1038/s41565-021-00965-6} {\bibfield  {journal} {\bibinfo  {journal} {Nature Nanotechnology}\ }\textbf {\bibinfo {volume} {16}},\ \bibinfo {pages} {1308} (\bibinfo {year} {2021})}\BibitemShut {NoStop}%
\bibitem [{\citenamefont {Ding}\ \emph {et~al.}(2025)\citenamefont {Ding}, \citenamefont {Guo}, \citenamefont {Xu}, \citenamefont {Liu}, \citenamefont {Zou}, \citenamefont {Zhao}, \citenamefont {Ge}, \citenamefont {Zhang}, \citenamefont {Liu}, \citenamefont {Wang}, \citenamefont {Chen}, \citenamefont {Wang}, \citenamefont {He}, \citenamefont {Huo}, \citenamefont {Lu},\ and\ \citenamefont {Pan}}]{DGX:NPhot2025}%
  \BibitemOpen
  \bibfield  {author} {\bibinfo {author} {\bibfnamefont {X.}~\bibnamefont {Ding}}, \bibinfo {author} {\bibfnamefont {Y.-P.}\ \bibnamefont {Guo}}, \bibinfo {author} {\bibfnamefont {M.-C.}\ \bibnamefont {Xu}}, \bibinfo {author} {\bibfnamefont {R.-Z.}\ \bibnamefont {Liu}}, \bibinfo {author} {\bibfnamefont {G.-Y.}\ \bibnamefont {Zou}}, \bibinfo {author} {\bibfnamefont {J.-Y.}\ \bibnamefont {Zhao}}, \bibinfo {author} {\bibfnamefont {Z.-X.}\ \bibnamefont {Ge}}, \bibinfo {author} {\bibfnamefont {Q.-H.}\ \bibnamefont {Zhang}}, \bibinfo {author} {\bibfnamefont {H.-L.}\ \bibnamefont {Liu}}, \bibinfo {author} {\bibfnamefont {L.-J.}\ \bibnamefont {Wang}}, \bibinfo {author} {\bibfnamefont {M.-C.}\ \bibnamefont {Chen}}, \bibinfo {author} {\bibfnamefont {H.}~\bibnamefont {Wang}}, \bibinfo {author} {\bibfnamefont {Y.-M.}\ \bibnamefont {He}}, \bibinfo {author} {\bibfnamefont {Y.-H.}\ \bibnamefont {Huo}}, \bibinfo {author} {\bibfnamefont {C.-Y.}\ \bibnamefont {Lu}},\ and\ \bibinfo {author} {\bibfnamefont {J.-W.}\ \bibnamefont
  {Pan}},\ }\bibfield  {title} {\bibinfo {title} {High-efficiency single-photon source above the loss-tolerant threshold for efficient linear optical quantum computing},\ }\href {https://doi.org/10.1038/s41566-025-01639-8} {\bibfield  {journal} {\bibinfo  {journal} {Nature Photonics}\ }\textbf {\bibinfo {volume} {19}},\ \bibinfo {pages} {387} (\bibinfo {year} {2025})}\BibitemShut {NoStop}%
\bibitem [{\citenamefont {Schweickert}\ \emph {et~al.}(2018)\citenamefont {Schweickert}, \citenamefont {Jöns}, \citenamefont {Zeuner}, \citenamefont {Covre~da Silva}, \citenamefont {Huang}, \citenamefont {Lettner}, \citenamefont {Reindl}, \citenamefont {Zichi}, \citenamefont {Trotta}, \citenamefont {Rastelli},\ and\ \citenamefont {Zwiller}}]{SJZ:APL2024}%
  \BibitemOpen
  \bibfield  {author} {\bibinfo {author} {\bibfnamefont {L.}~\bibnamefont {Schweickert}}, \bibinfo {author} {\bibfnamefont {K.~D.}\ \bibnamefont {Jöns}}, \bibinfo {author} {\bibfnamefont {K.~D.}\ \bibnamefont {Zeuner}}, \bibinfo {author} {\bibfnamefont {S.~F.}\ \bibnamefont {Covre~da Silva}}, \bibinfo {author} {\bibfnamefont {H.}~\bibnamefont {Huang}}, \bibinfo {author} {\bibfnamefont {T.}~\bibnamefont {Lettner}}, \bibinfo {author} {\bibfnamefont {M.}~\bibnamefont {Reindl}}, \bibinfo {author} {\bibfnamefont {J.}~\bibnamefont {Zichi}}, \bibinfo {author} {\bibfnamefont {R.}~\bibnamefont {Trotta}}, \bibinfo {author} {\bibfnamefont {A.}~\bibnamefont {Rastelli}},\ and\ \bibinfo {author} {\bibfnamefont {V.}~\bibnamefont {Zwiller}},\ }\bibfield  {title} {\bibinfo {title} {On-demand generation of background-free single photons from a solid-state source},\ }\href {https://doi.org/10.1063/1.5020038} {\bibfield  {journal} {\bibinfo  {journal} {Applied Physics Letters}\ }\textbf {\bibinfo {volume} {112}},\ \bibinfo
  {pages} {093106} (\bibinfo {year} {2018})}\BibitemShut {NoStop}%
\bibitem [{\citenamefont {Hanschke}\ \emph {et~al.}(2018)\citenamefont {Hanschke}, \citenamefont {Fischer}, \citenamefont {Appel}, \citenamefont {Lukin}, \citenamefont {Wierzbowski}, \citenamefont {Sun}, \citenamefont {Trivedi}, \citenamefont {Vu{\v{c}}kovi{\'c}}, \citenamefont {Finley},\ and\ \citenamefont {M{\"u}ller}}]{HF:npj2018}%
  \BibitemOpen
  \bibfield  {author} {\bibinfo {author} {\bibfnamefont {L.}~\bibnamefont {Hanschke}}, \bibinfo {author} {\bibfnamefont {K.~A.}\ \bibnamefont {Fischer}}, \bibinfo {author} {\bibfnamefont {S.}~\bibnamefont {Appel}}, \bibinfo {author} {\bibfnamefont {D.}~\bibnamefont {Lukin}}, \bibinfo {author} {\bibfnamefont {J.}~\bibnamefont {Wierzbowski}}, \bibinfo {author} {\bibfnamefont {S.}~\bibnamefont {Sun}}, \bibinfo {author} {\bibfnamefont {R.}~\bibnamefont {Trivedi}}, \bibinfo {author} {\bibfnamefont {J.}~\bibnamefont {Vu{\v{c}}kovi{\'c}}}, \bibinfo {author} {\bibfnamefont {J.~J.}\ \bibnamefont {Finley}},\ and\ \bibinfo {author} {\bibfnamefont {K.}~\bibnamefont {M{\"u}ller}},\ }\bibfield  {title} {\bibinfo {title} {Quantum dot single-photon sources with ultra-low multi-photon probability},\ }\bibfield  {journal} {\bibinfo  {journal} {npj Quantum Information}\ }\textbf {\bibinfo {volume} {4}},\ \href {https://doi.org/10.1038/s41534-018-0092-0} {10.1038/s41534-018-0092-0} (\bibinfo {year} {2018})\BibitemShut {NoStop}%
\bibitem [{\citenamefont {Tomm}\ \emph {et~al.}(2021)\citenamefont {Tomm}, \citenamefont {Javadi}, \citenamefont {Antoniadis}, \citenamefont {Najer}, \citenamefont {L{\"o}bl}, \citenamefont {Korsch}, \citenamefont {Schott}, \citenamefont {Valentin}, \citenamefont {Wieck}, \citenamefont {Ludwig},\ and\ \citenamefont {Warburton}}]{TJN:NNano2021}%
  \BibitemOpen
  \bibfield  {author} {\bibinfo {author} {\bibfnamefont {N.}~\bibnamefont {Tomm}}, \bibinfo {author} {\bibfnamefont {A.}~\bibnamefont {Javadi}}, \bibinfo {author} {\bibfnamefont {N.~O.}\ \bibnamefont {Antoniadis}}, \bibinfo {author} {\bibfnamefont {D.}~\bibnamefont {Najer}}, \bibinfo {author} {\bibfnamefont {M.~C.}\ \bibnamefont {L{\"o}bl}}, \bibinfo {author} {\bibfnamefont {A.~R.}\ \bibnamefont {Korsch}}, \bibinfo {author} {\bibfnamefont {R.}~\bibnamefont {Schott}}, \bibinfo {author} {\bibfnamefont {S.~R.}\ \bibnamefont {Valentin}}, \bibinfo {author} {\bibfnamefont {A.~D.}\ \bibnamefont {Wieck}}, \bibinfo {author} {\bibfnamefont {A.}~\bibnamefont {Ludwig}},\ and\ \bibinfo {author} {\bibfnamefont {R.~J.}\ \bibnamefont {Warburton}},\ }\bibfield  {title} {\bibinfo {title} {A bright and fast source of coherent single photons},\ }\href {https://doi.org/10.1038/s41565-020-00831-x} {\bibfield  {journal} {\bibinfo  {journal} {Nature Nanotechnology}\ }\textbf {\bibinfo {volume} {16}},\ \bibinfo {pages} {399}
  (\bibinfo {year} {2021})}\BibitemShut {NoStop}%
\bibitem [{\citenamefont {Hauser}\ \emph {et~al.}()\citenamefont {Hauser}, \citenamefont {Bayerbach}, \citenamefont {Kaupp}, \citenamefont {Reum}, \citenamefont {Peniakov}, \citenamefont {Michl}, \citenamefont {Kamp}, \citenamefont {Huber-Loyola}, \citenamefont {Pfenning}, \citenamefont {Höfling},\ and\ \citenamefont {Barz}}]{HBK:arxiv2025}%
  \BibitemOpen
  \bibfield  {author} {\bibinfo {author} {\bibfnamefont {N.}~\bibnamefont {Hauser}}, \bibinfo {author} {\bibfnamefont {M.}~\bibnamefont {Bayerbach}}, \bibinfo {author} {\bibfnamefont {J.}~\bibnamefont {Kaupp}}, \bibinfo {author} {\bibfnamefont {Y.}~\bibnamefont {Reum}}, \bibinfo {author} {\bibfnamefont {G.}~\bibnamefont {Peniakov}}, \bibinfo {author} {\bibfnamefont {J.}~\bibnamefont {Michl}}, \bibinfo {author} {\bibfnamefont {M.}~\bibnamefont {Kamp}}, \bibinfo {author} {\bibfnamefont {T.}~\bibnamefont {Huber-Loyola}}, \bibinfo {author} {\bibfnamefont {A.~T.}\ \bibnamefont {Pfenning}}, \bibinfo {author} {\bibfnamefont {S.}~\bibnamefont {Höfling}},\ and\ \bibinfo {author} {\bibfnamefont {S.}~\bibnamefont {Barz}},\ }\href@noop {} {\bibinfo {title} {Deterministic and highly indistinguishable single photons in the telecom c-band}},\ \Eprint {https://arxiv.org/abs/2505.09695} {arXiv:2505.09695 [quant-ph]} \BibitemShut {NoStop}%
\bibitem [{\citenamefont {Nowak}\ \emph {et~al.}(2014)\citenamefont {Nowak}, \citenamefont {Portalupi}, \citenamefont {Giesz}, \citenamefont {Gazzano}, \citenamefont {{Dal Savio}}, \citenamefont {Braun}, \citenamefont {Karrai}, \citenamefont {Arnold}, \citenamefont {Lanco}, \citenamefont {Sagnes}, \citenamefont {Lema{\^i}tre},\ and\ \citenamefont {Senellart}}]{NPG:NComms2014}%
  \BibitemOpen
  \bibfield  {author} {\bibinfo {author} {\bibfnamefont {A.~K.}\ \bibnamefont {Nowak}}, \bibinfo {author} {\bibfnamefont {S.~L.}\ \bibnamefont {Portalupi}}, \bibinfo {author} {\bibfnamefont {V.}~\bibnamefont {Giesz}}, \bibinfo {author} {\bibfnamefont {O.}~\bibnamefont {Gazzano}}, \bibinfo {author} {\bibfnamefont {C.}~\bibnamefont {{Dal Savio}}}, \bibinfo {author} {\bibfnamefont {P.-F.}\ \bibnamefont {Braun}}, \bibinfo {author} {\bibfnamefont {K.}~\bibnamefont {Karrai}}, \bibinfo {author} {\bibfnamefont {C.}~\bibnamefont {Arnold}}, \bibinfo {author} {\bibfnamefont {L.}~\bibnamefont {Lanco}}, \bibinfo {author} {\bibfnamefont {I.}~\bibnamefont {Sagnes}}, \bibinfo {author} {\bibfnamefont {A.}~\bibnamefont {Lema{\^i}tre}},\ and\ \bibinfo {author} {\bibfnamefont {P.}~\bibnamefont {Senellart}},\ }\bibfield  {title} {\bibinfo {title} {Deterministic and electrically tunable bright single-photon source},\ }\href {https://doi.org/10.1038/ncomms4240} {\bibfield  {journal} {\bibinfo  {journal} {Nature Communications}\
  }\textbf {\bibinfo {volume} {5}},\ \bibinfo {pages} {3240} (\bibinfo {year} {2014})}\BibitemShut {NoStop}%
\bibitem [{\citenamefont {Bennett}\ \emph {et~al.}(2010)\citenamefont {Bennett}, \citenamefont {Patel}, \citenamefont {Skiba-Szymanska}, \citenamefont {Nicoll}, \citenamefont {Farrer}, \citenamefont {Ritchie},\ and\ \citenamefont {Shields}}]{BPS:APL2010}%
  \BibitemOpen
  \bibfield  {author} {\bibinfo {author} {\bibfnamefont {A.~J.}\ \bibnamefont {Bennett}}, \bibinfo {author} {\bibfnamefont {R.~B.}\ \bibnamefont {Patel}}, \bibinfo {author} {\bibfnamefont {J.}~\bibnamefont {Skiba-Szymanska}}, \bibinfo {author} {\bibfnamefont {C.~A.}\ \bibnamefont {Nicoll}}, \bibinfo {author} {\bibfnamefont {I.}~\bibnamefont {Farrer}}, \bibinfo {author} {\bibfnamefont {D.~A.}\ \bibnamefont {Ritchie}},\ and\ \bibinfo {author} {\bibfnamefont {A.~J.}\ \bibnamefont {Shields}},\ }\bibfield  {title} {\bibinfo {title} {Giant stark effect in the emission of single semiconductor quantum dots},\ }\bibfield  {journal} {\bibinfo  {journal} {Applied Physics Letters}\ }\textbf {\bibinfo {volume} {97}},\ \href {https://doi.org/10.1063/1.3460912} {10.1063/1.3460912} (\bibinfo {year} {2010})\BibitemShut {NoStop}%
\bibitem [{\citenamefont {Hanschke}\ \emph {et~al.}()\citenamefont {Hanschke}, \citenamefont {Bracht}, \citenamefont {Schöll}, \citenamefont {Bauch}, \citenamefont {Berger}, \citenamefont {Kallert}, \citenamefont {Peter}, \citenamefont {Jr.}, \citenamefont {da~Silva}, \citenamefont {Manna}, \citenamefont {Rastelli}, \citenamefont {Schumacher}, \citenamefont {Reiter},\ and\ \citenamefont {Jöns}}]{HBS:arxiv2024}%
  \BibitemOpen
  \bibfield  {author} {\bibinfo {author} {\bibfnamefont {L.}~\bibnamefont {Hanschke}}, \bibinfo {author} {\bibfnamefont {T.~K.}\ \bibnamefont {Bracht}}, \bibinfo {author} {\bibfnamefont {E.}~\bibnamefont {Schöll}}, \bibinfo {author} {\bibfnamefont {D.}~\bibnamefont {Bauch}}, \bibinfo {author} {\bibfnamefont {E.}~\bibnamefont {Berger}}, \bibinfo {author} {\bibfnamefont {P.}~\bibnamefont {Kallert}}, \bibinfo {author} {\bibfnamefont {M.}~\bibnamefont {Peter}}, \bibinfo {author} {\bibfnamefont {A.~J.~G.}\ \bibnamefont {Jr.}}, \bibinfo {author} {\bibfnamefont {S.~F.~C.}\ \bibnamefont {da~Silva}}, \bibinfo {author} {\bibfnamefont {S.}~\bibnamefont {Manna}}, \bibinfo {author} {\bibfnamefont {A.}~\bibnamefont {Rastelli}}, \bibinfo {author} {\bibfnamefont {S.}~\bibnamefont {Schumacher}}, \bibinfo {author} {\bibfnamefont {D.~E.}\ \bibnamefont {Reiter}},\ and\ \bibinfo {author} {\bibfnamefont {K.~D.}\ \bibnamefont {Jöns}},\ }\href@noop {} {\bibinfo {title} {Experimental measurement of the reappearance of rabi
  rotations in semiconductor quantum dots}},\ \Eprint {https://arxiv.org/abs/2409.19167} {arXiv:2409.19167 [cond-mat.mes-hall]} \BibitemShut {NoStop}%
\bibitem [{\citenamefont {Tighineanu}\ \emph {et~al.}(2018)\citenamefont {Tighineanu}, \citenamefont {Dree\ss{}en}, \citenamefont {Flindt}, \citenamefont {Lodahl},\ and\ \citenamefont {S\o{}rensen}}]{TDF:PRL2018}%
  \BibitemOpen
  \bibfield  {author} {\bibinfo {author} {\bibfnamefont {P.}~\bibnamefont {Tighineanu}}, \bibinfo {author} {\bibfnamefont {C.~L.}\ \bibnamefont {Dree\ss{}en}}, \bibinfo {author} {\bibfnamefont {C.}~\bibnamefont {Flindt}}, \bibinfo {author} {\bibfnamefont {P.}~\bibnamefont {Lodahl}},\ and\ \bibinfo {author} {\bibfnamefont {A.~S.}\ \bibnamefont {S\o{}rensen}},\ }\bibfield  {title} {\bibinfo {title} {Phonon decoherence of quantum dots in photonic structures: Broadening of the zero-phonon line and the role of dimensionality},\ }\href {https://doi.org/10.1103/PhysRevLett.120.257401} {\bibfield  {journal} {\bibinfo  {journal} {Phys. Rev. Lett.}\ }\textbf {\bibinfo {volume} {120}},\ \bibinfo {pages} {257401} (\bibinfo {year} {2018})}\BibitemShut {NoStop}%
\bibitem [{\citenamefont {Liu}\ \emph {et~al.}(2018)\citenamefont {Liu}, \citenamefont {Konthasinghe}, \citenamefont {Davan\ifmmode~\mbox{\c{c}}\else \c{c}\fi{}o}, \citenamefont {Lawall}, \citenamefont {Anant}, \citenamefont {Verma}, \citenamefont {Mirin}, \citenamefont {Nam}, \citenamefont {Song}, \citenamefont {Ma}, \citenamefont {Chen}, \citenamefont {Ni}, \citenamefont {Niu},\ and\ \citenamefont {Srinivasan}}]{LKD:PRAppl2018}%
  \BibitemOpen
  \bibfield  {author} {\bibinfo {author} {\bibfnamefont {J.}~\bibnamefont {Liu}}, \bibinfo {author} {\bibfnamefont {K.}~\bibnamefont {Konthasinghe}}, \bibinfo {author} {\bibfnamefont {M.}~\bibnamefont {Davan\ifmmode~\mbox{\c{c}}\else \c{c}\fi{}o}}, \bibinfo {author} {\bibfnamefont {J.}~\bibnamefont {Lawall}}, \bibinfo {author} {\bibfnamefont {V.}~\bibnamefont {Anant}}, \bibinfo {author} {\bibfnamefont {V.}~\bibnamefont {Verma}}, \bibinfo {author} {\bibfnamefont {R.}~\bibnamefont {Mirin}}, \bibinfo {author} {\bibfnamefont {S.~W.}\ \bibnamefont {Nam}}, \bibinfo {author} {\bibfnamefont {J.~D.}\ \bibnamefont {Song}}, \bibinfo {author} {\bibfnamefont {B.}~\bibnamefont {Ma}}, \bibinfo {author} {\bibfnamefont {Z.~S.}\ \bibnamefont {Chen}}, \bibinfo {author} {\bibfnamefont {H.~Q.}\ \bibnamefont {Ni}}, \bibinfo {author} {\bibfnamefont {Z.~C.}\ \bibnamefont {Niu}},\ and\ \bibinfo {author} {\bibfnamefont {K.}~\bibnamefont {Srinivasan}},\ }\bibfield  {title} {\bibinfo {title} {Single self-assembled
  $\mathrm{InAs}/\mathrm{GaAs}$ quantum dots in photonic nanostructures: The role of nanofabrication},\ }\href {https://doi.org/10.1103/PhysRevApplied.9.064019} {\bibfield  {journal} {\bibinfo  {journal} {Phys. Rev. Appl.}\ }\textbf {\bibinfo {volume} {9}},\ \bibinfo {pages} {064019} (\bibinfo {year} {2018})}\BibitemShut {NoStop}%
\bibitem [{\citenamefont {Stockill}\ \emph {et~al.}(2016)\citenamefont {Stockill}, \citenamefont {{Le Gall}}, \citenamefont {Matthiesen}, \citenamefont {Huthmacher}, \citenamefont {Clarke}, \citenamefont {Hugues},\ and\ \citenamefont {Atat{\"u}re}}]{SGM:NComms2016}%
  \BibitemOpen
  \bibfield  {author} {\bibinfo {author} {\bibfnamefont {R.}~\bibnamefont {Stockill}}, \bibinfo {author} {\bibfnamefont {C.}~\bibnamefont {{Le Gall}}}, \bibinfo {author} {\bibfnamefont {C.}~\bibnamefont {Matthiesen}}, \bibinfo {author} {\bibfnamefont {L.}~\bibnamefont {Huthmacher}}, \bibinfo {author} {\bibfnamefont {E.}~\bibnamefont {Clarke}}, \bibinfo {author} {\bibfnamefont {M.}~\bibnamefont {Hugues}},\ and\ \bibinfo {author} {\bibfnamefont {M.}~\bibnamefont {Atat{\"u}re}},\ }\bibfield  {title} {\bibinfo {title} {Quantum dot spin coherence governed by a strained nuclear environment},\ }\href {https://doi.org/10.1038/ncomms12745} {\bibfield  {journal} {\bibinfo  {journal} {Nature Communications}\ }\textbf {\bibinfo {volume} {7}},\ \bibinfo {pages} {12745} (\bibinfo {year} {2016})}\BibitemShut {NoStop}%
\bibitem [{\citenamefont {Yoneda}\ \emph {et~al.}(2018)\citenamefont {Yoneda}, \citenamefont {Takeda}, \citenamefont {Otsuka}, \citenamefont {Nakajima}, \citenamefont {Delbecq}, \citenamefont {Allison}, \citenamefont {Honda}, \citenamefont {Kodera}, \citenamefont {Oda}, \citenamefont {Hoshi}, \citenamefont {Usami}, \citenamefont {Itoh},\ and\ \citenamefont {Tarucha}}]{YTO:NNano2018}%
  \BibitemOpen
  \bibfield  {author} {\bibinfo {author} {\bibfnamefont {J.}~\bibnamefont {Yoneda}}, \bibinfo {author} {\bibfnamefont {K.}~\bibnamefont {Takeda}}, \bibinfo {author} {\bibfnamefont {T.}~\bibnamefont {Otsuka}}, \bibinfo {author} {\bibfnamefont {T.}~\bibnamefont {Nakajima}}, \bibinfo {author} {\bibfnamefont {M.~R.}\ \bibnamefont {Delbecq}}, \bibinfo {author} {\bibfnamefont {G.}~\bibnamefont {Allison}}, \bibinfo {author} {\bibfnamefont {T.}~\bibnamefont {Honda}}, \bibinfo {author} {\bibfnamefont {T.}~\bibnamefont {Kodera}}, \bibinfo {author} {\bibfnamefont {S.}~\bibnamefont {Oda}}, \bibinfo {author} {\bibfnamefont {Y.}~\bibnamefont {Hoshi}}, \bibinfo {author} {\bibfnamefont {N.}~\bibnamefont {Usami}}, \bibinfo {author} {\bibfnamefont {K.~M.}\ \bibnamefont {Itoh}},\ and\ \bibinfo {author} {\bibfnamefont {S.}~\bibnamefont {Tarucha}},\ }\bibfield  {title} {\bibinfo {title} {A quantum-dot spin qubit with coherence limited by charge noise and fidelity higher than 99.9{\%}},\ }\href
  {https://doi.org/10.1038/s41565-017-0014-x} {\bibfield  {journal} {\bibinfo  {journal} {Nature Nanotechnology}\ }\textbf {\bibinfo {volume} {13}},\ \bibinfo {pages} {102} (\bibinfo {year} {2018})}\BibitemShut {NoStop}%
\bibitem [{\citenamefont {Zaporski}\ \emph {et~al.}(2023)\citenamefont {Zaporski}, \citenamefont {Shofer}, \citenamefont {Bodey}, \citenamefont {Manna}, \citenamefont {Gillard}, \citenamefont {Appel}, \citenamefont {Schimpf}, \citenamefont {{Da Covre Silva}}, \citenamefont {Jarman}, \citenamefont {Delamare}, \citenamefont {Park}, \citenamefont {Haeusler}, \citenamefont {Chekhovich}, \citenamefont {Rastelli}, \citenamefont {Gangloff}, \citenamefont {Atat{\"u}re},\ and\ \citenamefont {{Le Gall}}}]{ZSB:NNano2023}%
  \BibitemOpen
  \bibfield  {author} {\bibinfo {author} {\bibfnamefont {L.}~\bibnamefont {Zaporski}}, \bibinfo {author} {\bibfnamefont {N.}~\bibnamefont {Shofer}}, \bibinfo {author} {\bibfnamefont {J.~H.}\ \bibnamefont {Bodey}}, \bibinfo {author} {\bibfnamefont {S.}~\bibnamefont {Manna}}, \bibinfo {author} {\bibfnamefont {G.}~\bibnamefont {Gillard}}, \bibinfo {author} {\bibfnamefont {M.~H.}\ \bibnamefont {Appel}}, \bibinfo {author} {\bibfnamefont {C.}~\bibnamefont {Schimpf}}, \bibinfo {author} {\bibfnamefont {S.~F.}\ \bibnamefont {{Da Covre Silva}}}, \bibinfo {author} {\bibfnamefont {J.}~\bibnamefont {Jarman}}, \bibinfo {author} {\bibfnamefont {G.}~\bibnamefont {Delamare}}, \bibinfo {author} {\bibfnamefont {G.}~\bibnamefont {Park}}, \bibinfo {author} {\bibfnamefont {U.}~\bibnamefont {Haeusler}}, \bibinfo {author} {\bibfnamefont {E.~A.}\ \bibnamefont {Chekhovich}}, \bibinfo {author} {\bibfnamefont {A.}~\bibnamefont {Rastelli}}, \bibinfo {author} {\bibfnamefont {D.~A.}\ \bibnamefont {Gangloff}}, \bibinfo {author}
  {\bibfnamefont {M.}~\bibnamefont {Atat{\"u}re}},\ and\ \bibinfo {author} {\bibfnamefont {C.}~\bibnamefont {{Le Gall}}},\ }\bibfield  {title} {\bibinfo {title} {Ideal refocusing of an optically active spin qubit under strong hyperfine interactions},\ }\href {https://doi.org/10.1038/s41565-022-01282-2} {\bibfield  {journal} {\bibinfo  {journal} {Nature Nanotechnology}\ }\textbf {\bibinfo {volume} {18}},\ \bibinfo {pages} {257} (\bibinfo {year} {2023})}\BibitemShut {NoStop}%
\bibitem [{\citenamefont {Reindl}\ \emph {et~al.}(2019)\citenamefont {Reindl}, \citenamefont {Weber}, \citenamefont {Huber}, \citenamefont {Schimpf}, \citenamefont {Covre~da Silva}, \citenamefont {Portalupi}, \citenamefont {Trotta}, \citenamefont {Michler},\ and\ \citenamefont {Rastelli}}]{RWH:PRB2019}%
  \BibitemOpen
  \bibfield  {author} {\bibinfo {author} {\bibfnamefont {M.}~\bibnamefont {Reindl}}, \bibinfo {author} {\bibfnamefont {J.~H.}\ \bibnamefont {Weber}}, \bibinfo {author} {\bibfnamefont {D.}~\bibnamefont {Huber}}, \bibinfo {author} {\bibfnamefont {C.}~\bibnamefont {Schimpf}}, \bibinfo {author} {\bibfnamefont {S.~F.}\ \bibnamefont {Covre~da Silva}}, \bibinfo {author} {\bibfnamefont {S.~L.}\ \bibnamefont {Portalupi}}, \bibinfo {author} {\bibfnamefont {R.}~\bibnamefont {Trotta}}, \bibinfo {author} {\bibfnamefont {P.}~\bibnamefont {Michler}},\ and\ \bibinfo {author} {\bibfnamefont {A.}~\bibnamefont {Rastelli}},\ }\bibfield  {title} {\bibinfo {title} {Highly indistinguishable single photons from incoherently excited quantum dots},\ }\href {https://doi.org/10.1103/PhysRevB.100.155420} {\bibfield  {journal} {\bibinfo  {journal} {Phys. Rev. B}\ }\textbf {\bibinfo {volume} {100}},\ \bibinfo {pages} {155420} (\bibinfo {year} {2019})}\BibitemShut {NoStop}%
\bibitem [{\citenamefont {Gl{\"a}ssl}\ \emph {et~al.}(2013)\citenamefont {Gl{\"a}ssl}, \citenamefont {Barth},\ and\ \citenamefont {Axt}}]{GBA:PRL2013}%
  \BibitemOpen
  \bibfield  {author} {\bibinfo {author} {\bibfnamefont {M.}~\bibnamefont {Gl{\"a}ssl}}, \bibinfo {author} {\bibfnamefont {A.~M.}\ \bibnamefont {Barth}},\ and\ \bibinfo {author} {\bibfnamefont {V.~M.}\ \bibnamefont {Axt}},\ }\bibfield  {title} {\bibinfo {title} {Proposed robust and high-fidelity preparation of excitons and biexcitons in semiconductor quantum dots making active use of phonons},\ }\href {https://doi.org/10.1103/PhysRevLett.110.147401} {\bibfield  {journal} {\bibinfo  {journal} {Phys. Rev. Lett.}\ }\textbf {\bibinfo {volume} {110}},\ \bibinfo {pages} {147401} (\bibinfo {year} {2013})}\BibitemShut {NoStop}%
\bibitem [{\citenamefont {Vyvlecka}\ \emph {et~al.}(2023)\citenamefont {Vyvlecka}, \citenamefont {Jehle}, \citenamefont {Nawrath}, \citenamefont {Giorgino}, \citenamefont {Bozzio}, \citenamefont {Sittig}, \citenamefont {Jetter}, \citenamefont {Portalupi}, \citenamefont {Michler},\ and\ \citenamefont {Walther}}]{VJN:APL23}%
  \BibitemOpen
  \bibfield  {author} {\bibinfo {author} {\bibfnamefont {M.}~\bibnamefont {Vyvlecka}}, \bibinfo {author} {\bibfnamefont {L.}~\bibnamefont {Jehle}}, \bibinfo {author} {\bibfnamefont {C.}~\bibnamefont {Nawrath}}, \bibinfo {author} {\bibfnamefont {F.}~\bibnamefont {Giorgino}}, \bibinfo {author} {\bibfnamefont {M.}~\bibnamefont {Bozzio}}, \bibinfo {author} {\bibfnamefont {R.}~\bibnamefont {Sittig}}, \bibinfo {author} {\bibfnamefont {M.}~\bibnamefont {Jetter}}, \bibinfo {author} {\bibfnamefont {S.~L.}\ \bibnamefont {Portalupi}}, \bibinfo {author} {\bibfnamefont {P.}~\bibnamefont {Michler}},\ and\ \bibinfo {author} {\bibfnamefont {P.}~\bibnamefont {Walther}},\ }\bibfield  {title} {\bibinfo {title} {{Robust excitation of C-band quantum dots for quantum communication}},\ }\href {https://doi.org/10.1063/5.0166285} {\bibfield  {journal} {\bibinfo  {journal} {Appl. Phys. Lett.}\ }\textbf {\bibinfo {volume} {123}},\ \bibinfo {pages} {174001} (\bibinfo {year} {2023})}\BibitemShut {NoStop}%
\bibitem [{\citenamefont {Laccotripes}\ \emph {et~al.}()\citenamefont {Laccotripes}, \citenamefont {Huang}, \citenamefont {Shooter}, \citenamefont {Barbiero}, \citenamefont {Winnel}, \citenamefont {Ritchie}, \citenamefont {Shields}, \citenamefont {Muller},\ and\ \citenamefont {Stevenson}}]{LHS:arxiv2025}%
  \BibitemOpen
  \bibfield  {author} {\bibinfo {author} {\bibfnamefont {P.}~\bibnamefont {Laccotripes}}, \bibinfo {author} {\bibfnamefont {J.}~\bibnamefont {Huang}}, \bibinfo {author} {\bibfnamefont {G.}~\bibnamefont {Shooter}}, \bibinfo {author} {\bibfnamefont {A.}~\bibnamefont {Barbiero}}, \bibinfo {author} {\bibfnamefont {M.~S.}\ \bibnamefont {Winnel}}, \bibinfo {author} {\bibfnamefont {D.~A.}\ \bibnamefont {Ritchie}}, \bibinfo {author} {\bibfnamefont {A.~J.}\ \bibnamefont {Shields}}, \bibinfo {author} {\bibfnamefont {T.}~\bibnamefont {Muller}},\ and\ \bibinfo {author} {\bibfnamefont {R.~M.}\ \bibnamefont {Stevenson}},\ }\href@noop {} {\bibinfo {title} {An entangled photon source for the telecom c-band based on a semiconductor-confined spin}},\ \Eprint {https://arxiv.org/abs/2507.01648} {arXiv:2507.01648 [quant-ph]} \BibitemShut {NoStop}%
\bibitem [{\citenamefont {Bozzio}\ \emph {et~al.}(2022)\citenamefont {Bozzio}, \citenamefont {Vyvlecka}, \citenamefont {Cosacchi}, \citenamefont {Nawrath}, \citenamefont {Seidelmann}, \citenamefont {Loredo}, \citenamefont {Portalupi}, \citenamefont {Axt}, \citenamefont {Michler},\ and\ \citenamefont {Walther}}]{BVN:npj2022}%
  \BibitemOpen
  \bibfield  {author} {\bibinfo {author} {\bibfnamefont {M.}~\bibnamefont {Bozzio}}, \bibinfo {author} {\bibfnamefont {M.}~\bibnamefont {Vyvlecka}}, \bibinfo {author} {\bibfnamefont {M.}~\bibnamefont {Cosacchi}}, \bibinfo {author} {\bibfnamefont {C.}~\bibnamefont {Nawrath}}, \bibinfo {author} {\bibfnamefont {T.}~\bibnamefont {Seidelmann}}, \bibinfo {author} {\bibfnamefont {J.~C.}\ \bibnamefont {Loredo}}, \bibinfo {author} {\bibfnamefont {S.~L.}\ \bibnamefont {Portalupi}}, \bibinfo {author} {\bibfnamefont {V.~M.}\ \bibnamefont {Axt}}, \bibinfo {author} {\bibfnamefont {P.}~\bibnamefont {Michler}},\ and\ \bibinfo {author} {\bibfnamefont {P.}~\bibnamefont {Walther}},\ }\bibfield  {title} {\bibinfo {title} {Enhancing quantum cryptography with quantum dot single-photon sources},\ }\bibfield  {journal} {\bibinfo  {journal} {npj Quantum Information}\ }\textbf {\bibinfo {volume} {8}},\ \href {https://doi.org/10.1038/s41534-022-00626-z} {10.1038/s41534-022-00626-z} (\bibinfo {year} {2022})\BibitemShut {NoStop}%
\bibitem [{\citenamefont {Margaria}\ \emph {et~al.}()\citenamefont {Margaria}, \citenamefont {Pastier}, \citenamefont {Bennour}, \citenamefont {Billard}, \citenamefont {Ivanov}, \citenamefont {Hease}, \citenamefont {Stepanov}, \citenamefont {Adiyatullin}, \citenamefont {Singla}, \citenamefont {Pont}, \citenamefont {Descampeaux}, \citenamefont {Bernard}, \citenamefont {Pishchagin}, \citenamefont {Morassi}, \citenamefont {Lemaître}, \citenamefont {Volz}, \citenamefont {Giesz}, \citenamefont {Somaschi}, \citenamefont {Maring}, \citenamefont {Boissier}, \citenamefont {Au},\ and\ \citenamefont {Senellart}}]{MPB:arxiv2024}%
  \BibitemOpen
  \bibfield  {author} {\bibinfo {author} {\bibfnamefont {N.}~\bibnamefont {Margaria}}, \bibinfo {author} {\bibfnamefont {F.}~\bibnamefont {Pastier}}, \bibinfo {author} {\bibfnamefont {T.}~\bibnamefont {Bennour}}, \bibinfo {author} {\bibfnamefont {M.}~\bibnamefont {Billard}}, \bibinfo {author} {\bibfnamefont {E.}~\bibnamefont {Ivanov}}, \bibinfo {author} {\bibfnamefont {W.}~\bibnamefont {Hease}}, \bibinfo {author} {\bibfnamefont {P.}~\bibnamefont {Stepanov}}, \bibinfo {author} {\bibfnamefont {A.~F.}\ \bibnamefont {Adiyatullin}}, \bibinfo {author} {\bibfnamefont {R.}~\bibnamefont {Singla}}, \bibinfo {author} {\bibfnamefont {M.}~\bibnamefont {Pont}}, \bibinfo {author} {\bibfnamefont {M.}~\bibnamefont {Descampeaux}}, \bibinfo {author} {\bibfnamefont {A.}~\bibnamefont {Bernard}}, \bibinfo {author} {\bibfnamefont {A.}~\bibnamefont {Pishchagin}}, \bibinfo {author} {\bibfnamefont {M.}~\bibnamefont {Morassi}}, \bibinfo {author} {\bibfnamefont {A.}~\bibnamefont {Lemaître}}, \bibinfo {author} {\bibfnamefont
  {T.}~\bibnamefont {Volz}}, \bibinfo {author} {\bibfnamefont {V.}~\bibnamefont {Giesz}}, \bibinfo {author} {\bibfnamefont {N.}~\bibnamefont {Somaschi}}, \bibinfo {author} {\bibfnamefont {N.}~\bibnamefont {Maring}}, \bibinfo {author} {\bibfnamefont {S.}~\bibnamefont {Boissier}}, \bibinfo {author} {\bibfnamefont {T.~H.}\ \bibnamefont {Au}},\ and\ \bibinfo {author} {\bibfnamefont {P.}~\bibnamefont {Senellart}},\ }\href@noop {} {\bibinfo {title} {Efficient fiber-pigtailed source of indistinguishable single photons}},\ \Eprint {https://arxiv.org/abs/2410.07760} {arXiv:2410.07760 [quant-ph]} \BibitemShut {NoStop}%
\bibitem [{\citenamefont {Coste}\ \emph {et~al.}(2023)\citenamefont {Coste}, \citenamefont {Fioretto}, \citenamefont {Belabas}, \citenamefont {Wein}, \citenamefont {Hilaire}, \citenamefont {Frantzeskakis}, \citenamefont {Gundin}, \citenamefont {Goes}, \citenamefont {Somaschi}, \citenamefont {Morassi}, \citenamefont {Lema{\^i}tre}, \citenamefont {Sagnes}, \citenamefont {Harouri}, \citenamefont {Economou}, \citenamefont {Auffeves}, \citenamefont {Krebs}, \citenamefont {Lanco},\ and\ \citenamefont {Senellart}}]{CFB:NPhot2023}%
  \BibitemOpen
  \bibfield  {author} {\bibinfo {author} {\bibfnamefont {N.}~\bibnamefont {Coste}}, \bibinfo {author} {\bibfnamefont {D.~A.}\ \bibnamefont {Fioretto}}, \bibinfo {author} {\bibfnamefont {N.}~\bibnamefont {Belabas}}, \bibinfo {author} {\bibfnamefont {S.~C.}\ \bibnamefont {Wein}}, \bibinfo {author} {\bibfnamefont {P.}~\bibnamefont {Hilaire}}, \bibinfo {author} {\bibfnamefont {R.}~\bibnamefont {Frantzeskakis}}, \bibinfo {author} {\bibfnamefont {M.}~\bibnamefont {Gundin}}, \bibinfo {author} {\bibfnamefont {B.}~\bibnamefont {Goes}}, \bibinfo {author} {\bibfnamefont {N.}~\bibnamefont {Somaschi}}, \bibinfo {author} {\bibfnamefont {M.}~\bibnamefont {Morassi}}, \bibinfo {author} {\bibfnamefont {A.}~\bibnamefont {Lema{\^i}tre}}, \bibinfo {author} {\bibfnamefont {I.}~\bibnamefont {Sagnes}}, \bibinfo {author} {\bibfnamefont {A.}~\bibnamefont {Harouri}}, \bibinfo {author} {\bibfnamefont {S.~E.}\ \bibnamefont {Economou}}, \bibinfo {author} {\bibfnamefont {A.}~\bibnamefont {Auffeves}}, \bibinfo {author} {\bibfnamefont
  {O.}~\bibnamefont {Krebs}}, \bibinfo {author} {\bibfnamefont {L.}~\bibnamefont {Lanco}},\ and\ \bibinfo {author} {\bibfnamefont {P.}~\bibnamefont {Senellart}},\ }\bibfield  {title} {\bibinfo {title} {High-rate entanglement between a semiconductor spin and indistinguishable photons},\ }\href {https://doi.org/10.1038/s41566-023-01186-0} {\bibfield  {journal} {\bibinfo  {journal} {Nature Photonics}\ }\textbf {\bibinfo {volume} {17}},\ \bibinfo {pages} {582} (\bibinfo {year} {2023})}\BibitemShut {NoStop}%
\bibitem [{\citenamefont {Seidelmann}\ \emph {et~al.}(2022)\citenamefont {Seidelmann}, \citenamefont {Schimpf}, \citenamefont {Bracht}, \citenamefont {Cosacchi}, \citenamefont {Vagov}, \citenamefont {Rastelli}, \citenamefont {Reiter},\ and\ \citenamefont {Axt}}]{SSB:PRL2022}%
  \BibitemOpen
  \bibfield  {author} {\bibinfo {author} {\bibfnamefont {T.}~\bibnamefont {Seidelmann}}, \bibinfo {author} {\bibfnamefont {C.}~\bibnamefont {Schimpf}}, \bibinfo {author} {\bibfnamefont {T.~K.}\ \bibnamefont {Bracht}}, \bibinfo {author} {\bibfnamefont {M.}~\bibnamefont {Cosacchi}}, \bibinfo {author} {\bibfnamefont {A.}~\bibnamefont {Vagov}}, \bibinfo {author} {\bibfnamefont {A.}~\bibnamefont {Rastelli}}, \bibinfo {author} {\bibfnamefont {D.~E.}\ \bibnamefont {Reiter}},\ and\ \bibinfo {author} {\bibfnamefont {V.~M.}\ \bibnamefont {Axt}},\ }\bibfield  {title} {\bibinfo {title} {Two-photon excitation sets limit to entangled photon pair generation from quantum emitters},\ }\href {https://doi.org/10.1103/PhysRevLett.129.193604} {\bibfield  {journal} {\bibinfo  {journal} {Phys. Rev. Lett.}\ }\textbf {\bibinfo {volume} {129}},\ \bibinfo {pages} {193604} (\bibinfo {year} {2022})}\BibitemShut {NoStop}%
\bibitem [{\citenamefont {Sch\"oll}\ \emph {et~al.}(2020)\citenamefont {Sch\"oll}, \citenamefont {Schweickert}, \citenamefont {Hanschke}, \citenamefont {Zeuner}, \citenamefont {Sbresny}, \citenamefont {Lettner}, \citenamefont {Trivedi}, \citenamefont {Reindl}, \citenamefont {Covre~da Silva}, \citenamefont {Trotta}, \citenamefont {Finley}, \citenamefont {Vu\ifmmode \check{c}\else \v{c}\fi{}kovi\ifmmode~\acute{c}\else \'{c}\fi{}}, \citenamefont {M\"uller}, \citenamefont {Rastelli}, \citenamefont {Zwiller},\ and\ \citenamefont {J\"ons}}]{SSH:PRL2020}%
  \BibitemOpen
  \bibfield  {author} {\bibinfo {author} {\bibfnamefont {E.}~\bibnamefont {Sch\"oll}}, \bibinfo {author} {\bibfnamefont {L.}~\bibnamefont {Schweickert}}, \bibinfo {author} {\bibfnamefont {L.}~\bibnamefont {Hanschke}}, \bibinfo {author} {\bibfnamefont {K.~D.}\ \bibnamefont {Zeuner}}, \bibinfo {author} {\bibfnamefont {F.}~\bibnamefont {Sbresny}}, \bibinfo {author} {\bibfnamefont {T.}~\bibnamefont {Lettner}}, \bibinfo {author} {\bibfnamefont {R.}~\bibnamefont {Trivedi}}, \bibinfo {author} {\bibfnamefont {M.}~\bibnamefont {Reindl}}, \bibinfo {author} {\bibfnamefont {S.~F.}\ \bibnamefont {Covre~da Silva}}, \bibinfo {author} {\bibfnamefont {R.}~\bibnamefont {Trotta}}, \bibinfo {author} {\bibfnamefont {J.~J.}\ \bibnamefont {Finley}}, \bibinfo {author} {\bibfnamefont {J.}~\bibnamefont {Vu\ifmmode \check{c}\else \v{c}\fi{}kovi\ifmmode~\acute{c}\else \'{c}\fi{}}}, \bibinfo {author} {\bibfnamefont {K.}~\bibnamefont {M\"uller}}, \bibinfo {author} {\bibfnamefont {A.}~\bibnamefont {Rastelli}}, \bibinfo {author}
  {\bibfnamefont {V.}~\bibnamefont {Zwiller}},\ and\ \bibinfo {author} {\bibfnamefont {K.~D.}\ \bibnamefont {J\"ons}},\ }\bibfield  {title} {\bibinfo {title} {Crux of using the cascaded emission of a three-level quantum ladder system to generate indistinguishable photons},\ }\href {https://doi.org/10.1103/PhysRevLett.125.233605} {\bibfield  {journal} {\bibinfo  {journal} {Phys. Rev. Lett.}\ }\textbf {\bibinfo {volume} {125}},\ \bibinfo {pages} {233605} (\bibinfo {year} {2020})}\BibitemShut {NoStop}%
\bibitem [{\citenamefont {Cosacchi}\ \emph {et~al.}(2019)\citenamefont {Cosacchi}, \citenamefont {Ungar}, \citenamefont {Cygorek}, \citenamefont {Vagov},\ and\ \citenamefont {Axt}}]{CUC:PRL2019}%
  \BibitemOpen
  \bibfield  {author} {\bibinfo {author} {\bibfnamefont {M.}~\bibnamefont {Cosacchi}}, \bibinfo {author} {\bibfnamefont {F.}~\bibnamefont {Ungar}}, \bibinfo {author} {\bibfnamefont {M.}~\bibnamefont {Cygorek}}, \bibinfo {author} {\bibfnamefont {A.}~\bibnamefont {Vagov}},\ and\ \bibinfo {author} {\bibfnamefont {V.~M.}\ \bibnamefont {Axt}},\ }\bibfield  {title} {\bibinfo {title} {Emission-frequency separated high quality single-photon sources enabled by phonons},\ }\href {https://doi.org/10.1103/PhysRevLett.123.017403} {\bibfield  {journal} {\bibinfo  {journal} {Phys. Rev. Lett.}\ }\textbf {\bibinfo {volume} {123}},\ \bibinfo {pages} {017403} (\bibinfo {year} {2019})}\BibitemShut {NoStop}%
\bibitem [{\citenamefont {Rickert}\ \emph {et~al.}(2025)\citenamefont {Rickert}, \citenamefont {Vajner}, \citenamefont {von Helversen}, \citenamefont {Schall}, \citenamefont {Rodt}, \citenamefont {Reitzenstein}, \citenamefont {Liu}, \citenamefont {Li}, \citenamefont {Ni}, \citenamefont {Niu},\ and\ \citenamefont {Heindel}}]{RVH:ACSPhot2025}%
  \BibitemOpen
  \bibfield  {author} {\bibinfo {author} {\bibfnamefont {L.}~\bibnamefont {Rickert}}, \bibinfo {author} {\bibfnamefont {D.~A.}\ \bibnamefont {Vajner}}, \bibinfo {author} {\bibfnamefont {M.}~\bibnamefont {von Helversen}}, \bibinfo {author} {\bibfnamefont {J.}~\bibnamefont {Schall}}, \bibinfo {author} {\bibfnamefont {S.}~\bibnamefont {Rodt}}, \bibinfo {author} {\bibfnamefont {S.}~\bibnamefont {Reitzenstein}}, \bibinfo {author} {\bibfnamefont {H.}~\bibnamefont {Liu}}, \bibinfo {author} {\bibfnamefont {S.}~\bibnamefont {Li}}, \bibinfo {author} {\bibfnamefont {H.}~\bibnamefont {Ni}}, \bibinfo {author} {\bibfnamefont {Z.}~\bibnamefont {Niu}},\ and\ \bibinfo {author} {\bibfnamefont {T.}~\bibnamefont {Heindel}},\ }\bibfield  {title} {\bibinfo {title} {High purcell enhancement in quantum-dot hybrid circular bragg grating cavities for ghz clock rate generation of indistinguishable photons},\ }\href {https://doi.org/10.1021/acsphotonics.4c01873} {\bibfield  {journal} {\bibinfo  {journal} {ACS Photonics}\ }\textbf
  {\bibinfo {volume} {12}},\ \bibinfo {pages} {464} (\bibinfo {year} {2025})}\BibitemShut {NoStop}%
\bibitem [{\citenamefont {Barth}\ \emph {et~al.}(2016)\citenamefont {Barth}, \citenamefont {L\"uker}, \citenamefont {Vagov}, \citenamefont {Reiter}, \citenamefont {Kuhn},\ and\ \citenamefont {Axt}}]{BLV:PRB2016}%
  \BibitemOpen
  \bibfield  {author} {\bibinfo {author} {\bibfnamefont {A.~M.}\ \bibnamefont {Barth}}, \bibinfo {author} {\bibfnamefont {S.}~\bibnamefont {L\"uker}}, \bibinfo {author} {\bibfnamefont {A.}~\bibnamefont {Vagov}}, \bibinfo {author} {\bibfnamefont {D.~E.}\ \bibnamefont {Reiter}}, \bibinfo {author} {\bibfnamefont {T.}~\bibnamefont {Kuhn}},\ and\ \bibinfo {author} {\bibfnamefont {V.~M.}\ \bibnamefont {Axt}},\ }\bibfield  {title} {\bibinfo {title} {Fast and selective phonon-assisted state preparation of a quantum dot by adiabatic undressing},\ }\href {https://doi.org/10.1103/PhysRevB.94.045306} {\bibfield  {journal} {\bibinfo  {journal} {Phys. Rev. B}\ }\textbf {\bibinfo {volume} {94}},\ \bibinfo {pages} {045306} (\bibinfo {year} {2016})}\BibitemShut {NoStop}%
\bibitem [{\citenamefont {Kupko}\ \emph {et~al.}(2020)\citenamefont {Kupko}, \citenamefont {von Helversen}, \citenamefont {Rickert}, \citenamefont {Schulze}, \citenamefont {Strittmatter}, \citenamefont {Gschrey}, \citenamefont {Rodt}, \citenamefont {Reitzenstein},\ and\ \citenamefont {Heindel}}]{KHR:npjQI2020}%
  \BibitemOpen
  \bibfield  {author} {\bibinfo {author} {\bibfnamefont {T.}~\bibnamefont {Kupko}}, \bibinfo {author} {\bibfnamefont {M.}~\bibnamefont {von Helversen}}, \bibinfo {author} {\bibfnamefont {L.}~\bibnamefont {Rickert}}, \bibinfo {author} {\bibfnamefont {J.-H.}\ \bibnamefont {Schulze}}, \bibinfo {author} {\bibfnamefont {A.}~\bibnamefont {Strittmatter}}, \bibinfo {author} {\bibfnamefont {M.}~\bibnamefont {Gschrey}}, \bibinfo {author} {\bibfnamefont {S.}~\bibnamefont {Rodt}}, \bibinfo {author} {\bibfnamefont {S.}~\bibnamefont {Reitzenstein}},\ and\ \bibinfo {author} {\bibfnamefont {T.}~\bibnamefont {Heindel}},\ }\bibfield  {title} {\bibinfo {title} {Tools for the performance optimization of single-photon quantum key distribution},\ }\bibfield  {journal} {\bibinfo  {journal} {npj Quantum Information}\ }\textbf {\bibinfo {volume} {6}},\ \href {https://doi.org/10.1038/s41534-020-0262-8} {10.1038/s41534-020-0262-8} (\bibinfo {year} {2020})\BibitemShut {NoStop}%
\bibitem [{\citenamefont {Giorgino}\ \emph {et~al.}()\citenamefont {Giorgino}, \citenamefont {Zahálka}, \citenamefont {Jehle}, \citenamefont {Carosini}, \citenamefont {Hansen}, \citenamefont {Loredo},\ and\ \citenamefont {Walther}}]{GZJ:arxiv2025}%
  \BibitemOpen
  \bibfield  {author} {\bibinfo {author} {\bibfnamefont {F.}~\bibnamefont {Giorgino}}, \bibinfo {author} {\bibfnamefont {P.}~\bibnamefont {Zahálka}}, \bibinfo {author} {\bibfnamefont {L.}~\bibnamefont {Jehle}}, \bibinfo {author} {\bibfnamefont {L.}~\bibnamefont {Carosini}}, \bibinfo {author} {\bibfnamefont {L.~M.}\ \bibnamefont {Hansen}}, \bibinfo {author} {\bibfnamefont {J.~C.}\ \bibnamefont {Loredo}},\ and\ \bibinfo {author} {\bibfnamefont {P.}~\bibnamefont {Walther}},\ }\href@noop {} {\bibinfo {title} {Multi-photon emission from a resonantly pumped quantum dot}},\ \Eprint {https://arxiv.org/abs/2507.04843} {arXiv:2507.04843 [quant-ph]} \BibitemShut {NoStop}%
\bibitem [{\citenamefont {Cohen-Tannoudji}\ and\ \citenamefont {Reynaud}(1977)}]{CR:JPhysB1977}%
  \BibitemOpen
  \bibfield  {author} {\bibinfo {author} {\bibfnamefont {C.}~\bibnamefont {Cohen-Tannoudji}}\ and\ \bibinfo {author} {\bibfnamefont {S.}~\bibnamefont {Reynaud}},\ }\bibfield  {title} {\bibinfo {title} {Dressed-atom description of resonance fluorescence and absorption spectra of a multi-level atom in an intense laser beam},\ }\href {https://doi.org/10.1088/0022-3700/10/3/005} {\bibfield  {journal} {\bibinfo  {journal} {Journal of Physics B: Atomic and Molecular Physics}\ }\textbf {\bibinfo {volume} {10}},\ \bibinfo {pages} {345} (\bibinfo {year} {1977})}\BibitemShut {NoStop}%
\bibitem [{\citenamefont {Quilter}\ \emph {et~al.}(2015)\citenamefont {Quilter}, \citenamefont {Brash}, \citenamefont {Liu}, \citenamefont {Gl{\"a}ssl}, \citenamefont {Barth}, \citenamefont {Axt}, \citenamefont {Ramsay}, \citenamefont {Skolncik},\ and\ \citenamefont {Fox}}]{QBL:PRL2015}%
  \BibitemOpen
  \bibfield  {author} {\bibinfo {author} {\bibfnamefont {J.~H.}\ \bibnamefont {Quilter}}, \bibinfo {author} {\bibfnamefont {A.~J.}\ \bibnamefont {Brash}}, \bibinfo {author} {\bibfnamefont {F.}~\bibnamefont {Liu}}, \bibinfo {author} {\bibfnamefont {M.}~\bibnamefont {Gl{\"a}ssl}}, \bibinfo {author} {\bibfnamefont {A.~M.}\ \bibnamefont {Barth}}, \bibinfo {author} {\bibfnamefont {V.~M.}\ \bibnamefont {Axt}}, \bibinfo {author} {\bibfnamefont {A.~J.}\ \bibnamefont {Ramsay}}, \bibinfo {author} {\bibfnamefont {M.~S.}\ \bibnamefont {Skolncik}},\ and\ \bibinfo {author} {\bibfnamefont {A.~M.}\ \bibnamefont {Fox}},\ }\bibfield  {title} {\bibinfo {title} {Phonon-assisted population inversion of a single ingaas/gaas quantum dot by pulsed laser excitation},\ }\href {https://doi.org/10.1103/PhysRevLett.114.137401} {\bibfield  {journal} {\bibinfo  {journal} {Phys. Rev. Lett.}\ }\textbf {\bibinfo {volume} {114}} (\bibinfo {year} {2015})}\BibitemShut {NoStop}%
\bibitem [{\citenamefont {Unold}\ \emph {et~al.}(2004)\citenamefont {Unold}, \citenamefont {Mueller}, \citenamefont {Lienau}, \citenamefont {Elsaesser},\ and\ \citenamefont {Wieck}}]{UML:PRL2004}%
  \BibitemOpen
  \bibfield  {author} {\bibinfo {author} {\bibfnamefont {T.}~\bibnamefont {Unold}}, \bibinfo {author} {\bibfnamefont {K.}~\bibnamefont {Mueller}}, \bibinfo {author} {\bibfnamefont {C.}~\bibnamefont {Lienau}}, \bibinfo {author} {\bibfnamefont {T.}~\bibnamefont {Elsaesser}},\ and\ \bibinfo {author} {\bibfnamefont {A.~D.}\ \bibnamefont {Wieck}},\ }\bibfield  {title} {\bibinfo {title} {Optical stark effect in a quantum dot: Ultrafast control of single exciton polarizations},\ }\href {https://doi.org/10.1103/PhysRevLett.92.157401} {\bibfield  {journal} {\bibinfo  {journal} {Phys. Rev. Lett.}\ }\textbf {\bibinfo {volume} {92}},\ \bibinfo {pages} {157401} (\bibinfo {year} {2004})}\BibitemShut {NoStop}%
\bibitem [{\citenamefont {Muller}\ \emph {et~al.}(2008)\citenamefont {Muller}, \citenamefont {Fang}, \citenamefont {Lawall},\ and\ \citenamefont {Solomon}}]{MFL:PRL2008}%
  \BibitemOpen
  \bibfield  {author} {\bibinfo {author} {\bibfnamefont {A.}~\bibnamefont {Muller}}, \bibinfo {author} {\bibfnamefont {W.}~\bibnamefont {Fang}}, \bibinfo {author} {\bibfnamefont {J.}~\bibnamefont {Lawall}},\ and\ \bibinfo {author} {\bibfnamefont {G.~S.}\ \bibnamefont {Solomon}},\ }\bibfield  {title} {\bibinfo {title} {Emission spectrum of a dressed exciton-biexciton complex in a semiconductor quantum dot},\ }\href {https://doi.org/10.1103/PhysRevLett.101.027401} {\bibfield  {journal} {\bibinfo  {journal} {Phys. Rev. Lett.}\ }\textbf {\bibinfo {volume} {101}},\ \bibinfo {pages} {027401} (\bibinfo {year} {2008})}\BibitemShut {NoStop}%
\bibitem [{\citenamefont {Muller}\ \emph {et~al.}(2009)\citenamefont {Muller}, \citenamefont {Fang}, \citenamefont {Lawall},\ and\ \citenamefont {Solomon}}]{MFL:PRL2009}%
  \BibitemOpen
  \bibfield  {author} {\bibinfo {author} {\bibfnamefont {A.}~\bibnamefont {Muller}}, \bibinfo {author} {\bibfnamefont {W.}~\bibnamefont {Fang}}, \bibinfo {author} {\bibfnamefont {J.}~\bibnamefont {Lawall}},\ and\ \bibinfo {author} {\bibfnamefont {G.~S.}\ \bibnamefont {Solomon}},\ }\bibfield  {title} {\bibinfo {title} {Creating polarization-entangled photon pairs from a semiconductor quantum dot using the optical stark effect},\ }\href {https://doi.org/10.1103/PhysRevLett.103.217402} {\bibfield  {journal} {\bibinfo  {journal} {Phys. Rev. Lett.}\ }\textbf {\bibinfo {volume} {103}},\ \bibinfo {pages} {217402} (\bibinfo {year} {2009})}\BibitemShut {NoStop}%
\bibitem [{\citenamefont {Joos}\ \emph {et~al.}(2024)\citenamefont {Joos}, \citenamefont {Bauer}, \citenamefont {Rupp}, \citenamefont {Kolatschek}, \citenamefont {Fischer}, \citenamefont {Nawrath}, \citenamefont {Vijayan}, \citenamefont {Sittig}, \citenamefont {Jetter}, \citenamefont {Portalupi},\ and\ \citenamefont {Michler}}]{JBK:NanoLett24}%
  \BibitemOpen
  \bibfield  {author} {\bibinfo {author} {\bibfnamefont {R.}~\bibnamefont {Joos}}, \bibinfo {author} {\bibfnamefont {S.}~\bibnamefont {Bauer}}, \bibinfo {author} {\bibfnamefont {C.}~\bibnamefont {Rupp}}, \bibinfo {author} {\bibfnamefont {S.}~\bibnamefont {Kolatschek}}, \bibinfo {author} {\bibfnamefont {W.}~\bibnamefont {Fischer}}, \bibinfo {author} {\bibfnamefont {C.}~\bibnamefont {Nawrath}}, \bibinfo {author} {\bibfnamefont {P.}~\bibnamefont {Vijayan}}, \bibinfo {author} {\bibfnamefont {R.}~\bibnamefont {Sittig}}, \bibinfo {author} {\bibfnamefont {M.}~\bibnamefont {Jetter}}, \bibinfo {author} {\bibfnamefont {S.~L.}\ \bibnamefont {Portalupi}},\ and\ \bibinfo {author} {\bibfnamefont {P.}~\bibnamefont {Michler}},\ }\bibfield  {title} {\bibinfo {title} {Coherently and incoherently pumped telecom c-band single-photon source with high brightness and indistinguishability},\ }\href {https://doi.org/10.1021/acs.nanolett.4c01813} {\bibfield  {journal} {\bibinfo  {journal} {Nano Letters}\ }\textbf {\bibinfo {volume}
  {24}},\ \bibinfo {pages} {8626} (\bibinfo {year} {2024})}\BibitemShut {NoStop}%
\bibitem [{\citenamefont {Fischer}\ \emph {et~al.}(2017)\citenamefont {Fischer}, \citenamefont {Hanschke}, \citenamefont {Wierzbowski}, \citenamefont {Simmet}, \citenamefont {Dory}, \citenamefont {Finley}, \citenamefont {Vu{\v{c}}kovi{\'c}},\ and\ \citenamefont {M{\"u}ller}}]{FHW:NPhys2017}%
  \BibitemOpen
  \bibfield  {author} {\bibinfo {author} {\bibfnamefont {K.~A.}\ \bibnamefont {Fischer}}, \bibinfo {author} {\bibfnamefont {L.}~\bibnamefont {Hanschke}}, \bibinfo {author} {\bibfnamefont {J.}~\bibnamefont {Wierzbowski}}, \bibinfo {author} {\bibfnamefont {T.}~\bibnamefont {Simmet}}, \bibinfo {author} {\bibfnamefont {C.}~\bibnamefont {Dory}}, \bibinfo {author} {\bibfnamefont {J.~J.}\ \bibnamefont {Finley}}, \bibinfo {author} {\bibfnamefont {J.}~\bibnamefont {Vu{\v{c}}kovi{\'c}}},\ and\ \bibinfo {author} {\bibfnamefont {K.}~\bibnamefont {M{\"u}ller}},\ }\bibfield  {title} {\bibinfo {title} {Signatures of two-photon pulses from a quantum two-level system},\ }\href {https://doi.org/10.1038/nphys4052} {\bibfield  {journal} {\bibinfo  {journal} {Nature Physics}\ }\textbf {\bibinfo {volume} {13}},\ \bibinfo {pages} {649} (\bibinfo {year} {2017})}\BibitemShut {NoStop}%
\bibitem [{\citenamefont {Mollow}(1969)}]{Mollow:PRL1969}%
  \BibitemOpen
  \bibfield  {author} {\bibinfo {author} {\bibfnamefont {B.~R.}\ \bibnamefont {Mollow}},\ }\bibfield  {title} {\bibinfo {title} {Power spectrum of light scattered by two-level systems},\ }\href {https://doi.org/10.1103/PhysRev.188.1969} {\bibfield  {journal} {\bibinfo  {journal} {Physical Review Letters}\ }\textbf {\bibinfo {volume} {188}} (\bibinfo {year} {1969})}\BibitemShut {NoStop}%
\bibitem [{\citenamefont {Boos}\ \emph {et~al.}(2024)\citenamefont {Boos}, \citenamefont {Kim}, \citenamefont {Bracht}, \citenamefont {Sbresny}, \citenamefont {Kaspari}, \citenamefont {Cygorek}, \citenamefont {Riedl}, \citenamefont {Bopp}, \citenamefont {Rauhaus}, \citenamefont {Calcagno}, \citenamefont {Finley}, \citenamefont {Reiter},\ and\ \citenamefont {M\"uller}}]{BKB:PRL2024}%
  \BibitemOpen
  \bibfield  {author} {\bibinfo {author} {\bibfnamefont {K.}~\bibnamefont {Boos}}, \bibinfo {author} {\bibfnamefont {S.~K.}\ \bibnamefont {Kim}}, \bibinfo {author} {\bibfnamefont {T.}~\bibnamefont {Bracht}}, \bibinfo {author} {\bibfnamefont {F.}~\bibnamefont {Sbresny}}, \bibinfo {author} {\bibfnamefont {J.~M.}\ \bibnamefont {Kaspari}}, \bibinfo {author} {\bibfnamefont {M.}~\bibnamefont {Cygorek}}, \bibinfo {author} {\bibfnamefont {H.}~\bibnamefont {Riedl}}, \bibinfo {author} {\bibfnamefont {F.~W.}\ \bibnamefont {Bopp}}, \bibinfo {author} {\bibfnamefont {W.}~\bibnamefont {Rauhaus}}, \bibinfo {author} {\bibfnamefont {C.}~\bibnamefont {Calcagno}}, \bibinfo {author} {\bibfnamefont {J.~J.}\ \bibnamefont {Finley}}, \bibinfo {author} {\bibfnamefont {D.~E.}\ \bibnamefont {Reiter}},\ and\ \bibinfo {author} {\bibfnamefont {K.}~\bibnamefont {M\"uller}},\ }\bibfield  {title} {\bibinfo {title} {Signatures of dynamically dressed states},\ }\href {https://doi.org/10.1103/PhysRevLett.132.053602} {\bibfield  {journal}
  {\bibinfo  {journal} {Phys. Rev. Lett.}\ }\textbf {\bibinfo {volume} {132}},\ \bibinfo {pages} {053602} (\bibinfo {year} {2024})}\BibitemShut {NoStop}%
\bibitem [{\citenamefont {Liu}\ \emph {et~al.}(2024)\citenamefont {Liu}, \citenamefont {Gustin}, \citenamefont {Liu}, \citenamefont {Li}, \citenamefont {Yu}, \citenamefont {Ni}, \citenamefont {Niu}, \citenamefont {Hughes}, \citenamefont {Wang},\ and\ \citenamefont {Liu}}]{LGL:NPhot2024}%
  \BibitemOpen
  \bibfield  {author} {\bibinfo {author} {\bibfnamefont {S.}~\bibnamefont {Liu}}, \bibinfo {author} {\bibfnamefont {C.}~\bibnamefont {Gustin}}, \bibinfo {author} {\bibfnamefont {H.}~\bibnamefont {Liu}}, \bibinfo {author} {\bibfnamefont {X.}~\bibnamefont {Li}}, \bibinfo {author} {\bibfnamefont {Y.}~\bibnamefont {Yu}}, \bibinfo {author} {\bibfnamefont {H.}~\bibnamefont {Ni}}, \bibinfo {author} {\bibfnamefont {Z.}~\bibnamefont {Niu}}, \bibinfo {author} {\bibfnamefont {S.}~\bibnamefont {Hughes}}, \bibinfo {author} {\bibfnamefont {X.}~\bibnamefont {Wang}},\ and\ \bibinfo {author} {\bibfnamefont {J.}~\bibnamefont {Liu}},\ }\bibfield  {title} {\bibinfo {title} {Dynamic resonance fluorescence in solid-state cavity quantum electrodynamics},\ }\href {https://doi.org/10.1038/s41566-023-01359-x} {\bibfield  {journal} {\bibinfo  {journal} {Nature Photonics}\ }\textbf {\bibinfo {volume} {18}},\ \bibinfo {pages} {318} (\bibinfo {year} {2024})}\BibitemShut {NoStop}%
\bibitem [{\citenamefont {Olbrich}\ \emph {et~al.}(2017)\citenamefont {Olbrich}, \citenamefont {Kettler}, \citenamefont {Bayerbach}, \citenamefont {Paul}, \citenamefont {H{\"o}schele}, \citenamefont {Portalupi}, \citenamefont {Jetter},\ and\ \citenamefont {Michler}}]{OKB:APL2017}%
  \BibitemOpen
  \bibfield  {author} {\bibinfo {author} {\bibfnamefont {F.}~\bibnamefont {Olbrich}}, \bibinfo {author} {\bibfnamefont {J.}~\bibnamefont {Kettler}}, \bibinfo {author} {\bibfnamefont {M.}~\bibnamefont {Bayerbach}}, \bibinfo {author} {\bibfnamefont {M.}~\bibnamefont {Paul}}, \bibinfo {author} {\bibfnamefont {J.}~\bibnamefont {H{\"o}schele}}, \bibinfo {author} {\bibfnamefont {S.~L.}\ \bibnamefont {Portalupi}}, \bibinfo {author} {\bibfnamefont {M.}~\bibnamefont {Jetter}},\ and\ \bibinfo {author} {\bibfnamefont {P.}~\bibnamefont {Michler}},\ }\bibfield  {title} {\bibinfo {title} {Temperature-dependent properties of single long-wavelength ingaas quantum dots embedded in a strain reducing layer},\ }\bibfield  {journal} {\bibinfo  {journal} {Journal of Applied Physics}\ }\textbf {\bibinfo {volume} {121}},\ \href {https://doi.org/10.1063/1.4983362} {10.1063/1.4983362} (\bibinfo {year} {2017})\BibitemShut {NoStop}%
\bibitem [{\citenamefont {Ripin}\ \emph {et~al.}(2023)\citenamefont {Ripin}, \citenamefont {Peng}, \citenamefont {Zhang}, \citenamefont {Chakravarthi}, \citenamefont {He}, \citenamefont {Xu}, \citenamefont {Fu}, \citenamefont {Cao},\ and\ \citenamefont {Li}}]{RPZ:NNano2023}%
  \BibitemOpen
  \bibfield  {author} {\bibinfo {author} {\bibfnamefont {A.}~\bibnamefont {Ripin}}, \bibinfo {author} {\bibfnamefont {R.}~\bibnamefont {Peng}}, \bibinfo {author} {\bibfnamefont {X.}~\bibnamefont {Zhang}}, \bibinfo {author} {\bibfnamefont {S.}~\bibnamefont {Chakravarthi}}, \bibinfo {author} {\bibfnamefont {M.}~\bibnamefont {He}}, \bibinfo {author} {\bibfnamefont {X.}~\bibnamefont {Xu}}, \bibinfo {author} {\bibfnamefont {K.-M.}\ \bibnamefont {Fu}}, \bibinfo {author} {\bibfnamefont {T.}~\bibnamefont {Cao}},\ and\ \bibinfo {author} {\bibfnamefont {M.}~\bibnamefont {Li}},\ }\bibfield  {title} {\bibinfo {title} {Tunable phononic coupling in excitonic quantum emitters},\ }\href {https://doi.org/10.1038/s41565-023-01410-6} {\bibfield  {journal} {\bibinfo  {journal} {Nature Nanotechnology}\ }\textbf {\bibinfo {volume} {18}},\ \bibinfo {pages} {1020} (\bibinfo {year} {2023})}\BibitemShut {NoStop}%
\bibitem [{\citenamefont {Sbresny}\ \emph {et~al.}()\citenamefont {Sbresny}, \citenamefont {Calcagno}, \citenamefont {Kim}, \citenamefont {Boos}, \citenamefont {Rauhaus}, \citenamefont {Bopp}, \citenamefont {Riedl}, \citenamefont {Finley}, \citenamefont {Casalengua},\ and\ \citenamefont {Müller}}]{SCS:arxiv2025}%
  \BibitemOpen
  \bibfield  {author} {\bibinfo {author} {\bibfnamefont {F.}~\bibnamefont {Sbresny}}, \bibinfo {author} {\bibfnamefont {C.}~\bibnamefont {Calcagno}}, \bibinfo {author} {\bibfnamefont {S.~K.}\ \bibnamefont {Kim}}, \bibinfo {author} {\bibfnamefont {K.}~\bibnamefont {Boos}}, \bibinfo {author} {\bibfnamefont {W.}~\bibnamefont {Rauhaus}}, \bibinfo {author} {\bibfnamefont {F.}~\bibnamefont {Bopp}}, \bibinfo {author} {\bibfnamefont {H.}~\bibnamefont {Riedl}}, \bibinfo {author} {\bibfnamefont {J.~J.}\ \bibnamefont {Finley}}, \bibinfo {author} {\bibfnamefont {E.~Z.}\ \bibnamefont {Casalengua}},\ and\ \bibinfo {author} {\bibfnamefont {K.}~\bibnamefont {Müller}},\ }\href@noop {} {\bibinfo {title} {Selective filtering of multi-photon events from a single-photon emitter}},\ \Eprint {https://arxiv.org/abs/2506.22378} {arXiv:2506.22378 [quant-ph]} \BibitemShut {NoStop}%
\bibitem [{\citenamefont {Sittig}\ \emph {et~al.}(2022)\citenamefont {Sittig}, \citenamefont {Nawrath}, \citenamefont {Kolatschek}, \citenamefont {Bauer}, \citenamefont {Schaber}, \citenamefont {Huang}, \citenamefont {Vijayan}, \citenamefont {Pruy}, \citenamefont {Portalupi}, \citenamefont {Jetter},\ and\ \citenamefont {Michler}}]{SNK:Nanophot2022}%
  \BibitemOpen
  \bibfield  {author} {\bibinfo {author} {\bibfnamefont {R.}~\bibnamefont {Sittig}}, \bibinfo {author} {\bibfnamefont {C.}~\bibnamefont {Nawrath}}, \bibinfo {author} {\bibfnamefont {S.}~\bibnamefont {Kolatschek}}, \bibinfo {author} {\bibfnamefont {S.}~\bibnamefont {Bauer}}, \bibinfo {author} {\bibfnamefont {R.}~\bibnamefont {Schaber}}, \bibinfo {author} {\bibfnamefont {J.}~\bibnamefont {Huang}}, \bibinfo {author} {\bibfnamefont {P.}~\bibnamefont {Vijayan}}, \bibinfo {author} {\bibfnamefont {P.}~\bibnamefont {Pruy}}, \bibinfo {author} {\bibfnamefont {S.~L.}\ \bibnamefont {Portalupi}}, \bibinfo {author} {\bibfnamefont {M.}~\bibnamefont {Jetter}},\ and\ \bibinfo {author} {\bibfnamefont {P.}~\bibnamefont {Michler}},\ }\bibfield  {title} {\bibinfo {title} {{Thin-film InGaAs metamorphic buffer for telecom C-band InAs quantum dots and optical resonators on GaAs platform}},\ }\href {https://doi.org/10.1515/nanoph-2021-0552} {\bibfield  {journal} {\bibinfo  {journal} {Nanophotonics}\ }\textbf {\bibinfo {volume}
  {11}},\ \bibinfo {pages} {1109} (\bibinfo {year} {2022})}\BibitemShut {NoStop}%
\bibitem [{\citenamefont {{Cornelius Nawrath}}\ \emph {et~al.}(2023)\citenamefont {{Cornelius Nawrath}}, \citenamefont {{Raphael Joos}}, \citenamefont {{Sascha Kolatschek}}, \citenamefont {{Stephanie Bauer}}, \citenamefont {{Pascal Pruy}}, \citenamefont {{Florian Hornung}}, \citenamefont {{Julius Fischer}}, \citenamefont {{Jiasheng Huang}}, \citenamefont {{Ponraj Vijayan}}, \citenamefont {{Robert Sittig}}, \citenamefont {{Michael Jetter}}, \citenamefont {{Simone Luca Portalupi}},\ and\ \citenamefont {{Peter Michler}}}]{NJK:AdvQTech2023}%
  \BibitemOpen
  \bibfield  {author} {\bibinfo {author} {\bibnamefont {{Cornelius Nawrath}}}, \bibinfo {author} {\bibnamefont {{Raphael Joos}}}, \bibinfo {author} {\bibnamefont {{Sascha Kolatschek}}}, \bibinfo {author} {\bibnamefont {{Stephanie Bauer}}}, \bibinfo {author} {\bibnamefont {{Pascal Pruy}}}, \bibinfo {author} {\bibnamefont {{Florian Hornung}}}, \bibinfo {author} {\bibnamefont {{Julius Fischer}}}, \bibinfo {author} {\bibnamefont {{Jiasheng Huang}}}, \bibinfo {author} {\bibnamefont {{Ponraj Vijayan}}}, \bibinfo {author} {\bibnamefont {{Robert Sittig}}}, \bibinfo {author} {\bibnamefont {{Michael Jetter}}}, \bibinfo {author} {\bibnamefont {{Simone Luca Portalupi}}},\ and\ \bibinfo {author} {\bibnamefont {{Peter Michler}}},\ }\bibfield  {title} {\bibinfo {title} {Bright source of purcell-enhanced, triggered, single photons in the telecom c-band},\ }\href {https://advanced.onlinelibrary.wiley.com/action/showCitFormats?doi=10.1002%2Fqute.202300111} {\bibfield  {journal} {\bibinfo  {journal} {Advanced Quantum
  Technologies}\ }\textbf {\bibinfo {volume} {6}},\ \bibinfo {pages} {2300111} (\bibinfo {year} {2023})}\BibitemShut {NoStop}%
\end{thebibliography}%

\end{document}





\title{Asymmetric two-photon response of an incoherently driven quantum emitter}

\author{Lennart Jehle}\thanks{Address all correspondence to lennart.jehle@univie.ac.at}
\affiliation{University of Vienna, Faculty of Physics, Vienna Center for Quantum Science and Technology (VCQ), 1090 Vienna, Austria}
\affiliation{Christian Doppler Laboratory for Photonic Quantum Computer, University of Vienna, Faculty of Physics, 1090 Vienna, Austria}

\author{Lena M. Hansen}
\affiliation{University of Vienna, Faculty of Physics, Vienna Center for Quantum Science and Technology (VCQ), 1090 Vienna, Austria}
\affiliation{Christian Doppler Laboratory for Photonic Quantum Computer, University of Vienna, Faculty of Physics, 1090 Vienna, Austria}

\author{Patrik I. Sund}
\affiliation{University of Vienna, Faculty of Physics, Vienna Center for Quantum Science and Technology (VCQ), 1090 Vienna, Austria}
\affiliation{Christian Doppler Laboratory for Photonic Quantum Computer, University of Vienna, Faculty of Physics, 1090 Vienna, Austria}

\author{Thomas W. Sand{\o}}
\affiliation{University of Vienna, Faculty of Physics, Vienna Center for Quantum Science and Technology (VCQ), 1090 Vienna, Austria}
\affiliation{Christian Doppler Laboratory for Photonic Quantum Computer, University of Vienna, Faculty of Physics, 1090 Vienna, Austria}

\author{Raphael Joos}
\affiliation{Institut f\"ur Halbleiteroptik und Funktionelle Grenzfl\"achen, \\Center for Integrated Quantum Science and Technology (IQ\textsuperscript{ST}) and SCoPE, \\University of Stuttgart, Allmandring 3, 70569 Stuttgart, Germany}

\author{Michael Jetter}
\affiliation{Institut f\"ur Halbleiteroptik und Funktionelle Grenzfl\"achen, \\Center for Integrated Quantum Science and Technology (IQ\textsuperscript{ST}) and SCoPE, \\University of Stuttgart, Allmandring 3, 70569 Stuttgart, Germany}

\author{Simone L. Portalupi}
\affiliation{Institut f\"ur Halbleiteroptik und Funktionelle Grenzfl\"achen, \\Center for Integrated Quantum Science and Technology (IQ\textsuperscript{ST}) and SCoPE, \\University of Stuttgart, Allmandring 3, 70569 Stuttgart, Germany}

\author{Mathieu Bozzio}
\affiliation{University of Vienna, Faculty of Physics, Vienna Center for Quantum Science and Technology (VCQ), 1090 Vienna, Austria}

\author{Peter Michler}
\affiliation{Institut f\"ur Halbleiteroptik und Funktionelle Grenzfl\"achen, \\Center for Integrated Quantum Science and Technology (IQ\textsuperscript{ST}) and SCoPE, \\University of Stuttgart, Allmandring 3, 70569 Stuttgart, Germany}

\author{Philip Walther}
\affiliation{University of Vienna, Faculty of Physics, Vienna Center for Quantum Science and Technology (VCQ), 1090 Vienna, Austria}
\affiliation{Christian Doppler Laboratory for Photonic Quantum Computer, University of Vienna, Faculty of Physics, 1090 Vienna, Austria}
\affiliation{Institute for Quantum Optics and Quantum Information (IQOQI) Vienna, Austrian Academy of Sciences, Vienna, Austria}

\maketitle
\onecolumngrid
\nolinenumbers
\section{\label{sec:opt_setup} Experimental setup}
\begin{figure}
	\begin{center}
		 \includegraphics[width=180mm]{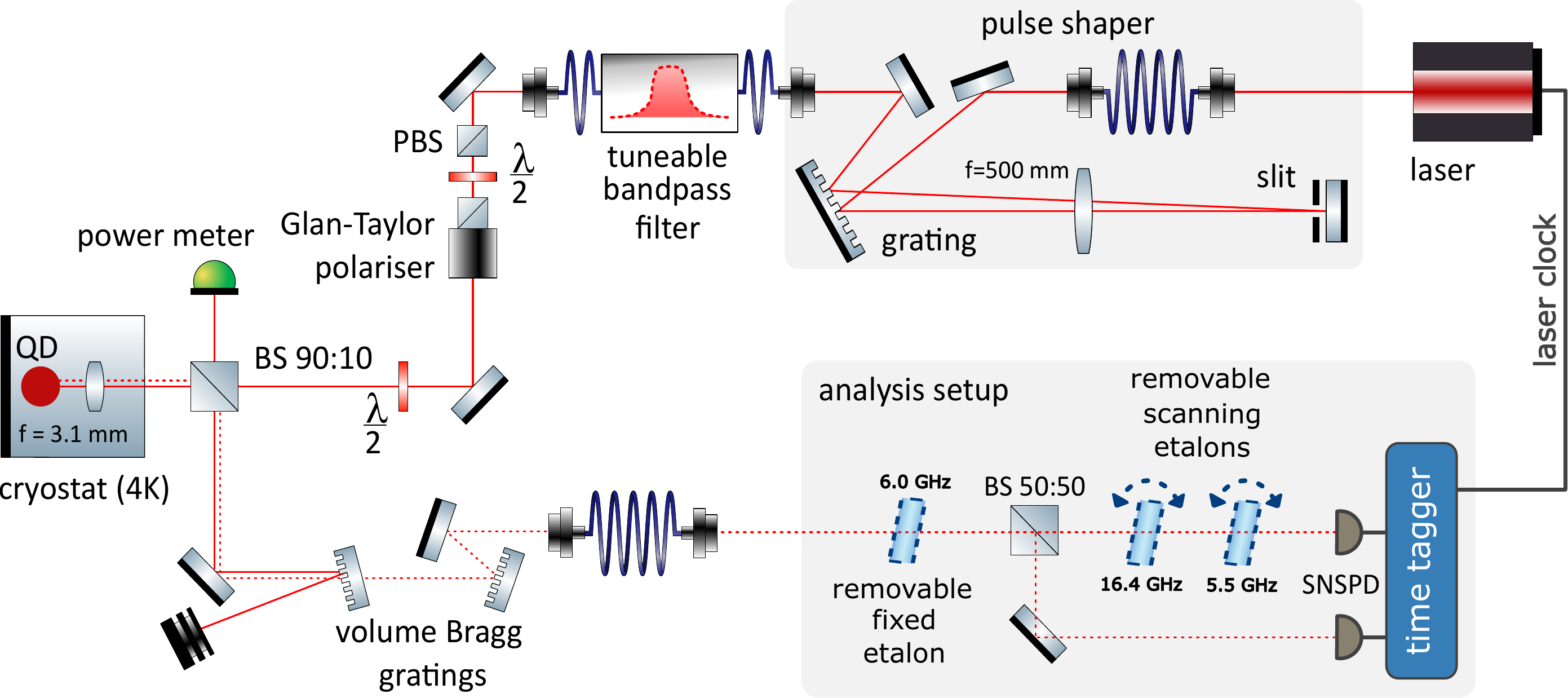}
		\caption{\textbf{Scheme of the experimental setup.} A schematic representation of the main building blocks of the experimental setup, such as the pulsed excitation laser, pulse shaper, tuneable bandwidth filter, excitation of the quantum dot (QD) sample, filtering of the QD transition line and the analysis stage. Solid (dashed) red lines indicate the path of the laser beam (QD emission). PBS: polarizing beam splitter, BS: beam splitter, SNSPD: superconducting nanowire single-photon detector.}
		\label{fig:setup}
	\end{center}
\end{figure}

The experimental setup is shown in Fig.~S\ref{fig:setup}.
\noindent
We use an Er-doped fiber mode-locked pulsed laser (\textit{PriTel UOC}) to excite the quantum dot (QD) at a repetition rate of $\nu_{\text{rep}}=\SI{75.95}{\MHz}$.
A filter with $\SI{1}{\nano\meter}$ bandwidth placed within the laser cavity stretches the generated pulses to a temporal pulse width of $\SI{9\pm3}{\ps}$ (spectral width FWHM $\approx\SI{450}{\pico\meter}$, the pulses are not fully Fourier-limited).
The laser provides wavelength tunability between $\SI{1530}{\nano\meter}$ and $\SI{1555}{\nano\meter}$ at an average output power of $\SI{200}{mW}$.
All remaining broadband amplifier noise is removed by a free-space $4\text{-}f$ pulse shaper based on a reflective grating (\textit{Spectrogon}, 1200 lines/mm, blaze at 1550 nm) with an efficiency of $\approx 90\,\%$, C-coated cylindrical lens with a focal length of $\SI{500}{\milli\meter}$ and a variable filtering slit.
Finally, a fiber-coupled tunable bandwidth filter (\textit{EXFO, XTM-50}) with a top-hat profile is used to stretch the pulses. 
The minimal bandwidth is $\SI{5.7\pm0.3}{\GHz}$ and results in a maximum pulse length of $\SI{80\pm1}{\ps}$ that is independently verified using a superconducting nanowire single-photon detector (SNSPD, \textit{Single Quantum EOS}) and a time-tagging device (\textit{Swabian Time Taggers TimeTagger X}).

\begin{figure}
	\begin{center}
		 \includegraphics[width=180mm]{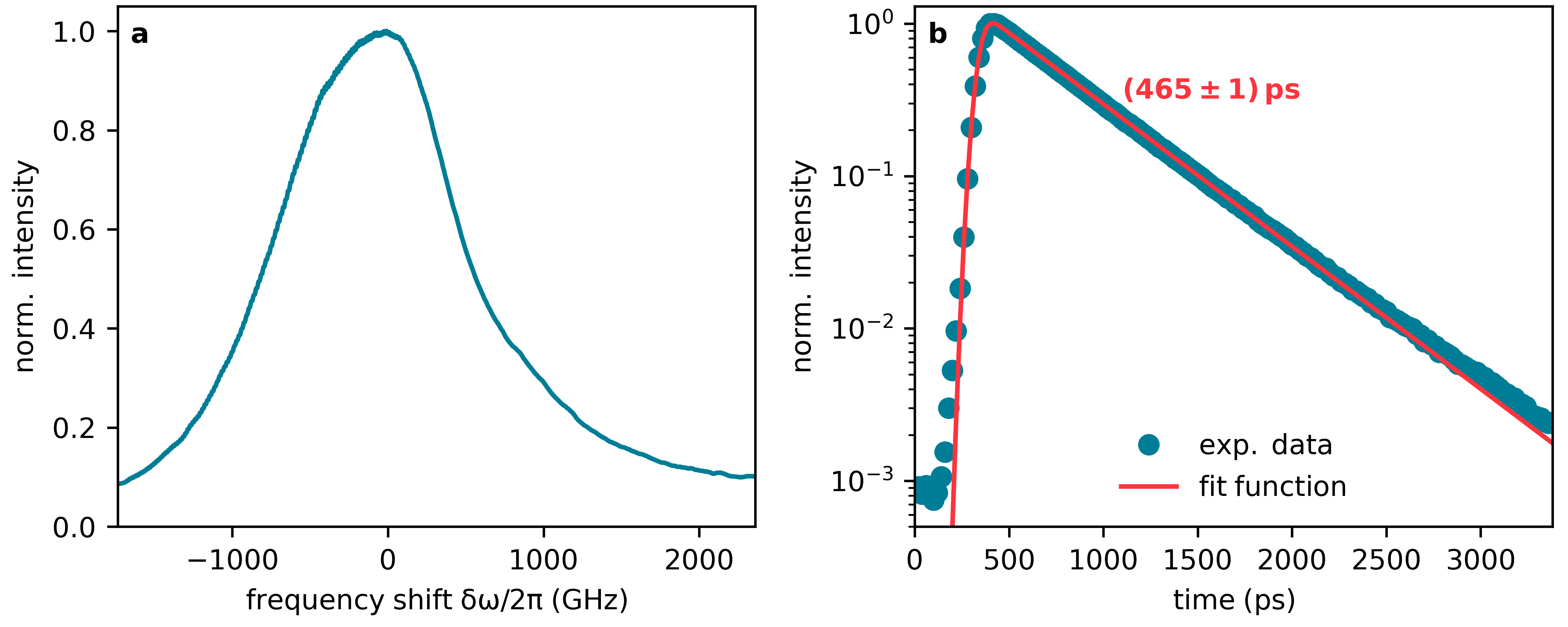}
		\caption{\textbf{Characteristics of the QD sample.}
        \textbf{(a)} The cavity mode is measured using a strong above band laser and collecting the emission of the sample on a spectrometer.
        The frequency is given as detuning from the spontaneous emission line, that is $\delta\omega = \omega - \omega_0$.
        The spectrometer resolution is $\SI{\approx10}{\GHz}$ per pixel.
        \textbf{(b)} The measurements are performed with pulsed LA excitation at $\SI{75.95}{\MHz}$, a blue-shifted detuning of $\SI{175}{\GHz}$, a laser spectral bandwidth of $\SI{26\pm1}{\GHz}$ and excitation power of $\SI{250\pm15}{\nW}$.
        The mono-exponential fit function (solid black line) yields a decay time of $\SI{465\pm1}{\ps}$.}
		\label{fig:QDchar}
	\end{center}
\end{figure}

\begin{figure}
	\begin{center}
		 \includegraphics[width=95mm]{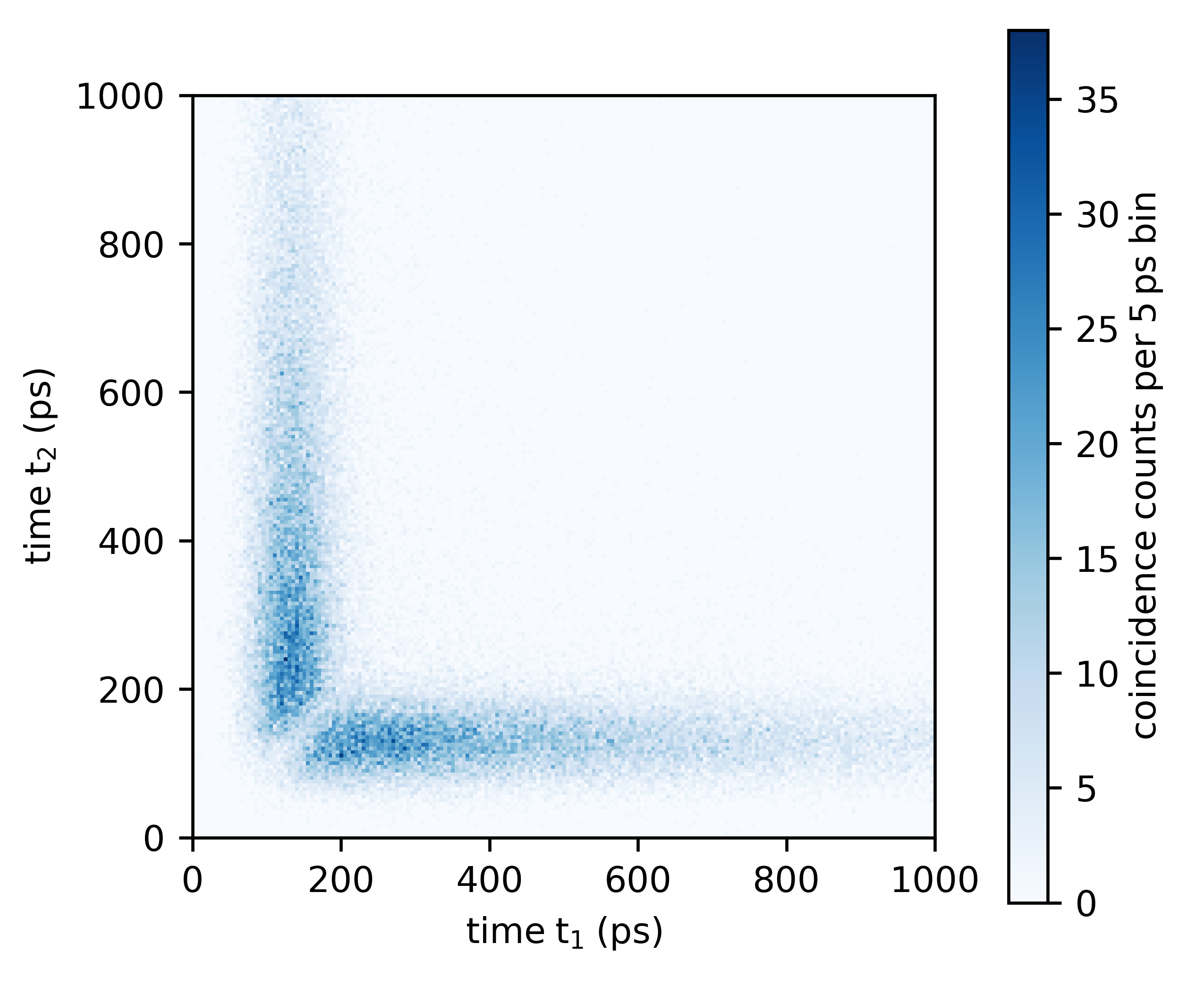}
		\caption{\textbf{2D histogram.}
        The two-photon coincidences are displayed in the 2D histogram with a bin width of $\SI{5}{\ps}$ per time axis.
        The long laser pulses are $\SI{80\pm1}{\ps}$ long, have a cw-power of $\SI{250}{\nW}$ and a detuning of $\delta\omega_L/2\pi=\SI{125}{\GHz}$,
        The emission from the QD was isolated from broadband background using a $\SI{120\pm1}{\GHz}$ VBG.
        }
		\label{fig:2Dhist}
	\end{center}
\end{figure}

The pulse-shaped excitation beam is collimated into free space by an $\SI{8}{\milli\meter}$ lens collimator (\textit{Schäfter+Kirchhoff 60FC-SF-0-M08-08}).
A Glan-Taylor polarizer (\textit{Leysop GTB10-M}, polarization extinction ratio $>\SI{60}{\dB}$) sets the excitation beam polarization to linear and the angle of the polarization is adjustable by a subsequent half-wave plate (\textit{Bernard Halle}).
Just before the cryostat chamber, a 90:10 beam splitter cube (BS, \textit{Thorlabs BS078}) is placed to separate the incoming excitation beam from the single photons emitted by the QD.
Approximately $90\,\%$ of the excitation beam (depending on its polarization state) is reflected by the BS to a power meter (used to control the QD excitation power) and only $\approx 10\,\%$ of the laser power is guided to the cryostat chamber (\textit{attocube attoDry800}) in which the QD sample is placed.
In Fig.~\ref{fig:QDchar}a, the cavity mode of the circular Bragg grating is shown and we infer a FWHM of $\approx\SI{1250}{\GHz}$.
Fitting the time-correlated photon-counting measurement presented in Fig.~\ref{fig:QDchar}b yields a lifetime of $\tau_{QD}=\SI{465\pm1}{\ps}$.
For more details on the QD sample see Methods of the Main Text.

The QD emission and the reflected laser light are collected by a lens ($f=\SI{3.1}{\milli\meter}$) with an NA of $0.68$.
Leaving through the reflection port of the 90:10 BS, QD emission and laser light are separated using a volume Bragg grating (VBG) notch filter (\textit{OptiGrate}) that blocks the laser wavelength with spectral bandwidth (FWHM) of $\SI{120\pm1}{\GHz}$ with an individual suppression of OD$6$.
Finally, the QD emission is filtered out from the remaining broadband spectrum by reflecting of another VBG (\textit{OptiGrate}) centered at $\approx\omega_0-\SI{40}{\GHz}$ and coupled by a $f=\SI{8}{\milli\meter}$ lens collimator into a single-mode fiber (SMF).\newline

The analysis setup is arranged in a Hanbury-Brown and Twiss (HBT) configuration and can be equipped with additional frequency filters, depending on the desired measurement type.
All spectrally resolved measurements employ a set of two etalons (FSR: $\SI{125\pm1}{\GHz}$ and $\SI{292\pm1}{\GHz}$; FWHM: $\SI{5.3\pm0.3}{\GHz}$ and $\SI{16.1\pm0.3}{\GHz}$), each mounted on a piezomoter-controlled rotation stage to scan their central frequency jointly, which are placed in one of the two HBT paths.
The cascaded arrangement of two etalons featuring FSRs that non-integer dividable greatly suppresses spectral components outside of the overlapped frequency peak.
For the spectrally filtered autocorrelation presented in Fig.~4 of the Main Text, the 50:50 BS is preceded by a $\SI{6.0\pm0.3}{\GHz}$ Lorentzian etalon fixed at $\omega_0$, but the scanning etalons are removed.



\section{\label{sec:2Dhist}Temporal modes of two-photon emission}

When using the wording \textit{two-photon emission} in the Main Text, we refer to the fact that two photons are emitted during or shortly after the interaction of QD with the laser pulse. 
However, these photons are never emitted simultaneously (that is, in the same time mode), and the overlap in arrival times of the \fst and \scnd photon (as observed in Fig.~2b of the Main Text) is the result of statistical averaging over many repetitions. 
These arrival-time distributions indicate that for some point in time, a detection could have been caused by either a \fst or a \scnd photon, but they do not imply that two photons have been emitted simultaneously. 

To demonstrate this, we display the time-resolved autocorrelation function $G(t_1, t_2)$ in Fig.~S\ref{fig:2Dhist}, where $t_1$ ($t_2$) is the arrival time measured in the first (second) optical channel with respect to a timing reference, typically the electronic clock of the pulsed laser.
Each coincidence is assigned to a bin in the histogram according to the arrival time of both photons.
The diagonal in the histogram corresponds to events that occur at $t_1=t_2$, such that it reflects the probability that two photons arrive at the same time (more accurately, within the binning width).
Importantly, the diagonal shows significantly lower counts than the surrounding bins, indicating that two photons are unlikely to be emitted at the same time.
Residual coincidences on the diagonal are caused by the finite system response function of $\SI{39\pm2}{\ps}$ of the detection setup, background emission and detector dark counts.


\section{\label{sec:extParam} Extended data set}

As discussed in the Main Text, other signatures of dynamically dressed states were recently observed \cite{BKB:PRL2024, LGL:NPhot2024}.
These experimental findings represent the extension of the well-known Mollow-triplet \cite{Mollow:PRL1969} into the regime of pulsed driving and bear some similarity to the spectra presented here. 
However, the underlying processes are fundamentally different.

\begin{figure}
	\begin{center}
		 \includegraphics[width=180mm]{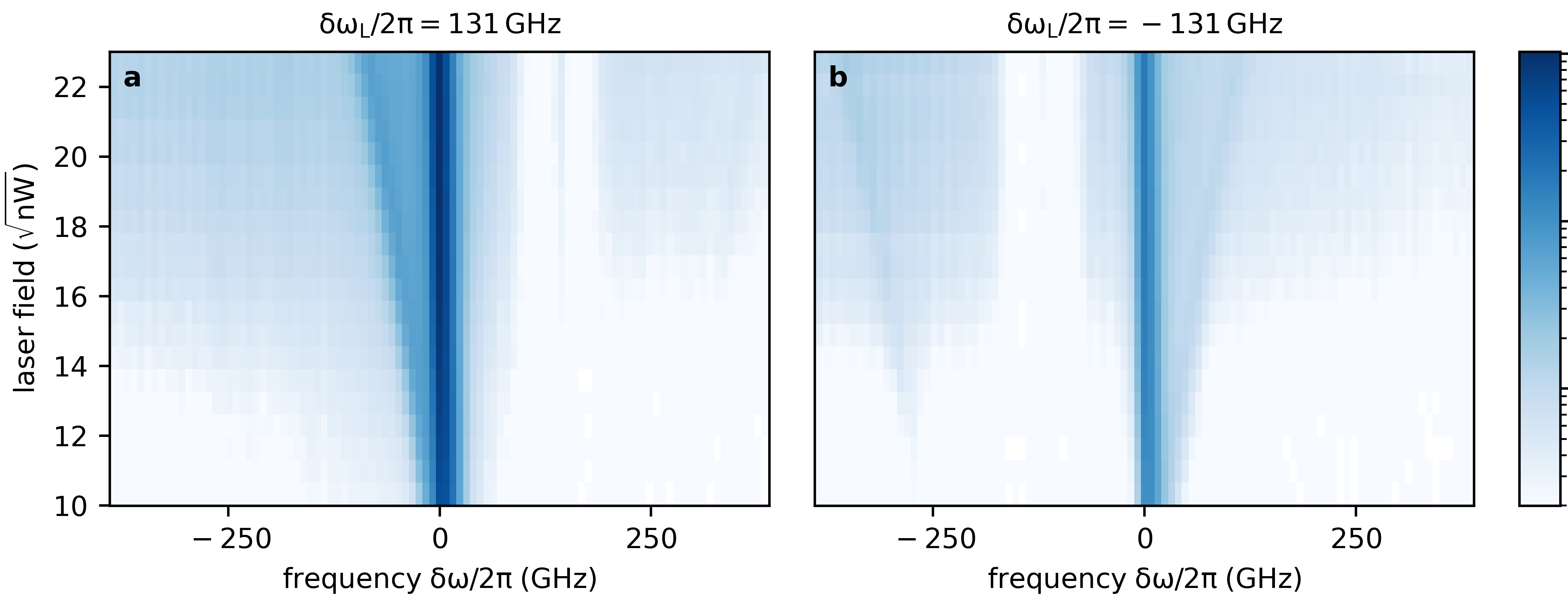}
		\caption{\textbf{QD spectrum for blue- and red-detuned laser at varying powers.}
        For $\SI{80\pm1}{\ps}$ long laser pulses, the QD emission is measured with a spectrometer after the scattered laser is filtered using a $\SI{120\pm1}{\GHz}$ VBG. 
        The spectrometer resolution is $\SI{\approx10}{\GHz}$ per pixel.
        \textbf{(a)} For a $\SI{131}{\GHz}$ blue-detuned laser, only the low-frequency side peak is visible, whereas the emission for a red-detuned excitation laser (\textbf{b}) features two side peaks, that are symmetric with respect to the laser frequency.}
		\label{fig:Mollow}
	\end{center}
\end{figure}

\begin{figure}
	\begin{center}
		 \includegraphics[width=180mm]{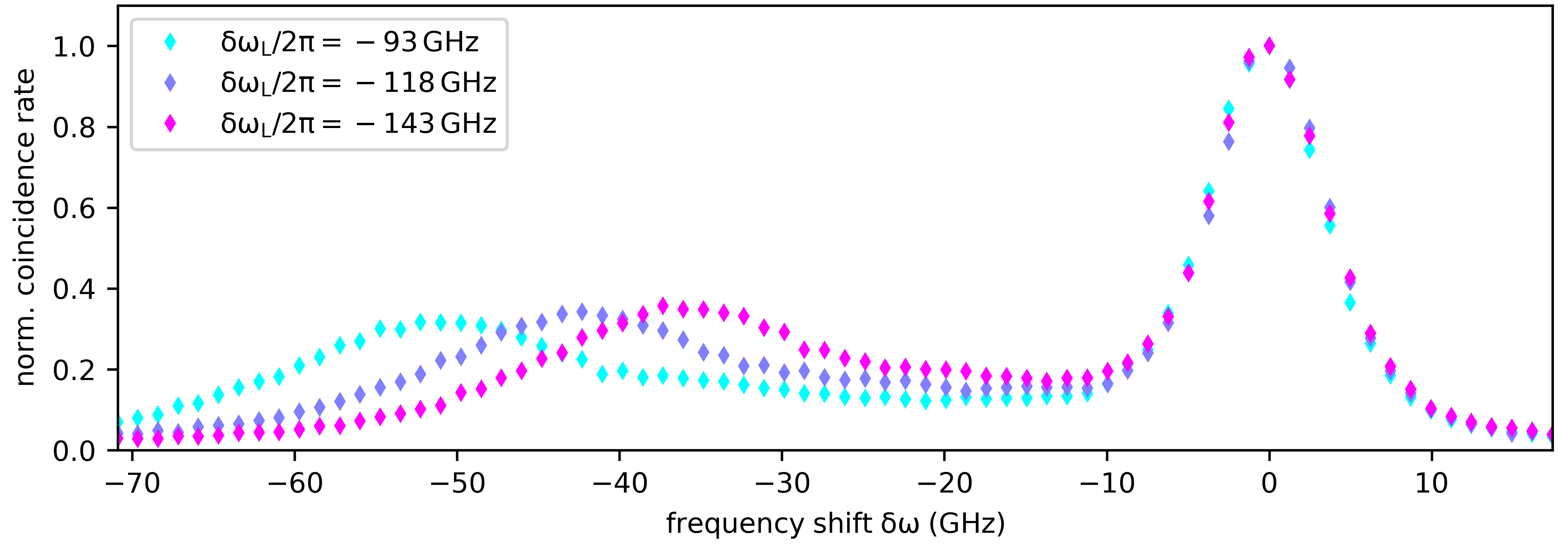}
		\caption{\textbf{Two-photon spectrum for varying laser--QD detunings.}
         For $\SI{80\pm1}{\ps}$ long laser pulses with $\SI{250}{\nW}$ and $\delta\omega_L/2\pi=\SI{125}{\GHz}$, the two-photon spectrum is measured in the heralded configuration explained in the Main Text.
         The three measurements represent different laser--QD detunings $\delta\omega_L$.
         For clarity, all error bars were omitted.}
		\label{fig:detscan}
	\end{center}
\end{figure}

There are four optical transitions between any two manifolds of laser-dressed states, all of which are allowed by optical selection rules \cite{CT:AtomPhotonBook}. 
Two of them are degenerate and match the frequency of the driving laser, the other two appear at $\omega_L\pm\Omega_\text{eff}(t)$ and are energetically only allowed for a phonon-photon or two-photon transition.
In a coherent process, the photons must appear in pairs of a low- and high-frequency photon (thus obeying the conversation of energy) and are the result of two-photon scattering from the laser field. 
In the absence of an asymmetric cavity or for a sufficiently large cavity mode, this leads to a symmetric spectral response. 

On the other hand, incoherent interactions involving the phononic environment facilitate also single-photon transitions at those frequencies (more accurately, single-phonon single-photon transitions).

However, depending on the sign of the laser--QD detuning $\delta\omega_L>0$ ($\delta\omega_L$<0), the emission (absorption) of a phonon is required and therefore only one of the transitions is supported.

In Fig.~\ref{fig:Mollow}, we present the spectra obtained from our QD, where the laser is either blue- or red-detuned and its power is scanned.
The scattered laser is filtered with a VBG.
For $\delta\omega_L/2\pi=\SI{131}{\GHz}$ (Fig.~\ref{fig:Mollow}a), a low-frequency side peak is clearly visible, and its separation increases with the laser field. 
The absence of any emission at $\omega_L+\Omega_\text{eff}(t)$, is also a strong indication toward the single-photon transition.
On the other hand, if $\delta\omega_L/2\pi=-\SI{131}{\GHz}$ (Fig.~\ref{fig:Mollow}b), we identify two side peaks at $\omega_L\pm\Omega_\text{eff}(t)$, which are much weaker in intensity than in Fig.~\ref{fig:Mollow}a.
Since at $T=\SI{4}{\K}$ the phonon occupation is low, the incoherent, phonon-assisted single-photon transition becomes significantly less efficient for a red-detuned laser pulse and symmetric two-photon scattering dominates.

Combining these findings, we conclude that the low-frequency side peak investigated in the Main Text for $\delta\omega_L>0$ is (almost exclusively) a result of phonon-assisted emission.

In addition to the laser power scan presented in the Main Text, Fig.~S\ref{fig:detscan} shows how the two-photon spectrum changes when $\delta\omega_L$ is varied at a constant power of $\SI{250}{\nW}$.
As expected from dressed-state emission (see Eq. 2 in the Main Text), a smaller laser--QD detuning induces a stronger level splitting, and thus a larger frequency shift of the \fst photon, that is, the side peak.

\section{\label{sec:freqfiltered} Frequency-filtered time traces }

\begin{figure}
	\begin{center}
		 \includegraphics[width=180mm]{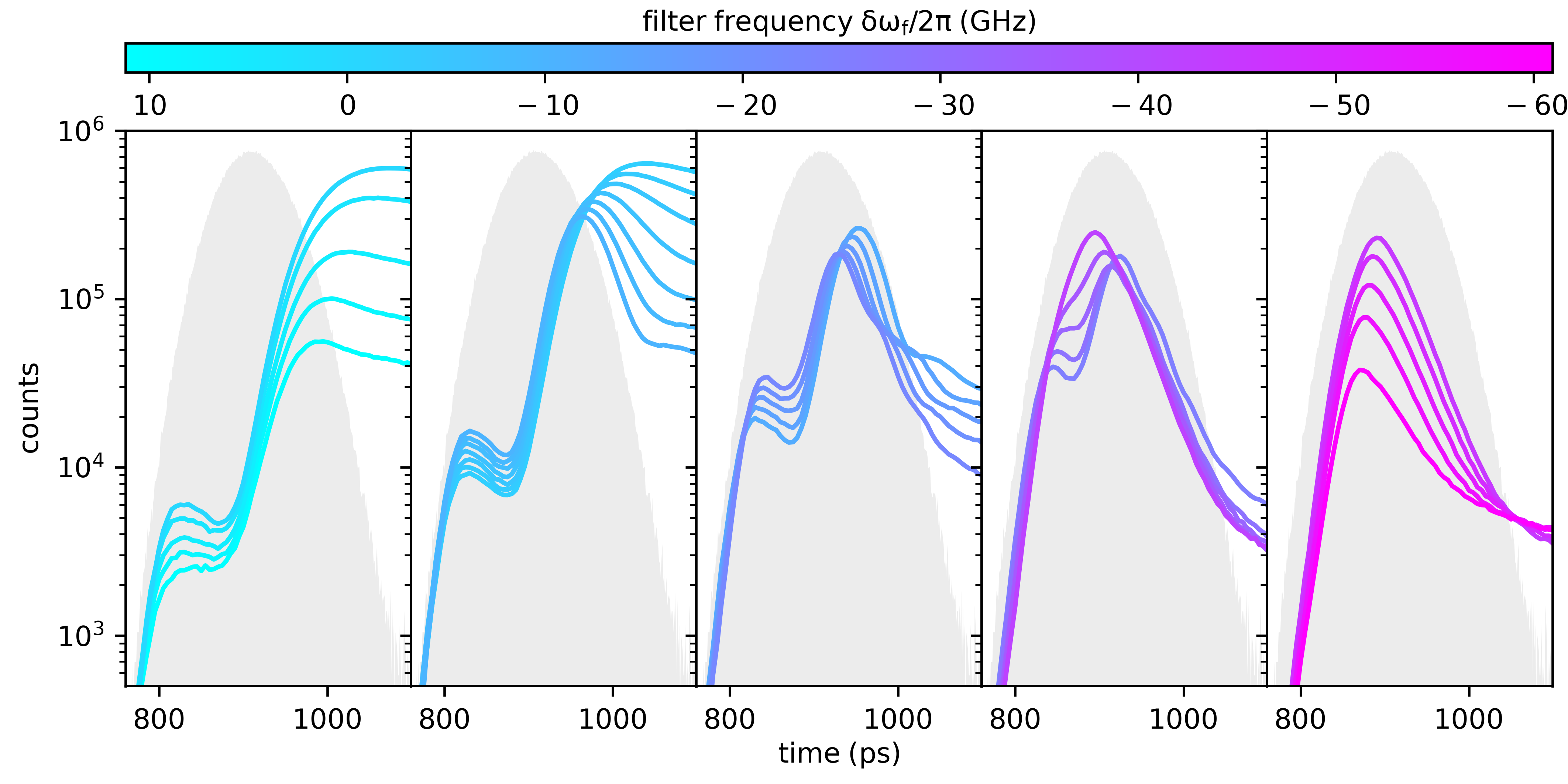}
		\caption{\textbf{Histogram of frequency filtered QD emission.}
        For $\SI{80\pm1}{\ps}$ long laser pulses with $\SI{250}{\nW}$ and $\delta\omega_L/2\pi=\SI{125}{\GHz}$, the QD emission is filtered using the set of scanning etalons shown in Fig.~\ref{fig:setup}.
        The filtered channel is correlated with the laser clock to construct the histograms for every filter position $\rm\delta\omega_f=\omega_f-\omega_0$. 
        The curves are colored according to $\rm\delta\omega_f$ as indicated by the color bar.
        Due to the Fourier-transform limit, the original temporal signal is exponentially broadened by the $\SI{5.3\pm0.3}{\GHz}$ wide Lorentzian filter.
        The errors were calculated using Poissonian statistics and are too small to be visible.
        }
		\label{fig:freqfiltered}
	\end{center}
\end{figure}

The dressed-state model suggests that the level splitting, and thus the emission frequency, changes throughout the laser pulse interaction, as captured in 
\begin{equation}
    \omega_\text{QD}(t) = \omega_L - \Omega_\text{eff}(t)\;,
\label{eq:emission_freq}
\end{equation}
where $\Omega_\text{eff}(t)=\sqrt{\Omega^2(t)+\delta\omega_L^2}$ is the effective Rabi frequency.
Using the setup depicted in Fig.~S\ref{fig:setup} without the fixed etalon but including the scanning etalons, we can measure the temporal response for different filter frequencies $\rm\omega_f$. 
Simply correlating the optical signal of the filtered HBT path with the laser clock yields a time trace for each specific filter frequency $\rm\delta\omega_f=\omega_f-\omega_0$ as depicted in Fig.~S\ref{fig:freqfiltered}.
Throughout the panels, the central frequency of the filter is shifted stepwise to larger detunings as indicated by the color bar. 

Initially, when the filter overlaps well with $\omega_0$, we find an early rise in signal counts at the very beginning of the laser pulse (indicated in gray) that plateaus first (or even drops slightly) and is then followed by a second much stronger signal increase and a slow exponential decay.
This behavior starts to change both qualitatively and quantitatively, when red-detuning the filter from $\omega_0$. 
The early rise lasts longer, reaching a higher signal level, and the following drop in counts in more pronounced. 
The second rising-edge peaks earlier, so that the maximum signal strength reduces and, more importantly, the falling edge becomes more Gaussian (see e.g. darkest curve in the second panel).
This trend continues as the red-shift enlarges and the signal drop after the first rise slowly washes out until there is only a single smooth rise and fall in detected counts (see curve with highest peak in the fourth panel).
Also note that the global peak height increases again throughout the fourth panel. 
In the last panel, the shape of the signal does not change anymore as it steadily declines in signal strength.

Recalling that a Gaussian laser pulse causes the QD transition frequency to first decrease until the pulse peaks and then increase again (see Eq.~\ref{eq:emission_freq}), the main features of our findings are readily explained.
For most filter positions, there are two points in time when they overlap with the time-varying emission frequency; once during the rising edge of the laser field and a second time when the field strength subsides.
Only the maximum frequency shift occurs for just a single point in time, namely when the laser pulse peaks.

This behavior is captured in Fig.~S\ref{fig:freqfiltered} where for all filter settings $\rm\delta\omega_f<\delta\omega_{max}$ (for these laser settings, $\delta\omega_\text{max}/2\pi\approx\SI{-45}{\GHz}$, see Fig.~S\ref{fig:detscan}), two peaks appear in the time trace.
The temporal separation of these peaks shrinks with increasing $\rm\delta\omega_f$ until they merge into one at $\rm\delta\omega_f=\delta\omega_{max}$.
Additionally, we clearly see that photons near $\omega_0$ exhibit an exponential decay as expected from spontaneous emission, whereas photons at significantly red-shifted frequencies must be emitted while the laser pulse is present and therefore inherit the temporal shape of the laser pulse.

To analyze and understand the process quantitatively, one must incorporate the time-dependent probability of the emitter being in the excited state. 
For phonon-assisted excitation, involving Non-Markovian processes, this is commonly done using numerical methods, including time--path integrals \cite{VCG:PRB2011}. 
However, some intuition about the excitation probability can be gained from considering the phonon spectral density $J(\omega)$ \cite{GH:AdvQTech2019, BLV:PRB2016}, which varies between materials and QD sizes but always features a single maximum.
In first approximation, the coupling efficiency of the exciton to the phonon bath can be estimated by evaluating $J[\Omega_\text{eff}(t)]$. 
Consequently, there is a Rabi frequency $\Omega_\text{eff}$ that maximizes the phonon coupling and overshooting causes less efficient phonon coupling, thus less efficient population inversion of the QD.
This non-monotonic evolution of the phonon coupling throughout the laser pulse can lead to strong phonon interactions at the beginning and end of a laser pulse but weak interaction at its peak field amplitude. 
The reduced phonon coupling at large effective Rabi frequencies was recently also shown to revive the Rabi oscillations of a coherently driven QD \cite{HBS:arxiv2024}.

\newpage
\bibliography{References}